\documentclass[tightenlines,twocolumn,superscriptaddress,preprintnumbers,amssymb,nofootinbib]{revtex4-1}
\usepackage{amsmath, amsthm, amssymb, mathrsfs, bm, mathtools} 
\usepackage{graphics}
\usepackage{epsfig}
\usepackage{graphicx}
\usepackage{wrapfig}
\usepackage{color}
\usepackage{tikz}
\usepackage[hidelinks,colorlinks=true,linkcolor=blue,citecolor=blue]{hyperref}

\RequirePackage{subfigure}

\newcommand{\beq}{\begin{equation}}
\newcommand{\eeq}{\end{equation}}
\newcommand{\bqa}{\begin{eqnarray}}
\newcommand{\eqa}{\end{eqnarray}}

\newcommand{\os}{\text{\tiny OS}}
\newcommand{\ms}{\overline{\text{\tiny MS}}}

\newcommand{\Tr}{{\mathrm{Tr}}}
\newcommand{\cep}{\text{\tiny CEP}}
\newcommand{\mev}{\text{MeV}}
\newcommand{\x}{\overline \sigma_x}
\newcommand{\y}{\overline \sigma_y}
\newcommand{\xms}{\overline \sigma_{x\ms}}
\newcommand{\yms}{\overline \sigma_{y\ms}}

\begin{document}

\title{Phase structure of the on-shell parametrized 2+1 flavor Polyakov quark-meson model}
\author{Suraj Kumar Rai}
\email{surajrai@allduniv.ac.in}
\affiliation{Department of Physics, University of Allahabad, Prayagraj, India-211002}
\affiliation{Department of Physics, Acharya Narendra Deo Kisan P.G. College, Babhnan Gonda, India-271313}
\author{Vivek Kumar Tiwari}
\email{vivekkrt@gmail.com}
\affiliation{Department of Physics, University of Allahabad, Prayagraj, India-211002}
\date{\today}

\begin{abstract}

    Augmenting the improved chiral effective potential of the on-shell renormalized 2+1 flavour quark-meson (RQM) model with the Polyakov-loop potential that accounts for the deconfinement transition,~we get the Quantum Chromodynamics (QCD) like framework of the renormalized Polyakov quark-meson (RPQM) model.~When the divergent quark one-loop vacuum term is included in the effective potential of the quark-meson (QM) model,~its tree level parameters or the parameters fixed by the use of meson curvature masses,~become inconsistent as the curvature masses involve the self energy evaluations at zero momentum.~Using the modified minimal subtraction method,~the consistent chiral effective potential for the RQM model has been calculated after relating the counterterms in the on-shell (OS) scheme to those in the $\overline{\text{MS}}$ scheme and finding the relations between the renormalized parameters of both the schemes where the physical (pole) masses of the $\pi, K, \eta$ and $\eta^{\prime}$ pseudo-scalar mesons and the scalar $\sigma$ meson,~the pion and kaon decay  constants,~have been put into the relation of the running couplings and mass parameter.~Using the RPQM model and the PQM Model with different forms for the Polyakov-loop potentials in the presence or the absence of the quark back-reaction,~we have computed and compared the effect of the consistent quark one-loop correction and the quark back-reaction on the scaled chiral order parameter,~the QCD phase diagrams and the different thermodynamic quantities.~The results have  been compared with the 2+1 flavor lattice QCD data from the Wuppertal-Budapest collaboration \{JHEP 09,73(2010); PLB 730,99(2014)\} and the HotQCD collaboration \{PRD 90,094503(2014)\}.
    
\end{abstract}
\keywords{ Dense QCD,
chiral transition,}
\maketitle
\section{Introduction}
The hadronic matter under the extreme conditions of high temperatures and/or densities,~gets dissolved into its quark and gluon constituents and the Quark Gluon Plasma (QGP) \cite{Cabibbo75,SveLer,Mull,Ortms,Riske} is formed as predicted by the strong interaction theory quantum chromodynamics (QCD).~The QCD phase diagram \cite{Cabibbo75} and the general properties of QGP are suject matter of intensive investigation for the  ultra-relativistic heavy ion collision experiments like the RHIC (BNL), LHC (CERN) and the upcoming CBM experiments at the FAIR facility (GSI-Darmstadt).~The first-principle lattice QCD simulations \cite{AliKhan:2001ek,Digal:01,Karsch:02,Fodor:03,Allton:05,Karsch:05,Aoki:06,Cheng:06,Cheng:08,JLange} give us important information and insights for the QCD phase transition that occurs on the temperature axis but when the baryon density is nonzero,~these calculations get severely hampered as the QCD action becomes complex due to the fermion sign problem \cite{Karsch:02}.~The QCD-like effective theory models \cite{Alf,Fukhat} built upon the symmetries of the QCD give us the much needed framework in which the QCD phase structure and its thermodynamics can be explored in great details.

The QCD Lagrangian has the global $SU_{\tiny{L+R}}(3) \times SU_{L-R}(3) $ symmetry for the three massless quarks.~The chiral (axial $A=L-R$) symmetry gets spontaneously broken in the low energy vacuum of the QCD
and one gets the non-strange and the strange chiral condensates as the order parameters with eight massless pseudoscalar bosons as Goldstone modes.~Since the small masses of the  $u$ and $d$ quarks cause a small explicit  
breaking while a relatively large mass of the $s$ quark generates a larger explicit breaking of the chiral symmetry,
the three pions are light while the four kaons and one eta are heavier in nature.~Due to the instanton effects,~the $U_A(1)$ axial symmetry also gets explicitly broken to the $Z_{A}(N_f)$ at the quantum level \cite{tHooft:76prl}.~The $\eta'$ meson does not remain a massless Goldstone boson even when the quarks are massless as it acquires a mass of about 1 GeV due to the $U_A(1)$ axial anomaly.~Coupling the nine scalar and nine pseudo-scalar mesons of the three flavor linear sigma model \cite{Rischke:00} with the two light quarks $u, d$  and the one heavier s quark,~one gets  the effective theory framework of the 2+1 flavor quark-meson (QM) model \cite{Schaefer:09}.

Several investigations of the QCD phase structure, have already been done in the chiral models \cite{Rischke:00,Roder,fuku11,grahl,jakobi,Herpay:05,Herpay:06,Herpay:07,Kovacs:2006ym,kahara,Bowman:2008kc,Fejos,Jakovac:2010uy,koch,marko}, two and three flavor QM model \cite{scav, mocsy,bj,Schaefer:2006ds,Schaefer:09}.~When the effect of the Dirac sea gets neglected in the standard mean field approximation (s-MFA),~the QM model studies look inconsistent because in the chiral limit,~the chiral phase transition at zero baryon density turns first-order which is ruled out by the general theoretical arguments \cite{rob,hjss}.~The inclusion of the quark one-loop vacuum fluctuation in the QM model \cite{vac},~removes the above
inconsistency.~In several of the QCD phase structure studies carried out in the ambit of the quark-meson model with the quark one-loop vacuum term \cite{lars,guptiw,schafwag12,chatmoh1,vkkr12,TranAnd,chatmoh2,vkkt13,Herbst,Weyrich,kovacs,zacchi1,zacchi2,Rai},~the model parameters are fixed by using the curvature masses of the mesons while the pion and kaon decay constant are fixed by the vacuum expectation value of the non-strange and strange chiral condensates.~This parameter fixing turns out to be inconsistent because the effective potential generates the n-point functions of the theory at vanishing external momenta and the curvature masses are defined by the self-energy evaluations at zero momentum \cite{laine,Adhiand1,BubaCar,Naylor,fix1}.

~The tree level relations between the physical quantities and the parameters of the Lagrangian,~are changed by the radiative corrections to the physical quantities.~Hence the use of the tree level parameters in the effective potential becomes inconsistent.~The running parameters in the $\overline{\text{MS}}$ scheme have the renormalization scale $\Lambda$ dependence while the on-shell parameters have their tree-level values.~Following the correct renormalization prescription,~when the counterterms calculated in the $\overline{\text{MS}}$ scheme are put into the relations of the counterterms in the on-shell scheme,~one finds the relations between the different renormalized parameters of the two schemes.~The relations between the on-shell parameters (physical quantities) and the running $\overline{\text{MS}}$ parameters are used as input when the effective potential is calculated using the modified minimal subtraction procedure \cite{Adhiand1}.~Adhikari and collaborators \cite{Adhiand1,Adhiand2,Adhiand3,asmuAnd} in a series of papers,~used the above renormalization prescription for including the quark one-loop vacuum correction in the two flavor QM model which uses the $O(4)$ sigma model with the $\sigma$ and $\vec{\pi}$ mesons.~In a recent QCD phase structure study,~we implemented \cite{RaiTiw} the on-shell parameter fixing method in the QM model where the two quark flavors are coupled to the  eight mesons of the $ SU_{L}(2) \times SU_{R}(2) $ linear sigma model.~Here,~we have calculated the consistent chiral effective potential for the on-shell renormalized 2+1 flavor quark-meson (RQM) model.

The introduction of the Polyakov-loop implements the physics of quark confinement inside the hadrons as the 
QCD confinement is mimicked in a statistical sense when the chiral models are coupled to a  constant background $SU(N_{c})$ gauge field $A_{\mu}^{a}$ \cite{SveLer,Polyakov:78plb,benji,BankUka,Pisarski:00prd,fuku,Vkt:06}.~Using the phenomenological Polyakov-loop potential \cite{ratti,Roesnr,fuku2},~when the free energy density from the gluons is added to the QM model,~it becomes the PQM model \cite{SchaPQM2F,SchaPQM3F,Mao,TiPQM3F}.~In the present work,~combining the Polyakov-loop potential with the consistent chiral effective potential calculated for the 2+1 flavor RQM Model,~we get the QCD like framework of the renormalized Polyakov quark-meson model (RPQM).~We have considered different parametrizations for the Polyakov-loop potentials along with the recent improvement in the form of the Polyakov-loop potential from a pure gauge potential to a unquenched glue potential in which the quark back-reaction effects are included \cite{Haas,Redlo,TkHerbst,BielichP}.~It is important to explore how the QCD phase structure computed in the present RPQM model framework,~gets modified by the unquenching of the Polyakov-loop potential because it leads to the linkage of the chiral and deconfinement phase transitions also at small temperatures and large chemical potentials \cite{BielichP}.~The above feature is also noticed in the studies having the functional renormalization (FRG) improvement of the PQM model when the Yang-Mills Polyakov-loop potential is used \cite{Herbst,THerbst2}.~In order to know the effect of the consistent quark one-loop vacuum correction,~we have calculated the QCD phase diagrams and the QCD thermodynamics in the RPQM model and compared the results with that of the PQM model.~Taking different forms of the Polyakov-loop potentials with and without quark back-reaction,~we have computed and compared how the phase structure gets influenced by the different implementations of the confinement-deconfinement physics in the RPQM model.

The organization of the paper is as follows.~Section~\ref{sec:II} presents a brief formulation of the $ SU_{L}(3) \times SU_{R}(3) $ PQM model.~The different forms of the Polyakov-loop potentials are described in the sub-section~\ref{sec:IIA},~while the sub-section~\ref{sec:IIB} gives the thermodynamic grand potential in the PQM model and the parameters of the PQM model are determined in the sub-section~\ref{sec:IIC}.~Section~\ref{sec:III} presents the technical details of the on-shell renormalization of the parameters in the renormalized Polyakov quark-meson model.~Sub-section~\ref{sec:IIIA} gives the description of the counterterms of the Lagrangian and the self-energies of the mesons,~sub-section~\ref{sec:IIIB} presents the renormalization of the model parameters and the effective potential of the RPQM model has been derived in the sub-section~\ref{sec:IIIC}.~Results and discussion are presented in the section~\ref{sec:IV}.~The comparisons of the thermodynamic quantities with the lattice QCD data,~have been discussed in the sub-section~\ref{sec:IVA} and the sub-section~\ref{sec:IVB} illustrates and compares the phase diagrams and their CEP position in different model scenarios.~Section~\ref{sec:V} provides the summary and conclusion.~The Appendix~\ref{appenA} contains the important integrals and expressions of the factors.    

\section{Model Formulation}
\label{sec:II}
In the model \cite{TiPQM3F,Schaefer:09wspax,Mao,SchaPQM3F},~three flavor of quarks are coupled to the $SU_V(3) \times SU_A(3)$ symmetric mesonic fields together with 
temporal component of gauge field represented by the Polyakov loop 
potential.~Thermal expectation value of color trace of Wilson loop in temporal 
direction defines the Polyakov loop field $\Phi$ as
\begin{equation}
\Phi(\vec{x}) = \frac{1}{N_c} \langle \Tr_c L(\vec{x})\rangle, \qquad \qquad  \bar\Phi(\vec{x}) =
\frac{1}{N_c} \langle \Tr_c L^{\dagger}(\vec{x}) \rangle
\end{equation}
where $L(\vec{x})$ is a matrix in the fundamental representation of the 
$SU_c(3)$ color gauge group.
\begin{equation}
\label{eq:Ploop}
L(\vec{x})=\mathcal{P}\mathrm{exp}\left[i\int_0^{\beta}d \tau
A_0(\vec{x},\tau)\right]
\end{equation}
Here $\mathcal{P}$ is path ordering,  $A_0$ is the temporal component of vector 
field and $\beta = T^{-1}$ \cite{Polyakov:78plb}.~As in the 
Ref. \cite{ratti,Roesnr},~the homogeneous Polyakov loop fields $\Phi(\vec{x})=\Phi$=constant and
$\bar\Phi(\vec{x})=\bar\Phi$=constant.
 
The Polyakov quark-meson (PQM) model Lagrangian is written in terms of the quarks, mesons, their couplings and the Polyakov loop potential ${\cal U} \left( \Phi, \bar\Phi, T \right)$  as
\bqa
{\cal L_{PQM}}&=&{\cal L_{QM}}-{\cal U} \big( \Phi , \bar\Phi , T \big),
\label{lag:PQM}
\eqa
\bqa
{\cal L_{QM}}&=&\bar{\psi}[i\gamma^\mu D_\mu- g\; T_a\big( \sigma_a 
+ i\gamma_5 \pi_a\big) ] \psi+\cal{L(M)},
\label{lag}
\eqa
here $\psi$ is a four-component Dirac spinor,~a flavor triplet and a color $N_c$-plet quark field. 
\bqa
\psi&=&
\left(
\begin{array}{c}
u\\
d\\
s
\end{array}\right)\;.
\eqa
The three flavor of quarks are coupled  to the nine scalar ($\sigma_a, J^{P}=0^{+}$) and nine pseudo-scalar ($\pi_a, J^{P}=0^{-}$) mesons by the flavor blind Yukawa coupling $g$.~Quarks couple with the uniform temporal background gauge field as the following $D_{\mu} = \partial_{\mu} -i A_{\mu}$ 
and  $A_{\mu} = \delta_{\mu 0} A_0$ (Polyakov gauge), where $A_{\mu} = g_s A^{a}_{\mu} \lambda^{a}/2$ with vector potential $A^{a}_{\mu}$ for color gauge field. $g_s$ is the $SU_c(3)$ gauge coupling.


The Lagrangian for the meson fields is written as \cite{Schaefer:09,Roder,TiPQM3F} 
\bqa
\nonumber
\label{lag11}
\hspace{-1.5 cm}\cal{L(M)}&=&\text{Tr} (\partial_\mu {\cal{M}}^{\dagger}\partial^\mu {\cal{M}}-m^{2}({\cal{M}}^{\dagger}{\cal{M}}))\\
\nonumber
&&-\lambda_1\left[\text{Tr}({\cal{M}}^{\dagger}{\cal{M}})\right]^2-\lambda_2\text{Tr}({\cal{M}}^{\dagger}{\cal{M}})^2\\
&&+c[\text{det}{\cal{M}}+\text{det}{\cal{M}}^\dagger]+\text{Tr}\left[H({\cal{M}}+{\cal{M}}^\dagger)\right]\;,
\label{lag1}
\eqa
the field ${\cal{M}}$ is a $3\times3$ complex matrix containing the nine scalars $\sigma_a$ and the nine pseudo-scalar $\pi_a$ mesons.
\bqa
{\cal{M}}&=&T_a\xi_a=T_a(\sigma_a+i\pi_a)\;.
\eqa
The $T_a$ in the above are 9 generators of $U(3)$ with $T_a = \frac{\lambda_a}{2}$ where $a=0,1 ,\dots ~,8$. The $\lambda_a$ are standard Gell-Mann matrices with $\lambda_0=\sqrt{\frac{2}{3}}\  {\mathbb I}_{3\times3}$. 
The generators follow the $U(3)$ algebra $\left[T_a, T_b\right]  = if_{abc}T_c$ and 
$\left\lbrace T_a, T_b\right\rbrace  = d_{abc}T_c$ where $f_{abc}$ and $d_{abc}$ are the standard 
antisymmetric and symmetric structure constants respectively with $f_{ab0}=0$ and 
$d_{ab0}=\sqrt{\frac{2}{3}}\ \delta_{ab}$ and matrices are normalized as 
$\text{Tr}(T_a T_b)=\frac{\delta_{ab}}{2}$.~The  $SU_L(3) \times SU_R(3)$ chiral symmetry is explicitly broken by  the following term
\bqa
H = T_a h_a\;.
\eqa
$H$ is a $3 \times 3$ matrix with 9 parameters.~Due to the spontaneous chiral symmetry breaking,~the field $\xi$ takes the nonzero vacuum expectation value, $\bar{\xi}$.~Since $\bar{\xi}$ must have the quantum numbers of the vacuum,~only three nonzero parameters $h_0$, $h_3$ and $h_8$ can cause the explicit breakdown of the chiral symmetry.~Neglecting isospin symmetry breaking,~we have taken $h_0$, $h_8  \neq 0$.~One gets the $2+1$ flavor symmetry breaking scenario having the two nonzero condensates $\bar{\sigma_0}$ and $\bar{\sigma_8}$.~In addition to the $h_0$ and $h_8$,~the five other parameters of the model at tree-level are the quartic coupling constants $\lambda_1$ and $\lambda_2$,~squared mass parameter  $m^2$,~ a Yukawa coupling $g$ and the coefficient of the t'Hooft determinant term $c$ that models the $U_A(1)$ axial anomaly of the QCD vacuum.

\subsection{Polyakov-loop potentials}
\label{sec:IIA}
Different forms of the Polyakov-loop effective potential $\mathcal{U}(\Phi,\bar{\Phi},T)$,~have been used in the literature to study the deconfinement phase transition.~One constructs its simplest form by finding a potential which respects all the given symmetries and accounts for the spontaneously broken $Z(3)$ symmetry for the system in the deconfined phase \cite{SveLer,benji,BankUka}.~The following polynomial form gives the minimal content of the Polyakov-loop potential

\bqa
\label{plykov_poly}
\hspace{-0.5 cm}\frac{\mathcal{U_{\rm Poly}}}{T^4}&=&-\frac{b_2(T)}{2}\Phi\bar{\Phi}-\frac{b_3}{6}(\Phi^3+\bar{\Phi}^3)+\frac{b_4}{4}(\Phi\bar{\Phi})^2\;,
\eqa
the coefficients of the Eq.~(\ref{plykov_poly}) are given by
\bqa
b_2(T)=a_0+a_1\left(\frac{T_0}{T}\right)+a_2\left(\frac{T_0}{T}\right)^2+a_3\left(\frac{T_0}{T}\right)^3\;,
\eqa
where $a_0=6.75$, $a_1=-1.95$, $a_2=2.625$, $a_3=-7.44$, $b_3=0.75$ and $b_4=7.5$ .\\

The Polyakov-loop potential in the above is improved by adding the contribution coming from  the integration of the $SU(3)$ group volume in the generating functional for the Euclidean action.~The Haar measure is used to perform this integration which takes the form of a Jacobian determinant.~Its logarithm is added as an effective potential to the action in the generating functional.~The positive coefficient of the logarithm term bounds the potential from below for large $\Phi$ and $\bar{\Phi}$ and the logarithmic form of the Polyakov-loop potential is written as \cite{fuku,Roesnr}:
\bqa
\label{plykov_log}
\nonumber
\hspace{-0.5 cm}\frac{\mathcal{U_{\rm Log}}}{T^4}&=&b(T)\ln[1-6\Phi\bar{\Phi}+4(\Phi^3+\bar{\Phi}^3)-3(\Phi\bar{\Phi})^2]\;\\&&-\frac{1}{2}a(T)\Phi\bar{\Phi}\;.
\eqa
The parameters of the polynomial and log form of the Polyakov-loop potential were determined \cite{ratti,Roesnr} by fitting the lattice data for pressure, entropy density  as well as energy density and the evolution of Polyakov-loop $<\Phi>$ on the lattice in pure gauge theory.~The coefficients of the Eq.~(\ref{plykov_log}) are the following \cite{Roesnr},
\bqa
&&a(T)=a_0+a_1\left(\frac{T_0}{T}\right)+a_2\left(\frac{T_0}{T}\right)^2\;,\\
&&b(T)=b_3\left(\frac{T_0}{T}\right)^3\;,
\eqa
where $a_0=3.51$, $a_1=-2.47$, $a_2=15.2$, $b_3=-1.75$. 
One should note that the log potential has qualitative consistency with the leading order result of the strong-coupling expansion \cite{JLange}.~Also,~because the potential diverges for $\Phi$,~$\bar{\Phi} \longrightarrow$  1, the Polyakov-loop always remains smaller than 1 and approaches this value asymptotically as $T \longrightarrow \infty$.

Accounting for the Polyakov-loop fluctuations,~the new Polyakov-loop effective potential was  constructed  in the Ref.~\cite{Redlo} where the parameters get so adjusted that in addition to the other existing lattice data,~the longitudinal as well as the transverse susceptibilities are also reproduced.~After the addition of the logarithmic term to the polynomial form of the Polyakov-loop potential,~the new expression of the PolyLog Polyakov-loop potential in the above work has been found as the following.
\bqa
\label{plykov_polylog}
\nonumber
\hspace{-0.5 cm}\frac{\mathcal{U_{\rm PolyLog}}}{T^4}&=&b(T)\ln[1-6\Phi\bar{\Phi}+4(\Phi^3+\bar{\Phi}^3)-3(\Phi\bar{\Phi})^2]\;\\ \nonumber
&&+a_2(T)\Phi\bar{\Phi}+a_3(T)(\Phi^3+\bar{\Phi}^3)+a_4(T)(\Phi\bar{\Phi})^2.\\
\eqa
The coefficients of the Eq.~(\ref{plykov_polylog}) PolyLog parametrization are defined as
\bqa
&&a_i(T)=\frac{a^{(i)}_0+a^{(i)}_1\left(\frac{T_0}{T}\right)+a^{(i)}_2\left(\frac{T_0}{T}\right)^2}{1+a^{(i)}_3\left(\frac{T_0}{T}\right)+a_4^{(i)}\left(\frac{T_0}{T}\right)^2}\;\\
&&b(T)=b_0\left(\frac{T_0}{T}\right)^{b_1}\left[1-e^{b_2\left(\frac{T_0}{T}\right)^{b_3}}\right].
\eqa
\\
The parameters are summarized in the Table~\ref{tab:plglg}.
\begin{table}[!htbp]
    \caption{Parameters of the PolyLog Polyakov-loop potential have been taken from the Ref.~\cite{Redlo}.}
    \label{tab:plglg}
    \resizebox{0.48\textwidth}{!}{
    \begin{tabular}{p{1.5cm} p{1.5cm} p{1.5cm} p{1.5cm} p{1.5cm} p{1.5cm} p{0.5 cm}}
      \toprule 
      PolyLog& $a^{(2)}_0$ & $a^{(2)}_1$ & $a^{(2)}_2$ & $a^{(2)}_3$ & $a^{(2)}_4$&\\
      & 22.07 & -75.7 & 45.03385 & 2.77173 & 3.56403&\\
      & $a^{(3)}_0$ & $a^{(3)}_1$ & $a^{(3)}_2$ & $a^{(3)}_3$ & $a^{(3)}_4$&\\
      &-25.39805&57.019&-44.7298&3.08718&6.72812\\
      & $a^{(4)}_0$ & $a^{(4)}_1$ & $a^{(4)}_2$ & $a^{(4)}_3$ & $a^{(4)}_4$&\\
      &27.0885&-56.0859&71.2225&2.9715&6.61433&\\
      &$b_0$&$b_1$&$b_2$&$b_3$& &\\
      &-0.32665&5.8559&-82.9823&3.0& &\\
      \hline      
      \hline 
    \end{tabular}}
\end{table}

For the pure gauge Yang-Mills theory,~the deconfinement phase transition is first order  and $T_c^{\rm YM}=T_0=270$ MeV.~When the dynamical quarks are present,~the first order transition becomes a crossover.~The parameter $T_0$ depends on the number of quark flavors and chemical potential in the full dynamical QCD~\cite{SchaPQM2F,Haas,kovacs,BielichP,THerbst2} as it is linked to the mass-scale $\Lambda_{\rm QCD}$ which gets modified by the effect of the fermionic matter fields.~The number of flavor and the chemical potential dependence of $T_0 \longrightarrow T_0(N_f,\mu)$ is written as,
\bqa
\label{t0_mu}
T_0(N_f,\mu)=\hat{T} \ e^{-1/(\alpha_0 b(N_f,\mu))}\;,
\eqa
with 
\bqa
b(N_f,\mu)=\frac{1}{6\pi}(11N_c-2N_f)-b_\mu\frac{\mu^2}{(\hat{\gamma}\hat{T})^2}.
\eqa
where the parameter $\hat{T}$ is fixed at the scale $\tau$,  $\hat{T}=T_\tau=1.77$ GeV and $\alpha_0=\alpha(\Lambda)$ at a UV scale $\Lambda$. The $T_0(N_f=0)$ = 270 MeV gives  $\alpha_0$ = 0.304 and $b_\mu\simeq\frac{16}{\pi}N_f$. The parameter $\hat{\gamma}$ governs the curvature of $T_0(\mu)$ with the systematic error estimation range $0.7\lesssim\hat{\gamma}\lesssim1$ \cite{SchaPQM2F,THerbst2}.~Massive flavors lead to suppression factors of the order $T_0^2/(T^2_0 +m^2)$ in the $\beta$-function.~For 2+1 flavors and a current strange quark mass $m_s\sim$ 150 MeV,~one obtains $T_0(2+ 1)$=187 MeV\cite{SchaPQM2F,THerbst2}.


When the back-reaction of the quarks is accounted for,~in the full QCD with dynamical quarks,~the Polyakov-loop potential gets replaced by the QCD glue potential.~The Ref.~\cite{Haas} applied the FRG equations to the QCD and compared the pure gauge potential $\mathcal{U}_{\rm YM}$ to the ``glue'' potential $\mathcal{U}_{\rm glue}$ where quark polarization was included in the gluon propagator and they found significant differences between the two potentials.~However,~it was observed that the two potentials are of the same shape and they can be mapped into each other by relating the temperatures of the two systems,~$T_{\rm YM}$ and $T_{\rm glue}$.~Denoting the previous equations of the Polyakov-loop potential by $\mathcal{U}_{\rm YM}$,~the improved Polyakov-loop potential $\mathcal{U}_{\rm glue}$ can be constructed as \cite{Haas}     

\bqa
\frac{\mathcal{U}_{\rm glue}}{T^4_{\rm glue}}(\Phi, \bar{\Phi}, T_{\rm glue})&=&\frac{\mathcal{U}_{\rm YM}}{T^4_{\rm YM}}(\Phi, \bar{\Phi}, T_{\rm YM})
\eqa
here the temperature $T_{\rm glue}$ is related to $T_{\rm YM}$ as 
\bqa
\frac{T_{\rm YM}-T^{\rm YM}_{\rm c}}{T^{\rm YM}_{\rm c}}=0.57 \
\frac{T_{\rm glue}-T^{\rm glue}_{\rm c}}{T^{\rm glue}_{\rm c}}
\eqa
The transition temperature for the unquenched case is the $T^{\rm glue}_{\rm c}$.~The coefficient 0.57 comes from the comparison of the two effective potentials.~$T^{\rm glue}_{\rm c}$ lies within a range $T^{\rm glue}_{\rm c} \in [180,270]$.~In practice,~one  uses
the replacement $T  \longrightarrow T^{\rm YM}_{\rm c}(1+0.57(\frac{T}{T^{\rm glue}_{\rm c}}-1))$ in the right-hand side of the Polyakov-loop potentials where $T_0$ means $T^{\rm YM}_{\rm c}$ and ($T\sim T_{\rm YM}$) on the left side of the arrow while ($T\sim T_{\rm glue}$) on the right side.~In our calculations,~we have taken $T^{\rm glue}_{\rm c}$ and $T_0$ both fixed at 187 MeV for the 2+1 quark flavor as in the Ref.~\cite{SchaPQM2F,Haas}.


\subsection{Grand Potential in the Mean Field Approach}
\label{sec:IIB}
We have the spatially uniform system that is in the thermal equilibrium at temperature $T$ and quark chemical potential $\mu_{f} (f=u,d,s)$.~The partition function is calculated by performing the path integral over the quark/antiquark and meson fields \cite{Schaefer:09,TiPQM3F}
\bqa
\nonumber
\mathcal{Z}&=& \mathrm{Tr\, exp}[-\beta (\hat{\mathcal{H}}-\sum_{f=u,d,s} 
\mu_{f} \hat{\mathcal{N}}_{f})]  \\ \nonumber
&=& \int\prod_a \mathcal{D} \sigma_a \mathcal{D} \pi_a \int
\mathcal{D}\psi \mathcal{D} \bar{\psi} \; \mathrm{exp} \bigg[- \int_0^{\beta}d\tau\int_Vd^3x   \\ \label{eq:partf}
&& \bigg(\mathcal{L_{QM}^{E}} 
 + \sum_{f=u,d,s} \mu_{f} \bar{\psi}_{f} \gamma^0 \psi_{f} \bigg) \bigg]\;. 
\eqa
Here $\beta= \frac{1}{T}$ and the volume of the system in the three dimension is $V$.~There will be three chemical potentials for the three flavor of quarks and the $SU_V(2)$ symmetry is assumed to be preserved as the small difference in the mass of $u$ and $d$ quark gets neglected.~Hence the quark chemical potential for the $u$ and $d$ quarks is equal $\mu_u = \mu_d$ and the strange quark chemical potential is $\mu_s$.  
   
In the standard mean-field approximation \cite{scav,Schaefer:09,TiPQM3F}, the partition function is calculated by replacing the meson fields with their vacuum expectation values
$\langle M \rangle =  T_0 \bar{\sigma_0} + T_8 \bar{\sigma_8}$ and neglecting the thermal as well as quantum fluctuations of the meson fields while retaining the quarks and antiquarks as quantum fields.~Now following the standard procedure as given in Refs.~\cite{Kapusta_Gale,SchaPQM2F,ratti,fuku},~one can obtain the expression of grand potential as the sum of pure gauge field contribution ${\cal U} \left(\Phi, \bar\Phi, T \right)$, 
meson contribution and quark/antiquark contribution evaluated in 
the presence of Polyakov loop, 
\bqa
\label{eq:grandp}
\nonumber
\Omega_{\rm MF}(T,\mu)&=&U(\overline{\sigma_0},\overline{\sigma_8})+\Omega_{q\bar{q}} (T,\mu;\overline{\sigma_0},\overline{\sigma_8},\Phi,\bar{\Phi})\;\\
&&+\mathcal{U}(T,\Phi,\bar{\Phi})\;.
\eqa 
The 2 + 1 flavor case can be explored after implementing the basis transformation of the condensates and the external fields from the original singlet octet (0, 8) basis to the nonstrange strange basis ($\x$, $\y$)
\bqa
\x &=&
\sqrt{\frac{2}{3}}\bar{\sigma}_0 +\frac{1}{\sqrt{3}}\bar{\sigma}_8, \\
\y &=&
\frac{1}{\sqrt{3}}\bar{\sigma}_0-\sqrt{\frac{2}{3}}\bar{\sigma}_8.
\eqa
The grand potential is written in $\x$, $\y$ basis as,
\bqa
\nonumber
\Omega_{\rm MF }(T,\mu) &=&U(\x,\y)+\mathcal{U}(T,\Phi,\bar{\Phi})\;\\
&&+\Omega_{q\bar{q}} (T,\mu;\x,\y,\Phi,\bar{\Phi})\;.
\label{Grandpxy}
\eqa
The external fields ($h_x$, $h_y$) can be written in terms of the ($h_0$, $h_8$) by similar expressions. Since the nonstrange and strange quark/antiquark decouple, the quark masses are written as,
\bqa
\label{mums}
m_{u} = g \frac{\x}{2}, \qquad m_{s} = g \frac{\y}{\sqrt{2}}\;.
\eqa 
One writes the tree level effective potential in the nonstrange-strange basis as the following.
\bqa
\label{eq:mesop}
\nonumber
 U(\x,\y) & = &\frac{m^{2}}{2}\left(\x^{2} +
  \y^{2}\right) -h_{x} \x -h_{y} \y
 - \frac{c}{2 \sqrt{2}} \x^2 \y \\ \nonumber 
 && + \frac{\lambda_{1}}{2} \x^{2} \y^{2}+
  \frac{1}{8}\left(2 \lambda_{1} +
    \lambda_{2}\right)\x^{4} \\
 && +\frac{1}{8}\left(2 \lambda_{1} +
    2\lambda_{2}\right) \y^{4}\ .
\eqa    

Applying the stationarity conditions $ \frac{\partial U(\x,\y)}{\partial \x}=0=\frac{\partial U(\x,\y)}{\partial \y}$  to the effective potential (\ref{eq:mesop}),~one gets 
\bqa
\label{hxhy}
h_{x}&=&\overline \sigma_x \ m^2_{\pi}\, \\
h_{y}&=&{\biggl\{\frac{\sqrt{2}}{2}(m_{K}^2-m_{\pi}^2) \overline  \sigma_x+m_{K}^2 \overline \sigma_y\biggr\}}\;. \quad
\eqa
The tree level curvature masses of the pions, kaons and other mesons in the QM model are given by the mass matrix $(m_{\alpha,ab})^2$ evaluated in Ref.~\cite{Rischke:00,Schaefer:09}. Here $\alpha=$ s, p; ``s'' stands for the scalar and ``p'' stands for the pseudo-scalar mesons and $a,b=0,1,2,\cdots,8$. In the scalar sector, the $a_{0}$ meson mass is given by the 11 element (degenerate with the 22 and 33 elements) and the $\kappa$ meson mass is given by the 44 element (degenerate with the 55, 66 and 77 elements). The $\sigma$ and $f_{0}$ meson masses are found by diagonalizing the (00)-(88) sector of the scalar mass matrix. In exactly analogous manner for the pseudoscalar sector $m^2_{\text{p},11}=m^2_{\text{p},22}=m^2_{\text{p},33}\equiv m^2_{\pi}$ and $m^2_{\text{p},44}=m^2_{\text{p},55}=m^2_{\text{p},66}=m^2_{\text{p},77} \equiv m^2_{K}$. Diagonalization of the pseudoscalar (00)-(88) sector of the  mass matrix gives us the masses of the physical $\eta$ and $\eta^{\prime} $ mesons.~The Table~\ref{tab:table1} contains  masses of all the mesons.

\begin{table*}[!htbp]
    \caption{Meson masses calculated from the second derivative of the grand potential at its minimum as given in Ref.\ \cite{Schaefer:09,Herpay:06}} 
    \label{tab:table1}
    \resizebox{1.0\hsize}{!}{
    \begin{tabular}{p{0.08\textwidth} p{0.47\textwidth} p{0.08\textwidth} p{0.47\textwidth}}
             \\ 
             \hline
            & Scalar meson masses \ \ \ \ \ \ \ \ \ \ \ &             & Pseudo-scalar meson masses  \\
      \hline 
      $(m_{a_{0}})^2$ & $m^2 +\lambda_1(\x^2+\y^2)+\frac{3\lambda_2}{2} \x^2+
      \frac{\sqrt{2}c}{2}\y $&$ (m_{\pi})^{2}$ & $m^2 + \lambda_1 (\x^2 + \y^2) +\frac{\lambda_2}{2} \x^2 -\frac{\sqrt{2} c}{2} \y$\\
      $(m_{\kappa})^{2}$ & $m^2+\lambda_1(\x^2+\y^2)+\frac{\lambda_{2}}{2}(\x^2+\sqrt{2}\x\y+2\y^2)+\frac{c}{2}\x$&$(m_{K})^{2}$ &$m^2 + \lambda_1 (\x^2 + \y^2) +\frac{\lambda_2}{2} (\x^2 - \sqrt{2} \x \y +2 \y^2) - \frac{c}{2} \x$\\
      $(m_{s,00})^2$ &  $m^2+\frac{\lambda_1}{3}(7\x^2+4\sqrt{2}\x\y+5\y^2)+\lambda_2(\x^2 + \y^2)-\frac{\sqrt{2}c}{3} (\sqrt{2} \x +\y)$&$(m_{p,00})^{2}$ & $m^2 + \lambda_1(\x^2 +\y^2) + \frac{\lambda_2}{3}(\x^2 +\y^2) + \frac{c}{3} (2\x + \sqrt{2} \y)$\\
      $(m_{s,88})^{2}$ & $m^2 +\frac{\lambda_1}{3}(5\x^2-4\sqrt{2}\x\y +7\y^2)+\lambda_2(\frac{\x^2}{2} +2\y^2)+\frac{\sqrt{2}c}{3} (\sqrt{2}\x-\frac{\y}{2})$&$(m_{p,88})^{2}$ & $m^2 +\lambda_1(\x^2 +\y^2) +\frac{\lambda_2}{6}(\x^2 +4\y^2)-\frac{c}{6}(4\x -\sqrt{2}\y)$\\
   $(m_{s,08})^{2}$ &  $\frac{2\lambda_1}{3}(\sqrt{2}\x^2 -\x\y -\sqrt{2}\y^2) +\sqrt{2}\lambda_2(\frac{\x^2}{2}-\y^2) +\frac{c}{3\sqrt{2}}(\x- \sqrt{2}\y)$&$(m_{p,08})^{2}$&$\frac{\sqrt{2}\lambda_2}{6}(\x^2-2\y^2)-\frac{c}{6}(\sqrt{2}\x -2\y)$\\
   $(m_{s,xx})^{2}$&$ m^2+3(\lambda_1+\frac{\lambda_2}{2})\x^2+\lambda_1 \y^2-\frac{c}{\sqrt{2}}\y  $ &$(m_{p,xx})^{2}$&$  m^2+(\lambda_1+\frac{\lambda_2}{2})\x^2+\lambda_1 \y^2+\frac{c}{\sqrt{2}}\y $ \\
   $(m_{s,yy})^{2}$&$ m^2+\lambda_1 \x^2+3(\lambda_1+\lambda_2)\y^2  $ &$(m_{p,yy})^{2}$&$m^2+\lambda_1 \x^2+(\lambda_1+\lambda_2)\y^2 $ \\
   $(m_{s,xy})^{2}$&$ 2\lambda_1 \x\y-\frac{c}{\sqrt{2}}\x  $ &$(m_{p,xy})^{2}$&$\frac{c}{\sqrt{2}}\x $ \\ 
   $m_{\sigma}^2$& $\frac{1}{2}(m_{s,00}^2+m_{s,88}^2)-\frac{1}{2} \sqrt{(m_{s,00}^2-m_{s,88}^2)^2+4m_{s,00}^4 }$
   & $m_{\eta}^2$ & $\frac{1}{2}(m_{p,00}^2+m_{p,88}^2)-\frac{1}{2} \sqrt{(m_{p,00}^2-m_{p,88}^2)^2+4m_{p,00}^4} $ \\ 
    $m_{f_{0}}^2$ & $\frac{1}{2}(m_{s,00}^2+m_{s,88}^2)+\frac{1}{2} \sqrt{(m_{s,00}^2-m_{s,88}^2)^2+4m_{s,00}^4}$
   &$m_{\eta^{\prime}}^2$ &$\frac{1}{2}(m_{p,00}^2+m_{p,88}^2)+\frac{1}{2} \sqrt{(m_{p,00}^2-m_{p,88}^2)^2 \ +4m_{p,00}^4 } $ \\
   \hline 
\end{tabular}}
\end{table*}

The quark/antiquark  contribution is given by
\bqa
\label{vac1}
&&\Omega_{q\bar{q}} (T,\mu;\x,\y,\Phi,\bar{\Phi})= \Omega_{q\bar{q}}^{vac}+\Omega_{q\bar{q}}^{T,\mu}\;,\\
&&\Omega_{q\bar{q}}^{vac} =- 2 N_c\sum_q  \int \frac{d^3 p}{(2\pi)^3} E_q \theta( \Lambda_c^2 - \vec{p}^{2})\;,\\
\label{vac2}
&&\Omega_{q\bar{q}}^{T,\mu}=- 2 N_c\sum_q \int \frac{d^3 p}{(2\pi)^3} T \left[ \ln g_f^{+}+\ln g_f^{-}\right].\;
\label{vac3}
\eqa

The first term of the Eq.~(\ref{vac1}) is the contribution of the  fermion vacuum fluctuation,~where the $\Lambda_c$ is the ultraviolet cutoff.In presence of the Polyakov loop potential, the $g^{+}_f$ and $g^{-}_f$  are specified by the trace in the color space.
\bqa
\hspace{-1 cm}g^{+}&=&\left[1+3\Phi e^{-E_{q}^{+}/T}+3\bar{\Phi}e^{-2E_{q}^{+}/T}+e^{-3E_{q}^{+}/T}\right]\;,\qquad\\
g_q^{-}&=&\left[1+3\bar{\Phi} e^{-E_{q}^{-}/T}+3\Phi e^{-2E_{q}^{-}/T}+e^{-3E_{q}^{-}/T}\right]\;.
\eqa
$E_{f}^{\pm} =E_f \mp \mu_{f} $ and $E_f=\sqrt{p^2 + m{_f}{^2}}$ is the flavor dependent single particle energy of the quark/antiquark, $m_{u}=m_{d}=\frac{g\x}{2}$ is the mass of the light quarks $u$, $d$ and strange quark mass is $m_{s}=\frac{g\y}{\sqrt{2}}$. For the present work, it is assumed that $\mu_{u}=\mu_{d}=\mu_{s}=\mu$.

The quark one-loop vacuum term of the Eq.~(\ref{vac1}) is neglected in the standard mean-field approximation (s-MFA),~and the grand potential of the PQM model is written as,
\bqa
\nonumber
\Omega_{\rm PQM}(T,\mu,\x,\y,\Phi,\bar{\Phi})&=&U(\x,\y)+\mathcal{U}(\rm T,\Phi,\bar{\Phi})\;\\
&&+\Omega_{q\bar{q}} (T,\mu;\x,\y,\Phi,\bar{\Phi})\;.
 \label{Omega_MF}
 \eqa
The grand minima of the thermodynamic potential given by the  Eq.~(\ref{Omega_MF}) gives the $ \x$,~$ \y$,~$\Phi$ and $\bar{\Phi}$ by,
\begin{equation}
\frac{\partial \Omega_{\rm PQM}}{\partial
      \x}= \frac{\partial \Omega_{\rm PQM}}{\partial
      \y}=\frac{\partial \Omega_{\rm PQM}}{\partial
      \Phi}=\frac{\partial \Omega_{\rm PQM}}{\partial
      \bar{\Phi}}=0.
\label{EoMMF1}
\end {equation}

\subsection{Parameters of the QM model}
\label{sec:IIC}
The experimental values of the pion and kaon mass,~the average squared mass of the pseudo-scalars $\eta$ and $\eta^{\prime}$ mesons ($m_{\eta}^2+m_{\eta^{\prime}}^2$),~the scalar $\sigma$ meson mass $m_{\sigma}$ and the pion and kaon decay constants $f_{\pi}$ and $f_K$,~are used as the input \cite{Rischke:00,Schaefer:09} for determining the six model parameters $m^2$, $\lambda_1 $, $\lambda_2$, $c$, $h_x$ and $h_y$.

The values of the condensates in the vacuum are $\overline \sigma_x=f_{\pi}$ and  $\overline \sigma_y=(2f_{K}-f_{\pi})/\sqrt{2}$ according to the partially conserved axial-vector
current relation (PCAC).~The above values of condensates at the $T=0,\mu=0$,~give the  minimum of the vacuum effective potential in the Eq.~(\ref{EoMMF1}).~The parameters $\lambda_2$ and $c$ in the vacuum are given by, 
\bqa 
\nonumber
\label{para1}
\lambda_2&=&\dfrac{2  \ }{(\x^2+4\y^2)(\sqrt{2} \y -\x)}\left[(3\sqrt{2}\y) m_{K}^2- \right. \\ 
&&\left.(\sqrt{2}\y+2\x)m_{\pi}^2-(\sqrt{2}\y-\x)(m_{\eta}^2+m_{\eta^{'}}^2)\right]\;, \\
c&=&\dfrac{2(m^2_K-m^2_\pi)}{(\sqrt{2}\y-\x)}-\sqrt{2}\ \y \ \lambda_2 \;.
\label{para2}
\eqa
The mass parameter $m^2$ gets cancelled in the difference of the $\sigma$ and $\pi$ mass squares $(m_{\sigma}^2 - m_{\pi}^2)$.~The difference $(m_{\sigma}^2 - m_{\pi}^2)$,~depends on the parameters $\lambda_1$, $\lambda_2$ and $c$.~When the $\lambda_2$ and $c$ calculated from the above equations are put into the expression of  $(m_{\sigma}^2 - m_{\pi}^2)$ with the $\overline \sigma_x=f_{\pi}$ and  $\overline \sigma_y=(2f_{K}-f_{\pi})/\sqrt{2}$,~one gets the vacuum value of the parameter $\lambda_1$.~The mass parameter $m^2$ can be obtained from the expression of the $m_{\pi}^2$ as the following.
\bqa
\label{para3}
m^2&=&m^2_\pi-\lambda_1 (\x^2 + \y^2)-\frac{\lambda_2}{2} \x^2 +\frac{c}{\sqrt{2} } \y \;.
\eqa
Putting the vacuum values of $m_{\pi}^2$, $\lambda_1$, $\lambda_2$, $c$, $\x$ and $\y$ in the Eq. (\ref{para3}),~one gets the value of the mass parameter $m^2$.~Putting the $\x$ and $\y$ values in the Eq.~(\ref{hxhy}),~one finds 
\bqa
\label{para4}
h_{x}&=&f_{\pi} m^2_{\pi}  \ \text{and} \ h_{y}=\sqrt{2} f_{K} \ m_{K}^2-\frac{1}{\sqrt{2}}f_{\pi} \ m_{\pi}^2\;.
\eqa
The Yukawa coupling gets fixed from the non-strange constituent quark mass as $g=\frac{2m_{u}}{f_{\pi}}$.~For the $f_{\pi}=92.4$ MeV and the $m_{u} \sim 300.3$ MeV,~one gets the $g \sim 6.5$.~The strange quark  constituent mass becomes $m_{s} \sim 334.34$ MeV.~The experimental value of $m_{\eta}=547.5$ MeV and $m_{\eta^{\prime}}=957.78$ MeV.~In Ref.~\cite{Schaefer:09}, the parameter $\lambda_{2}$ is determined by taking the $m_{\eta}=539$ MeV and $m_{\eta^{\prime}}=963$ MeV as input because the sum of the squared masses $m_{\eta}^2+m_{\eta^{\prime}}^2=(539)^2+(963)^2$ is almost equal to the $(547.5)^2+(957.78)^2$ and the calculated parameters reproduce $m_{\eta}=539$ MeV and $m_{\eta^{\prime}}=963$ MeV in the output.

\section{Renormalized Polyakov Quark Meson Model}
\label{sec:III}

Several of the recent QM/PQM model investigations with quark one-loop vacuum correction \cite {lars,guptiw, schafwag12,chatmoh1,vkkr12,TranAnd,chatmoh2,vkkt13,Herbst,Weyrich,kovacs,zacchi1,zacchi2,Rai} have fixed the model parameters by the use of the  curvature (or screening) masses of the $\pi, K, \eta, \eta^{\prime}$ and $\sigma$ mesons where the pion decay constant has been identified as the vacuum expectation value of  non-strange condensate and the vacuum strange condensate gets related to the pion and kaon decay constant.~However,~it is well known that the physical masses are given by the poles of the meson propagators and the residue of the pion propagator at its pole gets related to the pion decay constant \cite{BubaCar,Naylor,fix1}.~Furthermore,~the definition of the  curvature mass of the meson,~involves the evaluation of its self-energy at the zero momentum \cite{laine,Adhiand1,Adhiand2,Adhiand3} as one knowns that the effective potential is the generator of the n-point functions of the theory at zero external momenta.~Also one notes that the pole definition is the physical and gauge invariant one \cite{Kobes,Rebhan}.~In the absence of the Dirac sea contributions,~the  pole mass becomes equivalent to the curvature mass  for the model parameter fixing but when the quark one-loop vacuum correction is present,~the pole masses of the mesons become different from their screening masses \cite{BubaCar,fix1}. ~In view of the above considerations,~it becomes important to  use of the exact on-shell parameter fixing method for the renormalized Polyakov enhanced quark meson (RPQM) model where the physical (pole) masses of the mesons, the pion and kaon decay constants are put into the relation of the running mass parameter and couplings by using the on-shell and the minimal subtraction renormalization prescriptions \cite{Adhiand2,asmuAnd,RaiTiw}.~In a very recent work~\cite{vkkr22},~one of us has calculated the consistent effective potential and the on-shell renormalized parameters for the 2+1 flavor quark-meson (QM) where the quark one-loop vacuum fluctuation is properly renormalized.~The above calculation of the exact effective potential and the on-shell renormalization of the seven parameters $m^2, \ \lambda_1,\ \lambda_2, \ c, \ h_x, \ h_y$ for the 2+1 flavor RPQM model are presented and  reproduced below.

\subsection{Description of the counterterms and self-energies}
\label{sec:IIIA}
The tree level parameters of the Eqs.~(\ref{para1})--(\ref{para4}) become inconsistent after including the quark one-loop vacuum correction unless the on-shell renormalization scheme is used.~In the on-shell scheme,~one uses the dimensional regularization to regularize the divergent loop integrals but the 
counterterm  are chosen differently from the minimal subtraction scheme.~The appropriate choice of counterterms in the on-shell scheme enables the exact cancellation of the loop corrections to the self-energies.~The renormalized parameters become renormalization scale independent as the couplings are evaluated on-shell.~The wave functions/fields and parameters of the Eq.~(\ref{lag}) are bare quantities.

The counterterms $\delta Z_\pi $, $\delta Z_{K} $, $\delta Z_\sigma $, $\delta Z_\eta $, $\delta Z_\eta^{\prime} $, 
$\delta Z_\psi $ $\delta Z_{\x}$ and $\delta Z_{\y} $ for the wave functions/fields and the counterterms $\delta m^{2} $, $\delta \lambda_{1} $,  $\delta \lambda_{2} $, $\delta c$, $\delta h_{x}$, $\delta h_{y}$ and $\delta g^{2} $, for the parameters,~are introduced in the Lagrangian (\ref{lag}) where the couplings and renormalized fields are defined as,
\bqa
\label{ctrm1}
\pi^i_b&=&\sqrt{Z_\pi} \ \pi^i, \ K_b=\sqrt{Z_K} \ K , \ \eta_b=\sqrt{Z_\eta} \ \eta, \\ 
 \eta^{\prime}_b&=&\sqrt{Z_\eta^{\prime}} \ \eta^{\prime}, \ \sigma_b=\sqrt{Z_\sigma} \ \sigma,\ m^2_b=Z_m \ m^2\\
\psi_b&=&\sqrt{Z_\psi} \ \psi, \
\lambda_{1b}=Z_{\lambda_{1}} \ \lambda_{1}, \ \lambda_{2b}=Z_{\lambda_{2}} \ \lambda_{2}, \\ \
g_b&=&\sqrt{Z_g} \ g, \
 h_{xb}=Z_{h_{x}} \ h_{x},\ h_{yb}=Z_{h_{y}} \ h_{y},  \\ \label{ctrm5} c_b&=&Z_c \ c,\ {\x}_b=\sqrt{Z_{\x}} \ \x, \ {\y}_b=\sqrt{Z_{\y}} \ \y 
\;.
\eqa
Here the $Z_ {(\pi,K,\eta,\eta^{\prime},\sigma,\psi,\x,\y)}=1+\delta Z_{(\pi,K,\eta,\eta^{\prime},\sigma,\psi,\x,\y)} $, identify  the field strength renormalization constants while $Z_ {(m,\lambda_{1},\lambda_{2},g,h_{x},h_{y},c )}=1+\delta Z_{(m,\lambda_{1},\lambda_{2},g,h_{x},h_{y},c ) } $ signify the mass and coupling renormalization constants.~One-loop corrections to the quark fields and the quark masses are zero because in the large $N_{c}$ limit, the $\pi$ and $\sigma$ loops that may renormalize the quark propagators are of the order $N_{c}^0$. Hence the $Z_\psi=1$ and the respective quark self energy corrections for the non-strange and the strange quarks are $\delta m_{u}=0$ and  $\delta m_{s}=0$.~Since,~the pion-quark $\pi \overline{\psi} \psi$ vertex is of order $N_{c}^0$,~the one-loop correction gets neglected.~As a result,~one gets $Z_{\psi}  \ \sqrt{Z_{g} \ g^2}\sqrt{Z_{\pi}} \approx g(1+\frac{1}{2}\frac{\delta g^2}{g^2}+\frac{1}{2}  \delta Z_\pi )=g$. Thus $\frac{\delta g^2}{g^2} \ + \delta Z_\pi=0$.~The $\delta m_{u}=0$ and $\delta m_{s}=0$ implies that $\delta g \ \x/2 + g \ \delta \x/2 =0 $ and  $\delta g \ \y/\sqrt{2} + g \ \delta \y/\sqrt{2} =$ 0.~One gets $\delta \x/\x= \delta \y/\y=-\delta g/g$ which can be written as
\bqa
\label{Zpi}
\frac{\delta  \x^2}{ \x^2}&=&\frac{\delta  \y^2}{ \y^2}=-\frac{\delta g^2}{g^2}=\delta Z_\pi\;.
\eqa
Following the Refs.~\cite{Adhiand1,Adhiand2,Adhiand3,asmuAnd,RaiTiw,vkkr22} and using the Eqs.~(\ref{ctrm1})-(\ref{ctrm5}) with the Eqs.~(\ref{para1}) and (\ref{para2}),~the counterterm $\delta \lambda_{2} $ can be expressed in terms of the counterterms  $\delta m^{2}_{\pi} $, $\delta m^{2}_{K}$, $\delta m^{2}_{\eta}$ , $\delta m^{2}_{\eta^{\prime}}$ and $\delta Z_{\pi}$ while the $\delta c $ is expressed in terms of the $\delta m^{2}_{\pi} $, $\delta m^{2}_{K}$, $\delta Z_{\pi}$ and the preceding $\delta \lambda_{2}$.~One gets the following expressions for the $\delta \lambda_{2}$ and $\delta c$.

\bqa
\nonumber
\label{delta:lambda_2}
\delta\lambda_{2}&=&\frac{2}{(\x^2+4\y^2)(\sqrt{2} \ \y -\x)}\biggl[(3\sqrt{2}  \y)\delta m_{K}^2 - (\sqrt{2} \y+2\x)     \; \\ 
&&\delta m_{\pi}^2-(\sqrt{2} \y-\x)(\delta m_{\eta}^2+\delta m_{\eta^{'}}^2) \biggr]-\lambda_{2} \delta Z_{\pi} \;,
\eqa

\bqa
\nonumber
\delta c&=&\dfrac{2(\delta m^2_K-\delta  m^2_\pi)}{(\sqrt{2} \ \y-\x)}-\sqrt{2} \  \y \ \delta \lambda_{2} \; \\
&&-(2\sqrt{2} \ \y \ \lambda_{2}+c) \ \frac{\delta Z_{\pi}}{2} \;. 
\label{deltac}
\eqa 
Knowing the $\delta \lambda_{2}$ and $\delta c$,~and using the expression of ($\delta m_{\sigma}^2-\delta m_{\pi}^2$),~the counterterm  $\delta \lambda_{1}$ can be written after some algebraic manipulations as the following. 

\begin{widetext}
\begin{align}
\label{lam1}
&\delta\lambda_{1}=\frac{\delta \lambda_{1\text{{\tiny NUMI}}}}{\lambda_{1\text{{\tiny DENOM}}}}-\lambda_{1} \ \delta  Z_{\pi};\; \\ \nonumber
\label{lam1de}
&\lambda_{1\text{{\tiny DENOM}}}=\biggl(\sqrt{(m^2_{s,00}-m^2_{s,88})^2+4m^4_{s,08}}\biggr) \ (\x^2+\y^2)-\frac{(m^2_{s,00}-m^2_{s,88})}{3}(\x^2+4\sqrt{2}\x\y-\y^2)\; \\
&\qquad \quad  \quad -\frac{4m^2_{s,08}}{3} \ (\sqrt{2} \x^2-\x\y-\sqrt{2} \y^2)\; 
\\
\nonumber 
&\delta\lambda_{1\text{{\tiny NUMI}}}=\sqrt{(m^2_{s,00}-m^2_{s,88})^2+4m^4_{s,08}}\biggl(\delta m_{\sigma}^2-\delta m_{\pi}^2\biggr)-\biggl\{\delta\lambda_{2}\frac{(\x^2+6\y^2)}{4}+\delta c \frac{\sqrt{2}\ \y}{4}\biggr\}\sqrt{(m^2_{s,00}-m^2_{s,88})^2+4m^4_{s,08}} \; \\ \nonumber 
&\qquad \quad + \biggl\{\delta\lambda_{2}\frac{(\x^2-2\y^2)}{4}
-\delta c \frac{\sqrt{2}\ (4\sqrt{2}\x+\y)}{12}\biggr\}
(m^2_{s,00}-m^2_{s,88})+\biggl\{\delta\lambda_{2}\sqrt{2}(\x^2-2\y^2)+\delta c \frac{\sqrt{2}\ (\x-\sqrt{2}\y)}{3}\biggr\}m^2_{s,08}\; \\ \nonumber
&\qquad \quad-\lambda_{2} \ \delta  Z_{\pi} \Biggl\{\frac{(\x^2+6\y^2)}{4}\sqrt{(m^2_{s,00}-m^2_{s,88})^2+4m^4_{s,08}}-\frac{(\x^2-2\y^2)}{4}(m^2_{s,00}-m^2_{s,88})-\sqrt{2}(\x^2-2\y^2)m^2_{s,08}\Biggr\}\; \\
\label{lam2}
&\qquad \quad-c\frac{\delta  Z_{\pi}}{2}\Biggl\{\frac{\sqrt{2} \y}{4}\sqrt{(m^2_{s,00}-m^2_{s,88})^2+4m^4_{s,08}} +\frac{\sqrt{2}\ (4\sqrt{2}\x+\y)}{12}(m^2_{s,00}-m^2_{s,88})+ \frac{\sqrt{2}\ (\sqrt{2}\y-\x)}{3}m^2_{s,08}\Biggr\}\;
\\ \nonumber 
&\text{The  counterterm $\delta m^2$ is written in terms of  the $\delta m_{\pi}^2$, $\delta \lambda_{1}$, $\delta \lambda_{2}$, $\delta c$ and $\delta Z_{\pi} $} \\  
& \delta m^2=\delta  m^2_\pi-\delta \lambda_{1} \ (\x^2+\y^2)- \ \frac{(\delta \lambda_{2}) \ \x^2}{2}+\frac{\delta c \ \y}{\sqrt{2}} -\delta  Z_{\pi} \biggl\{\lambda_{1} \ (\x^2+\y^2)+ \ \frac{\lambda_{2} \ \x^2}{2}-
 \frac{ c \ \y}{2\sqrt{2}} \biggr\}\;.
\end{align}
\end{widetext}
The Feynman diagrams of the self energy and tadpole contributions for the scalar particles are presented in the Fig.~(\ref{sclarse}) while the Fig.~(\ref{counts1b}) presents the corresponding counter term diagrams.~The  Fig.~(\ref{psclarse}) presents the Feynman diagrams of the self energy and tadpol contributions for the pseudo-scalar particles and the corresponding diagrams for the counter terms are shown in the Fig.~(\ref{countp2b}).~The on-shell parameter fixing requires,~the self energies of the scalar sigma $\sigma$,~pseudo-scalar eta~($\eta$),~eta-prime~($\eta^{\prime}$), pion~($\pi$) and kaon~($K$).~The self energy correction of the scalar $\sigma$  is calculated in terms of the self energy corrections $\Sigma_{\text{s},00}(p^2)$, $\Sigma_{\text{s},88}(p^2)$ and $\Sigma_{\text{s},08}(p^2)$ while the pseudo-scalar $\eta$ and $\eta^{\prime}$ self energy corrections are found in terms of the self energy corrections $\Sigma_{\text{p},00}(p^2)$, $\Sigma_{\text{p},88}(p^2)$ and $\Sigma_{\text{p},08}(p^2)$.~The scalar and pseudo-scalar self energies expressions are given below.
\begin{widetext}
\begin{align}
&\Sigma_{\text{s},00}(p^2)=-\frac{2}{3}N_cg^2\left[2\mathcal{A}(m^2_u)-(p^2-4m^2_u)\mathcal{B}(p^2,m_{u})\right]-\frac{1}{3}N_cg^2\left[2\mathcal{A}(m^2_s)-(p^2-4m^2_s)\mathcal{B}(p^2,m_{s})\right]+\Sigma^{tad}_{\text{s},00} \quad \;,\\
&\Sigma_{\text{s},11}(p^2)=-N_cg^2\left[2\mathcal{A}(m^2_u)-(p^2-4m^2_u)\mathcal{B}(p^2,m_{u})\right]+\Sigma^{tad}_{\text{s},11}\;,\\
&\Sigma_{\text{s},44}(p^2)=-N_cg^2\left[\mathcal{A}(m^2_u)+\mathcal{A}(m^2_s)-(p^2-(m_u+m_s)^2)\mathcal{B}(p^2,m_u,m_s)\right]+\Sigma^{tad}_{\text{ s},44}\;,\\
&\Sigma_{\text{s},88}(p^2)=-\frac{1}{3}N_cg^2\left[2\mathcal{A}(m^2_u)-(p^2-4m^2_u)\mathcal{B}(p^2,m_{u})\right]-\frac{2}{3}N_cg^2\left[2\mathcal{A}(m^2_s)-(p^2-4m^2_s)\mathcal{B}(p^2,m_{s})\right]+\Sigma^{tad}_{\text{s},88}\;,\\
&\Sigma_{\text{s},08}(p^2)=-\frac{\sqrt{2}}{3}N_cg^2\left[2\mathcal{A}(m^2_u)-(p^2-4m^2_u)\mathcal{B}(p^2,m_{u})\right]+\frac{\sqrt{2}}{3}N_cg^2\left[2\mathcal{A}(m^2_s)-(p^2-4m^2_s)\mathcal{B}(p^2,m_{s})\right]+\Sigma^{tad}_{\text{s},08}\;,
\end{align}
\begin{figure*}[htb]
\subfigure[\ One-loop self energy and tadpole diagrams.]{
\label{sclarse} 
\begin{minipage}[b]{0.48\textwidth}
\centering \includegraphics[width=\linewidth]{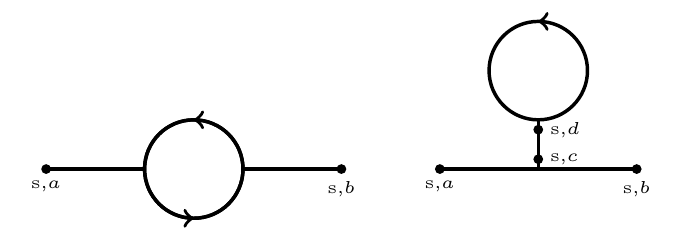}
\end{minipage}}
\hfill
\subfigure[\ One-loop self energy and tadpole counterterm diagrams.]{
\label{counts1b} 
\begin{minipage}[b]{0.48\textwidth}
\centering \includegraphics[width=\linewidth]{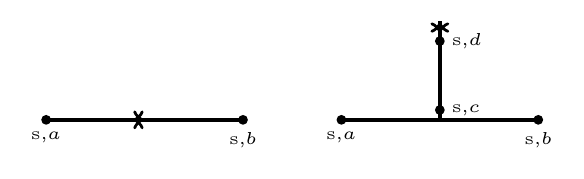}
\end{minipage}}
\caption{The solid line represents scalar particles  and an arrow on the solid line
denotes a quark.}
\label{fig:mini:fig1} 
\end{figure*}
\begin{figure*}[htb]
\subfigure[\ One-loop self energy and tadpole diagrams.]{
\label{psclarse} 
\begin{minipage}[b]{0.48\textwidth}
\centering \includegraphics[width=\linewidth]{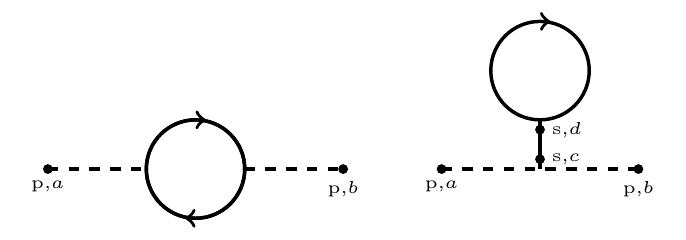}
\end{minipage}}
\hfill
\subfigure[\ One-loop self energy and tadpole counterterm diagrams.]{
\label{countp2b} 
\begin{minipage}[b]{0.48\textwidth}
\centering \includegraphics[width=\linewidth]{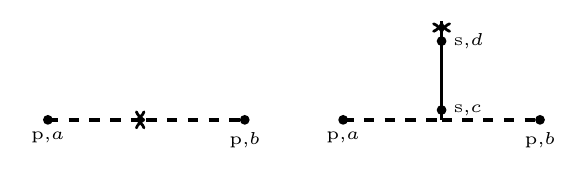}
\end{minipage}}
\caption{The dash line represents pseudo-scalar particles  and an arrow on the solid line
denotes a quark.}
\label{fig:mini:fig2} 
\end{figure*}
\bqa
\Sigma_{\text{p},00}(p^2)&=&-\frac{2}{3}N_cg^2\left[2\mathcal{A}(m^2_u)-p^2\mathcal{B}(p^2,m_{u})\right]-\frac{1}{3}N_cg^2\left[2\mathcal{A}(m^2_s)-p^2\mathcal{B}(p^2,m_{s})\right]+\Sigma^{tad}_{\text{p},00}\;,\\
\label{selfenpi}
\Sigma_{\text{p},11}(p^2)&=&\Sigma_{\pi}(p^2)=-N_cg^2\left[2\mathcal{A}(m^2_u)-p^2\mathcal{B}(p^2,m_{u})\right]+\Sigma^{tad}_{\text{ p},11}\;,\\
\label{selfenK}
\Sigma_{\text{ p},44}(p^2)&=&\Sigma_{K}(p^2)=-N_cg^2\left[\mathcal{A}(m^2_u)+\mathcal{A}(m^2_s)-\left\{p^2-(m_u-m_s)^2\right\} \mathcal{B}(p^2,m_u,m_s)\right]+\Sigma^{tad}_{\text{p},44}\;, \\
\Sigma_{\text{p},88}(p^2)&=&-\frac{1}{3}N_cg^2\left[2\mathcal{A}(m^2_u)-p^2\mathcal{B}(p^2,m_{u})\right]-\frac{2}{3}N_cg^2\left[2\mathcal{A}(m^2_s)-p^2\mathcal{B}(p^2,m_{s})\right]+\Sigma^{tad}_{\text{ p},88}\;,\\
\Sigma_{\text{ p},08}(p^2)&=&-\frac{\sqrt{2}}{3}N_cg^2\left[2\mathcal{A}(m^2_u)-p^2\mathcal{B}(p^2,m_{u})\right]+\frac{\sqrt{2}}{3}N_cg^2\left[2\mathcal{A}(m^2_s)-p^2\mathcal{B}(p^2,m_{s})\right]+\Sigma^{tad}_{\text{ p},08}\;.
\eqa
\end{widetext}

\begin{figure}[htb]
\begin{center}
\includegraphics[width=0.45\textwidth]{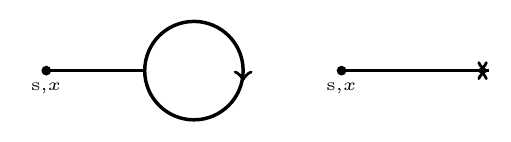}
\end{center}
\caption{One point diagram for the nonstrange scalar and its counterterm.}
\label{count33}
\end{figure}

\begin{figure}[htb]
\begin{center}
\includegraphics[width=0.45\textwidth]{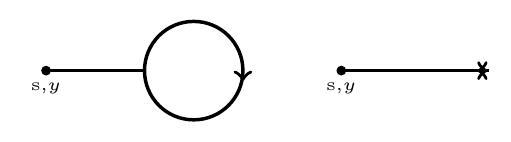}
\end{center}
\caption{One point diagram for the strange scalar and its counterterm.}
\label{count44}
\end{figure}
The Fig.~(\ref{count33}) shows the diagram of the one-point function and its counterterm for the quark one-loop correction to the non-strange component of the scalar $\sigma$.~One can write it as, 
\bqa
\delta\Gamma_x^{(1)}=-4 \ N_c \ g \ m_u \ \mathcal{A}(m_u^2)+i\delta t_{x}\;.
\eqa
The Fig.~(\ref{count44}) depicts the diagram of the one-point function and its counter-term for the quark one-loop correction to the 
strange component of the scalar $\sigma$.~It is written as,
\bqa
\delta\Gamma_y^{(1)}=-2\sqrt{2} \ N_c \ g \ m_s \ \mathcal{A}(m_s^2)+i\delta t_{y}\;.
\eqa
\subsection{Renormalization of the parameters}
\label{sec:IIIB}
The respective one-point functions $\Gamma^{(1)}_{x}=it_{x}=i(h_{x}-m_{\pi}^2 \x)$  and  $\Gamma^{(1)}_{y}=it_{y}=i{\tiny{\biggl\{h_{y}-\frac{\sqrt{2}}{2}(m_{K}^2-m_{\pi}^2) \  \x-m_{K}^2\y\biggr\}}}$ for the non-strange and the strange degree of freedom become zero and one gets two tree level equations of motion $t_{x}=0$ and $t_{y}=0$.~Hence the classical minimum of the effective potential gets fixed.~The first renormalization condition for the non-strange  $<\x>=0$ and the strange degree of freedom $<\y>=0$ demands that the respective one-loop corrections $\delta \Gamma^{(1)}_{x}$ and $\delta \Gamma^{(1)}_{y}$ to the one 
point functions,~are put to zero such that the minimum of the effective potential does not change.~Thus the $\delta\Gamma^{(1)}_{x}=0 $ and  $\delta\Gamma^{(1)}_{y}=0 $ give us 
\bqa
\delta t_{x}
&=&-4i \ N_c \ g \ m_u \ \mathcal{A}(m_u^2)  \;,
\\
\delta t_{y}
&=&-2\sqrt{2}i \ N_c \ g \ m_s \ \mathcal{A}(m_s^2)  \;. 
\eqa
The equation $h_{x}=t_{x}+m_{\pi}^2 \ \x $ and $h_{y}=t_{y}+{\tiny{\biggl\{\frac{\sqrt{2}}{2}(m_{K}^2-m_{\pi}^2) \  \x+m_{K}^2 \ \y\biggr\}}} $,~are used to write the counter-terms $\delta h_{x}$ and $\delta h_{y}$ in terms of the corresponding tadpole counter-terms  $\delta t_{x}$ and $\delta t_{y}$ as the following
\bqa
\label{delta:hx}
{\hskip -2.0 cm}\delta h_{x}&=&m^2_\pi \ \delta \x +\delta m^2_\pi \ \x+\delta t_{x},\qquad \qquad \qquad \qquad \; \\
\label{delta:hy}
\nonumber
{\hskip -2.0 cm}\delta h_{y}&=&{\biggl\{\frac{\sqrt{2}}{2}(m_{K}^2-m_{\pi}^2) \ \delta  \x+\frac{\sqrt{2}}{2}(\delta m_{K}^2-\delta m_{\pi}^2) \ \x \biggr\}} \\ 
&&+m_{K}^2 \ \delta \y +\delta m_{K}^2 \ \y +\delta t_{y}\;. 
\eqa

Using the Eq.~(\ref{Zpi}), one writes
\bqa
\label{delta:hxn}
\delta h_x&=&\frac{1}{2}m^2_\pi \ \x \ \delta Z_\pi+\delta m^2_\pi \ \x +\delta t_x, \qquad \qquad \qquad \qquad \;
\eqa
\bqa
\label{delta:hny}
\nonumber
{\hskip -1.5 cm}\delta h_{y}&=&{\biggl\{\frac{\sqrt{2}}{2}(m_{K}^2-m_{\pi}^2)\ \x \ \frac{\delta  Z_{\pi}}{2}+\frac{\sqrt{2}}{2}(\delta m_{K}^2-\delta m_{\pi}^2) \  \x \biggr\}} \\ 
&&+m_{K}^2 \ \y \ \frac{\delta Z_{\pi}}{2} +\delta m_{K}^2 \ \y +\delta t_{y}. 
\eqa
The pseudo-scalar $\pi$ and $K$ meson inverse propagators are defined as 
\bqa
p^2-m_{\pi,K}^2-i\Sigma_{\pi,K}(p^2)
{\rm +counterterms}
\;.
\label{definv}
\eqa
The physical states of the scalar $\sigma$ and $f_{0}$  and the pseudo-scalar $\eta$ and $\eta^{\prime}$,~are obtained by the mixing of the $00$ and $88$ components for the scalar (s) and pseudo-scalar (p)  particles.~The inverse propagator is given by the $2 \times 2$ matrix which results due to the mixing of the $00$ and $88$ components.~Solving the equation that one gets after putting the determinant of the matrix equal to zero,~the negative root of the solution gives the inverse propagator of the physical  $\sigma$ in the scalar and the physical $\eta$ in the pseudo-scalar channel.~The positive root of the solution gives the inverse propagator of the physically observed particles $f_{0}$ and  $\eta^{\prime}$ in the respective scalar and the pseudo-scalar channel.
\begin{widetext}
\begin{equation}
\textnormal{Det}\left.
\begin{pmatrix} 
p^2-m^2_{\text{s(p)},00}-i\Sigma_{\text{s(p)},00}(p^2) \,\,& 
-m^2_{\text{s(p)},08}-i\Sigma_{\text{s(p)},08}(p^2) \\ 
-m^2_{\text{s(p)},08}-i\Sigma_{\text{s(p)},08}(p^2) & 
p^2-m^2_{\text{s(p)},88}-i\Sigma_{\text{s(p)},88}(p^2) \,\,
\end{pmatrix}
\right.=0\; .
\end{equation}
One gets two solutions for the $p^{2}$ 
\begin{align}
\nonumber
& p^2=\frac{1}{2}\Biggl[{\tiny{\Biggl(\biggl\{ m^2_{\text{s(p)},00}+
i\Sigma_{\text{s(p)},00}(p^2)\biggr\}+\biggl\{m^2_{\text{s(p)},88}+i\Sigma_{\text{s(p)},88}(p^2)\biggr\} \Biggr)}} \mp  \\ 
& \qquad  {\tiny{ \sqrt{\Biggl( \biggl\{ m^2_{\text{s(p)},00}+
i\Sigma_{\text{s(p)},00}(p^2)\biggr\}-\biggl\{m^2_{\text{s(p)},88}+i\Sigma_{\text{s(p)},88}(p^2)\biggr\}\Biggr)^2+4\Biggl(m^2_{\text{s(p)},08}+i\Sigma_{\text{s(p)},08}(p^2) \Biggr)^2 }}}\Biggr]\;. \\ \nonumber
&\text{Neglecting the higher order ($N_{c}^2$) terms like $\tiny{ \biggl\{\Sigma_{\text{s(p)},00}(p^2))-\Sigma_{\text{s(p)},88}(p^2)\biggr\}^2 }$ and $ \Sigma_{\text{s(p)},08}^2(p^2)$ in self energy } 
\\ \nonumber
&\text{corrections, the above expression is written as } \\ 
\nonumber \\ \nonumber
& p^2=\frac{1}{2}\Biggl[\biggl( m^2_{\text{s(p)},00}+m^2_{\text{s(p)},88}\biggr)\mp {\tiny{ \sqrt{\Biggl( m^2_{\text{s(p)},00}-m^2_{\text{s(p)},88}\Biggr)^2+4\Biggl(m^2_{\text{s(p)},08}\Biggr)^2 }}}  \Biggr]+\frac{1}{2}\Biggl[\Biggl( i\Sigma_{\text{s(p)},00}(p^2)+i\Sigma_{\text{s(p)},88}(p^2)\Biggr) \mp \\ \nonumber & {\hskip -0.1 cm} 
\frac{1}{\sqrt{(m^2_{\text{s(p)},00}-m^2_{\text{s(p)},88})^2+4m^4_{\text{s(p)},08}}} \biggl\{\biggl(i\Sigma_{\text{s(p)},00}(p^2)-i\Sigma_{\text{s(p)},88}(p^2)\biggr)(m^2_{\text{s(p)},00}-m^2_{\text{s(p)},88})  +4i\Sigma_{\text{s(p)},08}(p^2) \   m^2_{\text{s(p)},08}  \biggr\}  \Biggr]\label{detroot}\;.\\
\\
\nonumber
&\text{Negative root of the Eq.~(\ref{detroot}) is equal to the sum of the mass and self energy correction for the scalar $\sigma$ (pseudoscalar $\eta$)}\\ \nonumber
& p^2=m_{\sigma(\eta)}^2 + i\Sigma_{\sigma(\eta)}(p^2) \  \text{where} \  m_{\sigma(\eta)}^2=\frac{1}{2}\Biggl[\biggl( m^2_{\text{s(p)},00}+m^2_{\text{s(p)},88}\biggr)-{\tiny{ \sqrt{\Biggl( m^2_{\text{s(p)},00}-m^2_{\text{s(p)},88}\Biggr)^2+4\Biggl(m^2_{\text{s(p)},08}\Biggr)^2 }}}  \Biggr] \ \text{and} \\
\nonumber
&\Sigma_{\sigma(\eta)}(p^2)=\frac{1}{2}\Biggl[\Sigma_{\text{s(p)},00}(p^2)+\Sigma_{\text{s(p)},88}(p^2)-\frac{1}{\sqrt{(m^2_{\text{s(p)},00}-m^2_{\text{s(p)},88})^2+4m^4_{\text{s(p)},08}}} \\ & \qquad \quad  \qquad \quad \biggl\{\biggl(\Sigma_{\text{s(p)},00}(p^2)-\Sigma_{\text{s(p)},88}(p^2)\biggr)(m^2_{\text{s(p)},00}-m^2_{\text{s(p)},88})  +4\Sigma_{\text{s(p)},08}(p^2) \   m^2_{\text{s(p)},08}  \biggr\}  \Biggr]\;.\label{selfensign}
\end{align}
Positive root of the Eq.~(\ref{detroot}) is equal to the sum of the mass and self energy correction for the scalar $f_{0}$ (pseudoscalar $\eta^{\prime}$)
\begin{align}
\nonumber
& p^2=m_{f_{0}(\eta^{\prime})}^2 + i\Sigma_{f_{0}(\eta^{\prime})}(p^2) \  \text{where} \  m_{f_{0}(\eta^{\prime})}^2=\frac{1}{2}\Biggl[\biggl( m^2_{\text{s(p)},00}+m^2_{\text{s(p)},88}\biggr)+{\tiny{ \sqrt{\Biggl( m^2_{\text{s(p)},00}-m^2_{\text{s(p)},88}\Biggr)^2+4\Biggl(m^2_{\text{s(p)},08}\Biggr)^2 }}}  \Biggr] \ \text{and} \\
\nonumber
&\Sigma_{f_{0}(\eta^{\prime})}(p^2)=\frac{1}{2}\Biggl[\Sigma_{\text{s(p)},00}(p^2)+\Sigma_{\text{s(p)},88}(p^2)+\frac{1}{\sqrt{(m^2_{\text{s(p)},00}-m^2_{\text{s(p)},88})^2+4m^4_{\text{s(p)},08}}} \\  & \qquad \quad  \qquad \quad \biggl\{\biggl(\Sigma_{\text{s(p)},00}(p^2)-\Sigma_{\text{s(p)},88}(p^2)\biggr)(m^2_{\text{s(p)},00}-m^2_{\text{s(p)},88})  +4\Sigma_{\text{s(p)},08}(p^2) \   m^2_{\text{s(p)},08}  \biggr\}  \Biggr]\;.\label{selfenf0n}
\end{align}
\end{widetext}
One can write the inverse propagator for the scalar $\sigma$ and the pseudo-scalar $ \eta, \ \eta^{\prime} $ mesons  as
\bqa
p^2-m_{\sigma,\eta,\eta^{\prime}}^2-i\Sigma_{\sigma,  \eta,  \eta^{\prime}}(p^2)
{\rm +counterterms}
\;.
\label{propmix}
\eqa
In the Lagrangian,~the renormalized mass is made equal to the physical mass, i.e.\ $m=m_{\rm pole}$ \footnote{The contributions of the imaginary parts of the self-energies for defining the mass are neglected.} when one implements the on-shell scheme.~One  writes
\bqa
\Sigma(p^2=m_{\sigma,\eta,  \eta^{\prime},\pi,K}^2)
{\rm +counterterms}
&=&0
\label{pole}
\;.
\eqa
~The propagator residue is set to 1 in the on-shell scheme and one gets 
\bqa
\label{res}
\nonumber
{\partial\over\partial p^2}\Sigma_{\sigma,\eta,  \eta^{\prime},\pi,K}(p^2)\Big|_{p^2=m_{\sigma,\eta,  \eta^{\prime},\pi,K}^2} \\
{\rm +counterterms}
&=&0\;.
\eqa
Using the diagrams of the Fig.~\ref{counts1b} and Fig.~\ref{countp2b}, the counterterms of the two point functions of the  scalar  and pseudo-scalar  mesons can be written as 
\bqa
\label{count1}
\Sigma_{\sigma}^{\rm ct1}(p^2)&=&i\left[\delta Z_{\sigma}(p^2-m_{\sigma}^2)-\delta m_{\sigma}^2\right]\;,
\\
\label{count2}
\Sigma_{\pi}^{\rm ct1}(p^2)&=&i\left[\delta Z_{\pi}(p^2-m_{\pi}^2)-\delta m_{\pi}^2\right]\;,
\\
\label{count3}
\Sigma_{K}^{\rm ct1}(p^2)&=&i\left[\delta Z_{K}(p^2-m_{K}^2)-\delta m_{K}^2\right]\;,
\\
\label{count4}
\Sigma_{\eta}^{\rm ct1}(p^2)&=&i\left[\delta Z_{\eta}(p^2-m_{\eta}^2)-\delta m_{\eta}^2\right]\;,
\\
\label{count5}
\Sigma_{\eta^{\prime}}^{\rm ct1}(p^2)&=&i\left[\delta Z_{\eta^{\prime}}(p^2-m_{\eta^{\prime}}^2)-\delta m_{\eta^{\prime}}^2\right]\;.
\eqa

The scalar and pseudo-scalar self energy tadpole contributions,~have two independent terms  proportional to $N_{c} g m_{u}\mathcal{A}(m_{u}^2)$  and $N_{c} g m_{s}\mathcal{A}(m_{s}^2)$ respectively as given in the Appendix B of the Ref.~\cite{vkkr22}.~The tadpole counterterms $\Sigma^{\rm ct2}$ for the scalar and pseudo-scalar particles are chosen (negative of the respective tadpole contributions to the scalar and pseudo-scalar self energies) such that they completely cancel the respective tadpole contributions to the self-energies.~When the self-energies and their derivatives are evaluated in the on-shell conditions,~one gets all the renormalization constants.~Combining the Eqs. (\ref{pole}), (\ref{res}) and (\ref{count1})--(\ref{count5}),~one finds the set of equations given below :  
\bqa
{\hskip -0.5 cm}
\delta m_{\pi}^2&=&-i\Sigma_{\pi}(m_{\pi}^2)\;;\delta Z_\pi= i{\partial\over\partial p^2}\Sigma_\pi(p^2)\Big|_{p^2=m_\pi^2}\;,
\\
{\hskip -0.5 cm}\delta m_{K}^2&=&-i\Sigma_{K}(m_{K}^2)\;;\delta Z_{K} =
i{\partial\over\partial p^2}\Sigma_{K}(p^2)\Big|_{p^2=m_{K}^2}\;,
\\
{\hskip -0.5 cm}\delta m_{\eta}^2&=&-i\Sigma_{\eta}(m_{\eta}^2)\;;\delta Z_\eta =i{\partial\over\partial p^2}\Sigma_\eta(p^2)\Big|_{p^2=m_\eta^2}\;,
\\
{\hskip -0.5 cm}\delta m_{\eta^{\prime}}^2&=&-i\Sigma_{\eta^{\prime}}(m_{\eta^{\prime}}^2)\;;\delta Z_{\eta^{\prime}} =i{\partial\over\partial p^2}\Sigma_{\eta^{\prime}}(p^2)\Big|_{p^2=m_{\eta^{\prime}}^2}\;,
\\
{\hskip -0.5 cm}\delta m_{\sigma}^2&=&-i\Sigma_{\sigma}(m_{\sigma}^2)\;;\delta Z_\sigma =
i{\partial\over\partial p^2}\Sigma_\sigma(p^2)\Big|_{p^2=m_\sigma^2}\;.
\eqa
When the self energy (neglecting the tadpole contributions) expressions from the Eqs.~(\ref{selfenpi}), (\ref{selfenK}), (\ref{selfensign}) and (\ref{selfenf0n}) are used, we get the following set of equations.

\begin{widetext}
\begin{align}
&\delta m_{\pi}^2=2ig^2N_c[\mathcal{A}(m_u^2)-\mbox{$1\over2$}{m_{\pi}^2}\mathcal{B}(m_{\pi}^2,m_u)]\;, \\
&\delta m_{K}^2=ig^2N_c\biggl[\mathcal{A}(m_u^2)+\mathcal{A}(m_s^2)-\{m_{K}^2-(m_{u}-m_{s})^{2}\} 
\mathcal{B}(m_{K}^2,m_{u},m_{s})\biggr]\;,
\end{align}

\begin{align}
\nonumber
&\delta m_{\eta}^2=\frac{-i}{2}\Biggl[\Sigma_{\text{p},00}(m_{\eta}^2)+\Sigma_{\text{p},88}(m_{\eta}^2)-\frac{1}{\sqrt{(m^2_{\text{p},00}-m^2_{\text{p},88})^2+4m^4_{\text{p},08}}} \\ 
&\qquad \quad \biggl\{\biggl(\Sigma_{\text{p},00}(m_{\eta}^2)-\Sigma_{\text{p},88}(m_{\eta}^2)\biggr)(m^2_{\text{p},00}-m^2_{\text{p},88})  +4\Sigma_{\text{p},08}(m_{\eta}^2) \   m^2_{\text{p},08}  \biggr\}  \Biggr]       \;, \\  \nonumber 
&\delta m_{\eta}^2=ig^2N_c\Biggl[ \biggl\{\mathcal{A}(m_u^2)+\mathcal{A}(m_s^2)-\mbox{$1\over2$}{m_{\eta}^2}\mathcal{B}(m_{\eta}^2,m_u)-\mbox{$1\over2$}{m_{\eta}^2}\mathcal{B}(m_{\eta}^2,m_s)\biggr\}  \\
&\qquad \quad -
\frac{\tiny{\biggl\{(m^2_{\text{p},00}-m^2_{\text{p},88})+4\sqrt{2}m^2_{\text{p},08}\biggr\}}}{3\sqrt{(m^2_{\text{p},00}-m^2_{\text{p},88})^2+4m^4_{\text{p},08}}} \biggl\{\mathcal{A}(m_u^2)-\mathcal{A}(m_s^2)-\mbox{$1\over2$}{m_{\eta}^2}\mathcal{B}(m_{\eta}^2,m_u)+\mbox{$1\over2$}{m_{\eta}^2}\mathcal{B}(m_{\eta}^2,m_s)\biggr\}\Biggr] \; , \\  \nonumber  
&\delta m_{\eta^{\prime}}^2=
\frac{-i}{2}\Biggl[\Sigma_{\text{p},00}(m_{\eta^{\prime}}^2)+\Sigma_{\text{p},88}(m_{\eta^{\prime}}^2)+\frac{1}{\sqrt{(m^2_{\text{p},00}-m^2_{\text{p},88})^2+4m^4_{\text{p},08}}} \\  
&\qquad \quad \biggl\{\biggl(\Sigma_{\text{p},00}(m_{\eta^{\prime}}^2)-\Sigma_{\text{p},88}(m_{\eta^{\prime}}^2)\biggr)(m^2_{\text{p},00}-m^2_{\text{p},88})  +4\Sigma_{\text{p},08}(m_{\eta^{\prime}}^2) \   m^2_{\text{p},08}  \biggr\}  \Biggr] \;,
\\
\nonumber
&\delta m_{\eta^{\prime}}^2=ig^2N_c\Biggl[ \biggl\{\mathcal{A}(m_u^2)+\mathcal{A}(m_s^2)-\mbox{$1\over2$}{m_{\eta^{\prime}}^2}\mathcal{B}(m_{\eta^{\prime}}^2,m_u)-\mbox{$1\over2$}{m_{\eta^{\prime}}^2}\mathcal{B}(m_{\eta^{\prime}}^2,m_s)\biggr\}         \\ 
&\qquad \quad + \frac{\tiny{\biggl\{(m^2_{\text{p},00}-m^2_{\text{p},88})+4\sqrt{2}m^2_{\text{p},08}\biggr\}}}{3\sqrt{(m^2_{\text{p},00}-m^2_{\text{p},88})^2+4m^4_{\text{p},08}}}  \biggl\{\mathcal{A}(m_u^2)-\mathcal{A}(m_s^2)-\mbox{$1\over2$}{m_{\eta^{\prime}}^2}\mathcal{B}(m_{\eta^{\prime}}^2,m_u)+\mbox{$1\over2$}{m_{\eta^{\prime}}^2}\mathcal{B}(m_{\eta^{\prime}}^2,m_s)\biggr\}\Biggr] \; , \\  \nonumber 
&\delta m_{\sigma}^2=
\frac{-i}{2}\Biggl[\Sigma_{\text{s},00}(m_{\sigma}^2)+\Sigma_{\text{s},88}(m_{\sigma}^2)-\frac{1}{\sqrt{(m^2_{\text{s},00}-m^2_{\text{s},88})^2+4m^4_{\text{s},08}}} \\ 
&\qquad \quad \biggl\{\biggl(\Sigma_{\text{s},00}(m_{\sigma}^2)-\Sigma_{\text{s},88}(m_{\sigma}^2)\biggr)(m^2_{\text{s},00}-m^2_{\text{s},88})  +4\Sigma_{\text{s},08}(m_{\sigma}^2) \  m^2_{\text{s},08}  \biggr\}  \Biggr] \;, \\
\nonumber  
&\delta m_{\sigma}^2=ig^2N_c\Biggl[ \biggl\{\mathcal{A}(m_u^2)+\mathcal{A}(m_s^2)-\mbox{$1\over2$}{(m_{\sigma}^2-4m_{u}^2)}
\mathcal{B}(m_{\sigma}^2,m_u)-\mbox{$1\over2$}{(m_{\sigma}^2-4m_{s}^2)}\mathcal{B}(m_{\sigma}^2,m_s)\biggr\}-\\
&\qquad \quad  \frac{\tiny{\biggl\{(m^2_{\text{s},00}-m^2_{\text{s},88})+4\sqrt{2}m^2_{\text{s},08}\biggr\}}}{3\sqrt{(m^2_{\text{s},00}-m^2_{\text{s},88})^2+4m^4_{\text{s},08}}} \biggl\{\mathcal{A}(m_u^2)-\mathcal{A}(m_s^2)-\frac{(m_{\sigma}^2-4m_{u}^2)}{2}\mathcal{B}(m_{\sigma}^2,m_u)+\frac{(m_{\sigma}^2-4m_{s}^2)}{2}\mathcal{B}(m_{\sigma}^2,m_s)\biggr\}\Biggr]\;,
\\
&\delta Z_{\pi}=
ig^2N_c\left[\mathcal{B}(m_{\pi}^2,m_u)+m_{\pi}^2\mathcal{B}^{\prime}(m_{\pi}^2,m_u)
\right]\;,\\
&\delta Z_{K}=
ig^2N_c\left[\mathcal{B}(m_{K}^2,m_{u},m_{s})+(m_{K}^2-(m_{u}-m_{s})^2)\mathcal{B}^{\prime}(m_{K}^2,m_{u},m_{s})\right]\;,
\\
\nonumber
&\delta Z_{\eta}=\frac{ig^2N_c}{2}\Biggl[\biggl\{
\mathcal{B}(m_{\eta}^2,m_{u})+
\mathcal{B}(m_{\eta}^2,m_{s})+m_{\eta}^2  \  \mathcal{B}^{\prime}(m_{\eta}^2,m_{u}) + m_{\eta}^2 \ \mathcal{B}^{\prime}(m_{\eta}^2,m_{s})\biggr\}  \\ 
&\qquad \quad +\frac{\tiny{\biggl\{(m^2_{\text{p},00}-m^2_{\text{p},88})+4\sqrt{2}m^2_{\text{p},08}\biggr\}}}{3\sqrt{(m^2_{\text{p},00}-m^2_{\text{p},88})^2+4m^4_{\text{p},08}}}\biggl\{-\mathcal{B}(m_{\eta}^2,m_{u})+\mathcal{B}(m_{\eta}^2,m_{s})-m_{\eta}^2 \ \mathcal{B}^{\prime}(m_{\eta}^2,m_{u}) + m_{\eta}^2 \ \mathcal{B}^{\prime}(m_{\eta}^2,m_{s})\biggr\}\Biggr]\;,
\\
\nonumber
&\delta Z_{\eta^{\prime}}=\frac{ig^2N_c}{2}\Biggl[\biggl\{
\mathcal{B}(m_{\eta^{\prime}}^2,m_{u})+
\mathcal{B}(m_{\eta^{\prime}}^2,m_{s})+m_{\eta^{\prime}}^2  \  \mathcal{B}^{\prime}(m_{\eta^{\prime}}^2,m_{u}) + m_{\eta^{\prime}}^2 \ \mathcal{B}^{\prime}(m_{\eta^{\prime}}^2,m_{s})\biggr\}  \\ 
&\qquad \quad -\frac{\tiny{\biggl\{(m^2_{\text{p},00}-m^2_{\text{p},88})+4\sqrt{2}m^2_{\text{p},08}\biggr\}}}{3\sqrt{(m^2_{\text{p},00}-m^2_{\text{p},88})^2+4m^4_{\text{p},08}}}\biggl\{-\mathcal{B}(m_{\eta^{\prime}}^2,m_{u})+\mathcal{B}(m_{\eta^{\prime}}^2,m_{s})-m_{\eta^{\prime}}^2 \ \mathcal{B}^{\prime}(m_{\eta^{\prime}}^2,m_{u}) + m_{\eta^{\prime}}^2 \ \mathcal{B}^{\prime}(m_{\eta^{\prime}}^2,m_{s})\biggr\}\Biggr]\;,
\end{align}
\begin{align}
\nonumber
&\delta Z_{\sigma}=\frac{ig^2N_c}{2}\Biggl[\biggl\{
\mathcal{B}(m_{\sigma}^2,m_{u})+
\mathcal{B}(m_{\sigma}^2,m_{s})+(m_{\sigma}^2-4m_u^2)\mathcal{B}^{\prime}(m_{\sigma}^2,m_{u}) + (m_{\sigma}^2-4m_s^2)\mathcal{B}^{\prime}(m_{\sigma}^2,m_{s})\biggr\}+
\\
&\qquad \quad  \frac{\tiny{\biggl\{(m^2_{\text{s},00}-m^2_{\text{s},88})+4\sqrt{2}m^2_{\text{s},08}\biggr\}}}{3\sqrt{(m^2_{\text{s},00}-m^2_{\text{s},88})^2+4m^4_{\text{s},08}}}\biggl\{\mathcal{B}(m_{\sigma}^2,m_{s})-\mathcal{B}(m_{\sigma}^2,m_{u})-(m_{\sigma}^2-4m_u^2)\mathcal{B}^{\prime}(m_{\sigma}^2,m_{u}) + (m_{\sigma}^2-4m_s^2)\mathcal{B}^{\prime}(m_{\sigma}^2,m_{s})\biggr\}\Biggr]\;.
\end{align}

The expressions of the field renormalization constants for the $\pi$, $K$, $\eta$, $\eta^{\prime}$ and $\sigma$  are given above.~However,~one needs to have the simplified expression of the $\delta Z_\pi$ only in the calculations below.~Substituting the expressions of $\delta Z_{\pi},  \delta m_{K}^2, \ \delta m_{\pi}^2,\ \delta m_{\eta}^2$ and $ \delta m_{\eta^{\prime}}^2$ from the above in the Eq.(\ref{delta:lambda_2}), the $\delta \lambda_{2}$ can be  written as
\begin{align}
&\delta\lambda_{2\os}=\dfrac{2 \ iN_{c} \ g^2 \ }{(\x^2+4\y^2)(\sqrt{2}\y -\x)}\Biggl[(3\sqrt{2}\y) \biggl\{A(m_{u}^2)+A(m_{s}^2)-\biggl( m_{K}^2-(m_{s}-m_{u})^2 \biggr) \mathcal{B}(m^2_K,m_u,m_s)  \biggr\}\; \\ \nonumber
&\qquad \quad-(\sqrt{2}\y+2\x) \biggl\{2 A(m_{u}^2)  -m_{\pi}^2\mathcal{B}(m^2_\pi,m_u) \biggr\}-(\sqrt{2}\y-\x)\Biggl\{2 A(m_{u}^2)+2 A(m_{s}^2)-\frac{m_{\eta}^2}{2}\biggl\{ \mathcal{B}(m^2_\eta \ , m_u) \; \\ \nonumber 
&\qquad \quad + \mathcal{B}(m^2_\eta \ , \ m_s)\biggr\}-\frac {m_{\eta^{'}}^2}{2}\biggl\{\mathcal{B}(m^2_{\eta^{'}} \ , \ m_u)+\mathcal{B}(m^2_{\eta^{'}}\ , \ m_s)    \biggr\}-\frac{\biggl(m^2_{p,00}-m^2_{p,88}+4\sqrt{2}m^2_{p,08}\biggr)}{6 \ \biggl( \sqrt{(m^2_{p,00}-m^2_{p,88})^2+4m^4_{p,08}\biggr) }} \\ 
&\qquad \quad\Biggl\{ m_{\eta}^2\biggl\{- \mathcal{B}(m^2_\eta \ , m_u)+\mathcal{B}(m^2_\eta \ , \ m_s)\biggr\} -m_{\eta^{'}}^2\biggl\{-\mathcal{B}(m^2_{\eta^{'}} \ , \ m_u)+\mathcal{B}(m^2_{\eta^{'}}\ , \ m_s)\biggr\} \Biggr\}\Biggr]\;,
\\
\label{lam2os}
&\delta\lambda_{2\os}=\delta \lambda_{2\text{div}}+\lambda_{2\text{{\tiny FIN}}}\;,
\\
\nonumber
&\lambda_{2\text{{\tiny FIN}}}= \dfrac{N_{c} g^2}{(4\pi)^2}(\lambda_2-g^2)\ln\biggl(\frac{\Lambda^2}{m_u^2}\biggr) +  \dfrac{N_cg^2}{(4\pi)^2} \dfrac{2}{(\x^2+4\y^2)}\Biggl[\dfrac{(\sqrt{2} \y+2\x)}{(\sqrt{2}\y -\x)}\Biggl\{ m^2_{u}-m^2_{s} \biggl\{1-2\ln\biggl(\frac{m_s}{m_u}\biggr)\biggr\} \\ \nonumber
&\qquad \quad-m_{\pi}^2\mathcal{C}(m^2_\pi,m_u)\Biggr\}+\dfrac{3\sqrt{2} \ \y}{(\sqrt{2}\y -\x)} \biggl\{ m_{K}^2-(m_{s}-m_{u})^2 \biggr\} \mathcal{C}(m^2_K,m_u,m_s)- \frac{ m_{\eta}^2}{2}\biggl\{ \mathcal{C}(m^2_\eta \ , m_u)+\mathcal{C}(m^2_\eta \ , \ m_s)\; \\ \nonumber
&\qquad \quad-2\ln\biggl(\frac{m_s}{m_u}\biggr)\biggr\}-\frac {m_{\eta^{'}}^2}{2}\biggl\{\mathcal{C}(m^2_{\eta^{'}} \ , \ m_u)+\mathcal{C}(m^2_{\eta^{'}}\ , \ m_s)-2\ln\biggl(\frac{m_s}{m_u}\biggr)\biggr\} +\frac{\biggl(m^2_{p,00}-m^2_{p,88}+4\sqrt{2}m^2_{p,08}\biggr)}{6 \ \biggl( \sqrt{(m^2_{p,00}-m^2_{p,88})^2+4m^4_{p,08}}\biggr) } \\ 
&\qquad \quad \Biggl\{ m_{\eta}^2\biggl\{ \mathcal{C}(m^2_\eta \ , m_u)-\mathcal{C}(m^2_\eta \ , \ m_s)+2\ln\biggl(\frac{m_s}{m_u}\biggr)\biggr\} -m_{\eta^{'}}^2\biggl\{\mathcal{C}(m^2_{\eta^{'}} \ , \ m_u)-\mathcal{C}(m^2_{\eta^{'}}\ , \ m_s)+2\ln\biggl(\frac{m_s}{m_u}\biggr)\biggr\} \Biggr\}\Biggr]\;,
\\
\nonumber
& \text{Substituting the expressions of $\delta Z_{\pi},\  \delta m_{K}^2 ,\  \delta m_{\pi}^2, \ \delta \lambda_{2}$ in the Eq.~(\ref{deltac}), the $\delta c$ is written as} \\ \nonumber
&\delta c_{\os}= \dfrac{2 \ i \ N_{c} \ g^{2}}{(\sqrt{2} \ \y-\x)}\biggl\{A(m_{u}^2)+A(m_{s}^2)-\biggl( m_{K}^2-(m_{s}-m_{u})^2 \biggr) \mathcal{B}(m^2_K,m_u,m_s)-2 A(m_{u}^2)  + m_{\pi}^2\mathcal{B}(m^2_\pi,m_u) \biggr\} \\  \label{cos}
&\qquad \quad \ \  -\sqrt{2} \ \y \ \delta \lambda_{2\os} - (2\sqrt{2}\ \y \ \lambda_{2}+c)\frac{\delta Z_{\pi}}{2} \ ;  \ \ \ \ \   \delta c_{\os}=\delta c_{\text{div}}+c_{\text{\tiny{FINTOT}}} \ \ \ ; \ \ \ \ \ c_{\text{\tiny{FINTOT}}}=-\sqrt{2} \ \y \ \lambda_{2\text{{\tiny FIN}}}+c_{\text{{\tiny FIN}}}\;,
\\ \nonumber 
&c_{\text{{\tiny FIN}}}
= \dfrac{N_cg^2}{(4\pi)^2}\Biggl[\biggl\{ c+\sqrt{2} \y (\lambda_2-g^2) \biggr\}  \ln\biggl(\frac{\Lambda^2}{m_u^2}\biggr)   +     \dfrac{2}{(\sqrt{2}\y -\x)} \Biggl\{ \biggl\{ m_{K}^2-(m_{s}-m_{u})^2 \biggr\} \mathcal{C}(m^2_K,m_u,m_s) 
-m_{\pi}^2\mathcal{C}(m^2_\pi,m_u) \Biggr\}\; \\ 
&\qquad \quad-\dfrac{ g^2}{2}(\sqrt{2} \y+\x)+\dfrac{2 g^2 \y^2 }{(\sqrt{2}\y -\x)}\ln\biggl(\frac{m_s}{m_u}\biggr)  \Biggr]\;.
\end{align}
Using the Eq.~(\ref{lam1}) and substituting the expressions of $\delta Z_{\pi},\  \delta m_{\sigma}^2 ,\ \delta m_{\pi}^2, \ \delta \lambda_{2}$ and $\delta c$ in the Eq.~(\ref{lam2}), the $\delta \lambda_{1}$ is written as
\begin{align}
&\delta\lambda_{1\os}=\frac{\lambda_{1\text{{\tiny NUMOS}}}}{\lambda_{1\text{{\tiny DENOM}}}}-\lambda_{1} \delta Z_{\pi}\;, \\ \nonumber
&\lambda_{1\text{{\tiny NUMOS}}}=iN_{c}g^2\Biggl[\biggl(\sqrt{(m^2_{s,00}-m^2_{s,88})^2+4m^4_{s,08}}\biggr)\biggl\{A(m_{u}^2)+A(m_{s}^2)-\biggl( \frac{m_{\sigma}^2-4m_{u}^2}{2} \biggr) \mathcal{B}(m^2_{\sigma},m_u)-\biggl( \frac{m_{\sigma}^2-4m_{s}^2}{2} \biggr) \\ \nonumber
&\qquad \quad \ \ \ \ \mathcal{B}(m^2_{\sigma},m_s)-2A(m_{u}^2)+m_{\pi}^2 \mathcal{B}(m^2_{\pi},m_u) \biggr\}-\biggl(\frac{m^2_{s,00}-m^2_{s,88}+4\sqrt{2}m^2_{s,08}}{3}\biggr) \biggl\{A(m_{u}^2)-A(m_{s}^2)-\biggl( \frac{m_{\sigma}^2-4m_{u}^2}{2} \biggr) \\ \nonumber
&\qquad \quad \ \ \  \mathcal{B}(m^2_{\sigma},m_u)+\biggl( \frac{m_{\sigma}^2-4m_{s}^2}{2} \biggr) \mathcal{B}(m^2_{\sigma},m_s) \biggr\}  \Biggr]+\frac{(m^2_{s,00}-m^2_{s,88})}{12}\biggl\{ (3\x^2-6\y^2)\delta \lambda_{2\os}-\sqrt{2}(4\sqrt{2}\x+\y)\delta c_{\os}\biggr\} \\  \nonumber
&\qquad \quad \ \  -\frac{\sqrt{2}m^2_{s,08}}{3}\biggl\{(2\y^2-\x^2) \delta \lambda_{2\os}+(\sqrt{2} \y-\x) \delta c_{\os} \biggl\}-\frac{1}{4}\sqrt{(m^2_{s,00}-m^2_{s,88})^2+4m^4_{s,08}}\biggl\{(\x^2+6\y^2) \delta \lambda_{2\os}+\sqrt{2} \y \delta c_{\os} \biggl\} \\ \nonumber 
&\qquad \quad \ \  + \delta Z_{\pi} \Biggl[ \frac{(m^2_{\sigma_{00}}-m^2_{s,88})}{12}\biggl\{ (3\x^2-6\y^2) \lambda_{2}-\sqrt{2}(4\sqrt{2}\x+\y) \frac{c}{2}\biggr\} -\frac{\sqrt{2}m^2_{s,08}}{3}\biggl\{(2\y^2-\x^2)  \lambda_2+(\sqrt{2} \y-\x)  \frac{c}{2} \biggl\} \\  
&\qquad \quad \ \  - \frac{1}{4}\biggl(\sqrt{(m^2_{s,00}-m^2_{s,88})^2+4m^4_{s,08}}\biggr)\biggl\{(\x^2+6\y^2) \lambda_2+\sqrt{2} \y \frac {c}{2} \biggl\}\Biggr]\;, \\ 
\label{lam1os}
&\delta\lambda_{1\os}=\delta \lambda_{1\text{div}}+\lambda_{1\text{{\tiny FIN}}} ;\qquad \lambda_{1\text{{\tiny FIN}}}=\frac{\lambda_{1\text{{\tiny NUMF}}}}{\lambda_{1\text{{\tiny DENOM}}}} ;\qquad \lambda_{1\text{{\tiny NUMF}}}=\lambda_{1\text{{\tiny NUMF-I}}}+\lambda_{1\text{{\tiny NUMF-II}}} \;, \\   \nonumber
&\text{expression of }\ \lambda_{1\text{{\tiny DENOM}}} \ \text{is given in the Eq.~(\ref{lam1de})}\;, \\ \nonumber
&\lambda_{1\text{{\tiny NUMF-I}}}=\frac{(m^2_{s,00}-m^2_{s,88})}{12}\biggl\{ (3\x^2+8\sqrt{2}\x\y-4\y^2)\lambda_{2\text{{\tiny FIN}}}-\sqrt{2}(4\sqrt{2}\x+\y)c_{\text{{\tiny FIN}}}\biggr\}-\frac{\sqrt{2}m^2_{s,08}}{3}\biggl\{(\sqrt{2}\y-\x) c_{\text{{\tiny FIN}}} \; \\ 
&\qquad \quad \ \ \  + (4\y^2+\sqrt{2}\x\y-3\x^2)\lambda_{2\text{{\tiny FIN}}} \biggr\}-\biggl(\frac{1}{4}\sqrt{(m^2_{s,00}-m^2_{s,88})^2+4m^4_{s,08}}\biggr) \ \biggl\{ (\x^2+4\y^2)\lambda_{2\text{{\tiny FIN}}}+\sqrt{2}\y c_{\text{{\tiny FIN}}}\biggr\}\;,
\\
\nonumber 
&\lambda_{1\text{{\tiny NUMF-II}}}=\dfrac{N_cg^2}{(4\pi)^2}\Biggl[\biggl(\sqrt{(m^2_{s,00}-m^2_{s,88})^2+4m^4_{s,08}}\biggr)\biggl\{\dfrac{g^2}{4}(\x^2-2 \y^2)+(m^2_{\sigma}-m^2_{\pi}-m^2_{u}-3m^2_{s}) \ln\biggl(\frac{\Lambda^2}{m_q^2}\biggr)\;\\ \nonumber
&\qquad \quad \ \ +2m^2_{s}\ln\biggl(\frac{m_s}{m_u}\biggr) + \frac{ (m_{\sigma}^2-4m^2_{u})}{2} \ \mathcal{C}(m^2_\sigma \ , m_u)+\frac{ (m_{\sigma}^2-4m^2_{s})}{2} \ \biggl( \mathcal{C}(m^2_\sigma \ , m_s)-2\ln\biggl(\frac{m_s}{m_u}\biggr)\biggr)-m_{\pi}^2 \mathcal{C}(m^2_\pi \ , m_u)\biggr\}\; \\ \nonumber
&\qquad \quad \ \ - \biggl(\dfrac{(m^2_{s,00}-m^2_{s,88})+4\sqrt{2}m^2_{s,08}}{3}\biggr) \biggl\{ \dfrac{g^2}{4}(2 \y^2-\x^2)\biggl(1+3\ln\biggl(\frac{\Lambda^2}{m_q^2}\biggr)\biggr)+(m^2_{\sigma}-6 \ m^2_{s} \ ) \ln\biggl(\frac{m_s}{m_u}\biggr)\; \\ 
&\qquad \quad \ \ + \frac{ (m_{\sigma}^2-4m^2_{u})}{2} \ \mathcal{C}(m^2_\sigma \ , m_u)-\frac{ (m_{\sigma}^2-4m^2_{s})}{2} \  \mathcal{C}(m^2_\sigma \ , m_s) \biggr\} \Biggl] \;,  
\\
\nonumber
&\delta m^2_{\os}=iN_{c}g^2\biggl\{2A(m_{u}^2)-m_{\pi}^2 \mathcal{B}(m^2_{\pi},m_u)  \biggl\}-\delta \lambda_{1\os}  (\x^2+\y^2)-\delta \lambda_{2\os}  \frac{  \x^2}{2}+\frac{\delta c_{\os} \y}{\sqrt{2}} -\delta  Z_{\pi} \biggl\{\lambda_{1} (\x^2+\y^2)+\lambda_{2}  \frac{  \x^2}{2}-\frac{ c  \y}{2\sqrt{2}} \biggr\}\;, \\
\label{m2os}
&\delta m^2_{\os}=\delta m^2_{\text{div}}+ m^2_{\text{\tiny{FIN}}} \;, \\
&m^2_{\text{\tiny{FIN}}}=\dfrac{N_cg^2}{(4\pi)^2} \Biggl[-2 m_{u}^2+(m_{\pi}^2-2 m_{u}^2)\ln\biggl(\frac{\Lambda^2}{m_u^2}\biggr)+m_{\pi}^2 \mathcal{C}(m^2_\pi \ , m_u)\Biggr]-\Biggl[\lambda_{1\text{\tiny{FIN}}}(\x^2+\y^2)+\lambda_{2\text{\tiny{FIN}}}\frac{\x^2}{2}-c_{\text{\tiny{FINTOT}}}\frac{\y}{\sqrt{2}}\Biggr]\;, \\ 
&\delta h_{x\os}=-\frac{i}{2}N_cg^2 m^2_\pi \ \x \left[\mathcal{B}(m^2_\pi,m_u)-m^2_\pi \mathcal{B}^\prime(m^2_\pi,m_u)\right]\;,
\\
\label{hxos}
&\delta h_{x\os}=\delta h_{x\text{div}}+ h_{x\text{\tiny{FIN}}} \;, \\ 
&h_{x\text{\tiny{FIN}}}=\dfrac{N_cg^2}{2(4\pi)^2}h_x\left[\ln\left(\frac{\Lambda^2}{m_u^2}\right)+\mathcal{C}(m^2_\pi,m_u)-m^2_\pi \mathcal{C}^{\prime}(m^2_\pi,m_u)\right]\;,
\\
\nonumber
&\delta h_{y\os}=iN_cg^2\Biggl[ \left(\frac{\sqrt{2}}{2}\x-\y\right)\left\{\mathcal{A}(m^2_s)-\mathcal{A}(m^2_u)\right\}-\left(\frac{\sqrt{2}}{2}\x+\y\right)\left\{m^2_{K}-(m_u-m_s)^2\right\}\mathcal{B}(m^2_K,m_u,m_s) \\ 
&\qquad \quad \ + \left(\frac{\sqrt{2}}{2}\x+\y\right)\frac{m^2_K}{2}\left[\mathcal{B}(m^2_\pi,m_u) +m^2_\pi \mathcal{B}^\prime(m^2_\pi,m_u)\right] +
\frac{\sqrt{2}}{4}\x \ m^2_\pi\left[\mathcal{B}(m^2_\pi,m_u)-m^2_\pi \mathcal{B}^\prime(m^2_\pi,m_u)\right] \Biggr]\;,
\end{align}
\begin{align}
\label{hyos}
&\delta h_{y\os}=\delta h_{y\text{div}}+ h_{y\text{\tiny{FIN}}} \;, \\ \nonumber
&h_{y\text{\tiny{FIN}}}=\dfrac{N_cg^2}{(4\pi)^2}\Biggl[ \dfrac{h_y}{2} \biggl\{  \ln\left(\frac{\Lambda^2}{m_u^2}\right) - \mathcal{C}(m^2_\pi,m_u)-m^2_\pi \mathcal{C}^{\prime}(m^2_\pi,m_u) \biggr\}-\dfrac{\sqrt{2} \  h_x}{2}\mathcal{C}(m^2_\pi,m_u) \\
& \qquad \quad +\left(\dfrac{\sqrt{2} \  \x}{2}-\y\right) \ \biggl\{m_{u}^2-m_{s}^2+2m_{s}^2 \ln\left(\frac{m_s}{m_u}\right)\biggr\}+ \left( \dfrac{\sqrt{2} \  \x}{2}+\y \right)  \biggl\{ m_{K}^2-(m_{s}-m_{u})^2 \biggr\} \mathcal{C}(m^2_K,m_u,m_s) \Biggr]\;, \\
\label{zpi}
&\delta Z^{\os}_{\pi}=\delta Z_{\pi,\rm div}-
\frac{N_cg^2}{(4\pi)^2}\left[\ln\left(\frac{\Lambda^2}{m_u^2}\right)+\mathcal{C}(m_{\pi}^2,m_u)+m_{\pi}^2\mathcal{C}^{\prime}(m_{\pi}^2,m_u)
\right]\;, \\  
&\delta g^2_{\os}=-iN_cg^4\left[m^2_\pi \mathcal{B}^\prime(m^2_\pi,m_u)+\mathcal{B}(m^2_\pi,m_u)\right]=\delta g^2_{\text{div}}+\dfrac{N_cg^4}{(4\pi)^2}\left[\ln\left(\frac{\Lambda^2}{m_u^2}\right)+\mathcal{C}(m^2_\pi,m_u)+m^2_\pi \mathcal{C}^{\prime}(m^2_\pi,m_u)\right]  \;,\\ 
\end{align}
\begin{align}
&\delta {\x}^2_{\os}=iN_cg^2 \x^2\left[m^2_\pi \mathcal{B}^\prime(m^2_\pi,m_u)+\mathcal{B}(m^2_\pi,m_u)\right]=\delta {\x}^2_{\text{div}}-\dfrac{N_cg^2 \x^2}{(4\pi)^2}\left[\ln\left(\frac{\Lambda^2}{m_u^2}\right)+\mathcal{C}(m^2_\pi,m_u)+m^2_\pi \mathcal{C}^{\prime}(m^2_\pi,m_u)\right]\;,\\
\label{y2os}
&\delta {\y}^2_{\os}=iN_cg^2 \y^2\left[m^2_\pi \mathcal{B}^\prime(m^2_\pi,m_u)+\mathcal{B}(m^2_\pi,m_u)\right]=\delta {\y}^2_{\text{div}}-\dfrac{N_cg^2 {\y}^2}{(4\pi)^2}\left[\ln\left(\frac{\Lambda^2}{m_u^2}\right)+\mathcal{C}(m^2_\pi,m_u)+m^2_\pi \mathcal{C}^{\prime}(m^2_\pi,m_u)\right]\;.\\ \nonumber
&\text{The common factor in the r.h.s. of the above four equations is defined as} \\ 
&\text{SCF}=\left[\ln\left(\frac{\Lambda^2}{m_u^2}\right)+\mathcal{C}(m^2_\pi,m_u)+m^2_\pi \mathcal{C}^{\prime}(m^2_\pi,m_u)\right]\;.
\end{align}

The $\mathcal{A}(m^2_f), \ \mathcal{B}(m^2,m_f), \ \mathcal{B}(m^2,m_u,m_s), \ \mathcal{B}^{\prime}(m^2,m_f)$, $\mathcal{C}(m^2,m_f)$, $\mathcal{C}(m^2,m_u,m_s)$, $\mathcal{C}^{\prime}(m^2,m_f)$ and $\mathcal{C}^{\prime}(m^2,m_u,m_s)$ are defined in the Appendix~(\ref{appenA}). The divergent part of the counterterms are  $ \delta \lambda_{2\text{div}}=\frac{N_cg^2}{(4\pi)^2\epsilon}(2\lambda_2-g^2)\;$,\ $\delta c_{\text{div}}=\frac{3N_cg^2 c}{2(4\pi)^2\epsilon}\;$, \ $\delta \lambda_{1\text{div}}=\frac{N_cg^2\lambda_1}{(4\pi)^2\epsilon}\;$, \ $\delta m^2_{\text{div}}=\frac{N_cg^2 m^2}{(4\pi)^2\epsilon}\;$, \ $ \delta h_{x\text{div}}=\frac{N_cg^2 h_x}{2(4\pi)^2\epsilon}\;$,\ $ \delta h_{y\text{div}}=\frac{N_cg^2 h_y}{2(4\pi)^2\epsilon}\;$, \ $\delta g^2_{\text{div}}=\frac{N_cg^4}{(4\pi)^2\epsilon}\; $, \ $\delta {\x}^2_{\text{div}}=-\frac{N_cg^2\x^2}{(4\pi)^2\epsilon}\; $, $\delta {\y}^2_{\text{div}}=-\frac{N_cg^2\y^2}{(4\pi)^2\epsilon}\;$, $\delta Z_{\pi,\rm div}=-\frac{N_cg^2}{(4\pi)^2\epsilon}$ . For both, the on-shell and the $\overline{\text{MS}}$ schemes, the divergent part of the counterterms are the same, i.e. $\delta \lambda_{1\text{div}}=\delta \lambda_{1\ms}$, $\delta \lambda_{2\text{div}}=\delta \lambda_{2\ms}$ etc.

\end{widetext}
Due to the renormalization scheme independence of the bare parameters,~one can  write down the relations between the renormalized parameters in the on-shell and the $\overline{\text{MS}}$ schemes as given below
\bqa
\lambda_{2\ms}&=&\lambda_2+\delta \lambda_{2\os}-\delta \lambda_{2\ms}\;,
\eqa
\bqa
c_{\ms}&=&c+\delta c_{\os}-\delta c_{\ms}\;,
\eqa
\bqa
\lambda_{1\ms}&=&\lambda_1+\delta \lambda_{1\os}-\delta \lambda_{1\ms}\;,
\eqa
\bqa
m^2_{\ms}&=&m^2+\delta m^2_{\os}-\delta m^2_{\ms}\;,
\eqa
\bqa
h_x{\ms}&=&h_{x}+\delta h_x{\os}-\delta h_x{\ms}\;,
\eqa
\bqa
h_y{\ms}&=&h_{y}+\delta h_y{\os}-\delta h_y{\ms}\;,
\eqa
\bqa
g^2_{\ms}&=&g^2+\delta g^2_{\os}-\delta g^2_{\ms}\;,
\eqa
\bqa
{\x}^2_{\ms}&=&\x^2+\delta {\x}^2_{\os}-\delta {\x}^2_{\ms}\;,
\eqa
\bqa
{\y}^2_{\ms}&=&\y^2+\delta {\y}^2_{\os}-\delta {\y}^2_{\ms}\;.
\eqa

The vacuum effective potential minimum lies at $\overline{x}=f_\pi$ and $\overline{y}=\frac{(2f_{K}-f_\pi)}{\sqrt{2}}$.~Using the above set of equations together with the Eqs.~(\ref{lam2os}), (\ref{cos}), (\ref{lam1os}), (\ref{m2os}), (\ref{hxos}), (\ref{hyos}) and (\ref{zpi})--(\ref{y2os}),~one  writes the scale $\Lambda$ dependent running parameters in the $\overline{\text{MS}}$ scheme as the following

\bqa
\label{params1}
\lambda_{2\ms}(\Lambda)
&=&\lambda_2+\lambda_{2\text{\tiny{FIN}}}\;,
\eqa
\bqa
c_{\ms}(\Lambda)
&=&c+c_{\text{\tiny{FINTOT}}}\;,
\eqa
\bqa
\lambda_{1\ms}(\Lambda)
&=&\lambda_1+\lambda_{2\text{\tiny{FIN}}}\;,
\eqa
\bqa
m^2_{\ms}(\Lambda)
&=&m^2+m^2_{\text{\tiny{FIN}}}\;,
\eqa
\bqa
h_x{\ms}(\Lambda)&=&h_{x}+h_{x\text{\tiny{FIN}}}\;,
\eqa
\bqa
\label{params5}
h_y{\ms}(\Lambda)&=&h_{y}+h_{y\text{\tiny{FIN}}}\;,
\eqa
\bqa
\label{params6}
g^2_{\ms}(\Lambda)
&=&g^2+\frac{N_cg^4}{(4\pi)^2} \text{SCF} \;,
\eqa
\bqa
\label{params7}
{\x}^2_{\ms}(\Lambda)
&=&f_\pi^2-\dfrac{4N_cm^2_u}{(4\pi)^2} \text{SCF} \;,
\eqa
\bqa
\label{params8}
{\y}^2_{\ms}(\Lambda)
&=&\left(\frac{2f_{K}-f_\pi}{\sqrt{2}}\right)^2-\dfrac{2N_cm^2_s}{(4\pi)^2} \text{SCF}\;.
\eqa
In the Eqs.~(\ref{params1})--(\ref{params6}),~the parameters $\lambda_2$, $c$, $\lambda_1$, $m^2$, $h_x$, $h_y$ and $g^2$,~have the tree level values of the QM model when the  $\x=f_\pi$ and  the  $\y=\frac{(2f_{K}-f_\pi)}{\sqrt{2}}$ in the vacuum.

~The parameters $\lambda_{2\ms}$, $c_{\ms}$, $\lambda_{1\ms}$, $m^2_{\ms}$, $h_{x\ms}$, $h_{y\ms}$ and $g^2_{\ms}$ in the large-$N_c$ limit,~are running with the scale $\Lambda$ and satisfy the following set of simultaneous renormalization group equations 
\bqa
\label{diffpara1}
\dfrac{d\lambda_{2\ms}(\Lambda)}{d\ln(\Lambda)}&=&\dfrac{2N_c}{(4\pi)^2}\left[2\lambda_{2\ms}g^2_{\ms}-g^4_{\ms}\right]\;,
\eqa
\bqa
\dfrac{d c_{\ms}(\Lambda)}{d\ln(\Lambda)}&=&\dfrac{2N_c}{(4\pi)^2}g^2_{\ms}c_{\ms}\;,
\eqa
\bqa
\dfrac{d\lambda_{1\ms}(\Lambda)}{d\ln(\Lambda)}&=&\dfrac{4N_c}{(4\pi)^2}g^2_{\ms}\lambda_{1\ms}\;,
\eqa
\bqa
\dfrac{dm^2_{\ms}(\Lambda)}{d\ln(\Lambda)}&=&\dfrac{2N_c}{(4\pi)^2}g^2_{\ms}m^2_{\ms}\;,
\eqa
\bqa
\dfrac{d h_{x\ms}(\Lambda)}{d\ln(\Lambda)}&=&\dfrac{N_c}{(4\pi)^2}g^2_{\ms}h_{x\ms}\;,
\eqa
\bqa
\dfrac{d h_{y\ms}(\Lambda)}{d\ln(\Lambda)}&=&\dfrac{N_c}{(4\pi)^2}g^2_{\ms}h_{y\ms}\;,
\eqa
\bqa
\dfrac{d g^2_{\ms}}{d\ln(\Lambda)}&=&\dfrac{2N_c}{(4\pi)^2}g^4_{\ms}\;,
\eqa
\bqa
\label{diffpara7}
\dfrac{d {\x}^2_{\ms}}{d\ln(\Lambda)}&=&-\dfrac{2N_c}{(4\pi)^2}g^2_{\ms}{\x}^2_{\ms}\;,
\eqa
\bqa
\label{diffpara8}
\dfrac{d {\y}^2_{\ms}}{d\ln(\Lambda)}&=&-\dfrac{2N_c}{(4\pi)^2}g^2_{\ms}{\y}^2_{\ms}\;.
\eqa

The differential the Eqs.~(\ref{diffpara1})--(\ref{diffpara8}) have the following solutions.

\bqa
\label{para01}
{\hskip -0.5 cm}\lambda_{2\ms}(\Lambda)&=&\frac{\lambda_{20}-\dfrac{N_c g^4_0}{(4\pi)^2}\ln\left(\dfrac{\Lambda^2}{\Lambda^2_0}\right)}{\left(1-\dfrac{N_c g^2_0}{(4\pi)^2}\ln\left(\dfrac{\Lambda^2}{\Lambda^2_0}\right)\right)^2}\;,  
\\
{\hskip -0.5 cm}c_{\ms}(\Lambda)&=&\frac{c_0}{\sqrt{\left[1-\dfrac{N_c g^2_0}{(4\pi)^2}\ln\left(\dfrac{\Lambda^2}{\Lambda^2_0}\right)\right]^3}}\;, \\
{\hskip -0.5 cm}\lambda_{1\ms}(\Lambda)&=&\frac{\lambda_{10}}{\left(1-\dfrac{N_c g^2_0}{(4\pi)^2}\ln\left(\dfrac{\Lambda^2}{\Lambda_0^2}\right)\right)^2}\;,
\eqa
\bqa
m^2_{\ms}(\Lambda)&=&\frac{m^2_0}{1-\dfrac{N_c g^2_0}{(4\pi)^2}\ln\left(\dfrac{\Lambda^2}{\Lambda^2_0}\right)}\;,
\\
\label{para05}
{\hskip -0.5 cm}h_{x\ms}(\Lambda)&=&\frac{h_{x0}}{\sqrt{1-\dfrac{N_c g^2_0}{(4\pi)^2}\ln\left(\dfrac{\Lambda^2}{\Lambda^2_0}\right)}}\;, \\ 
\label{para06}
{\hskip -0.5 cm}h_{y\ms}(\Lambda)&=&\frac{h_{y0}}{\sqrt{1-\dfrac{N_c g^2_0}{(4\pi)^2}\ln\left(\dfrac{\Lambda^2}{\Lambda^2_0}\right)}}\;, \\
\label{para07}
{\hskip -0.5 cm}g^2_{\ms}(\Lambda)&=&\frac{g^2_0}{1-\dfrac{N_c g^2_0}{(4\pi)^2}\ln\left(\dfrac{\Lambda^2}{\Lambda^2_0}\right)}\;,
\\
\label{para08}
{\hskip -0.5 cm}{\x^2}_{\ms}&=&f^2_\pi\left[1-\frac{N_cg^2_0}{(4\pi)^2}\ln\left(\frac{\Lambda^2}{\Lambda^2_0}\right)\right]\;, \\
\label{para09}
{\hskip -0.5 cm}{\y}^2_{\ms}&=&\frac{(2f_{K}-f_\pi)^{2}}{2}\left[1-\frac{N_cg^2_0}{(4\pi)^2}\ln\left(\frac{\Lambda^2}{\Lambda^2_0}\right)\right]\;.
\eqa
Here the $\lambda_{10}$,\ $\lambda_{20}$,\ $g^2_0$,\ $m^2_0$,\ $c_0$, $h_{x0}$ and $h_{y0}$ are the values of the running parameters at the scale $\Lambda_0$.~The $\Lambda_0$ can be chosen to satisfy the relation,
\bqa
\ln\left(\frac{\Lambda^2_0}{m_u^2}\right)+\mathcal{C}(m^2_\pi)+m^2_\pi \mathcal{C}^{\prime}(m^2_\pi)&=&0\;.
\eqa
The parameters of the Eqs.~(\ref{params1})--(\ref{params8}) are obtained at the scale $\Lambda=\Lambda_0$.

\subsection{Derivation of the effective potential}
\label{sec:IIIC}
The values of the parameters in the Eqs.~(\ref{para01})--(\ref{para07}) can be used to find expression of the vacuum effective potential in the $\overline{\text{MS}}$ scheme as,  

\bqa
\label{omegarqm}
\Omega_{vac}&=&U({\x}_{\ms},{\y}_{\ms})+\Omega^{q,vac}_{\ms}+\delta U({\x}_{\ms},{\y}_{\ms})\;,
\eqa
\begin{widetext}
where
\bqa
\label{omegams1}
\nonumber
U({\x}_{\ms},{\y}_{\ms})&=&\frac{m_{\ms}^{2}}{2}\left({\x}_{\ms}^{2} +
  {\y}_{\ms}^{2}\right) -h_{x \ms} \ {\x}_{\ms}-h_{y \ms} {\y}_{\ms} 
- \frac{c_{\ms}}{2 \sqrt{2}} {\x}_{\ms}^2 {\y}_{\ms} 
+ \frac{\lambda_{1\ms}}{2} {\x}_{\ms}^{2} {\y}_{\ms}^{2}
\\ 
&&+\frac{2 \lambda_{1\ms}+\lambda_{2\ms}}{8} \ {\x}_{\ms}^{4}+\frac{2 \lambda_{1\ms}+2\lambda_{2\ms}}{8} \ {\y}_{\ms}^{4}\;, 
\\
\nonumber
\delta U({\x}_{\ms},{\y}_{\ms})&=&\frac{\delta m_{\ms}^{2}}{2}\left(\xms^{2}+\yms^{2}\right)+\frac{m_{\ms}^{2}}{2}\left(
\delta \xms^2 + \delta \yms^2\right)-\delta h_{x\ms} \xms
-h_{x \ms} \ \delta {\x}_{\ms} \quad \quad \quad\;\\ \nonumber
&&-\delta h_{y \ms} \ \yms- h_{y \ms} \ \delta  {\yms}-\frac{\delta c_{\ms}}{2 \sqrt{2}} \ {\x}_{\ms}^2 \ {\y}_{\ms}- \frac{c_{\ms}}{2 \sqrt{2}} \ (\delta {\x}_{\ms}^2  \ {\y}_{\ms} +  {\x}_{\ms}^2  \ \delta {\y}_{\ms}) 
 \\ \nonumber
&&+ \frac{\delta \lambda_{1\ms}}{2} {\x}_{\ms}^{2} \ {\y}_{\ms}^{2}+ \frac{\lambda_{1\ms}}{2} (\delta {\x}_{\ms}^{2} \ {\y}_{\ms}^{2}+{\x}_{\ms}^{2} \ \delta {\y}_{\ms}^{2})+(\frac{2 \delta \lambda_{1\ms}+\delta \lambda_{2\ms}}{8}) \ {\x}_{\ms}^{4}\; \\ \nonumber
&&+(\frac{2 \lambda_{1\ms}+\lambda_{2\ms}}{8}) \ \delta {\x}_{\ms}^{4}
  +(\frac{2 \delta \lambda_{1\ms}+2 \delta \lambda_{2\ms}}{8}) \ {\y}_{\ms}^{4}
  +(\frac{2 \lambda_{1\ms}+2\lambda_{2\ms}}{8}) \ \delta {\y}_{\ms}^{4}\;, \\ 
\eqa
Dropping the  two loop terms ($\mathcal{O}(N^2_c)$),~we get
\bqa
\delta U({\x}_{\ms},{\y}_{\ms})&=&-\frac{N_cg^4_{\ms} ({\x}^4_{\ms}+2{\y}^4_{\ms})}{8(4\pi)^2}\frac{1}{\epsilon}=-\frac{N_c(2 \Delta_{x}^4+\Delta_{y}^4)}{(4\pi)^2}\frac{1}{\epsilon}\;.
\eqa
One writes the quark one-loop  vacuum correction for the two non-strange quark and the one strange quark flavor as, 
\bqa
\nonumber
\label{omegavac}
\Omega^{q,vac}_{\ms}&=&\frac{N_cg^4_{\ms} {\x}^4_{\ms}}{8(4\pi)^2}\left[\frac{1}{\epsilon}+\frac{3}{2}+\ln\left(\frac{4\Lambda^2}{g^2_{\ms} {\x}_{\ms}^{2}}\right)\right]+\frac{N_cg^4_{\ms} 2 {\y}^4_{\ms}}{8(4\pi)^2}\left[\frac{1}{\epsilon}+\frac{3}{2}+\ln\left(\frac{2\Lambda^2}{g^2_{\ms}{\y}^2_{\ms}}\right)\right] \\ \nonumber \\
&=&\frac{2N_c\Delta_{x}^4}{(4\pi)^2}\left[\frac{1}{\epsilon}+\frac{3}{2}+\ln\left(\frac{\Lambda^2}{\Delta_{x}^2}\right)\right]+\frac{N_c\Delta_{y}^4}{(4\pi)^2}\left[\frac{1}{\epsilon}+\frac{3}{2}+\ln\left(\frac{\Lambda^2}{\Delta_{y}^2}\right)\right]\;.
\eqa
The scale $\Lambda$ independent parameters $\Delta_{x}=\frac{g_{\ms} \ {\x}_{\ms}}{2}$ and $\Delta_{y}=\frac{g_{\ms} \ {\y}_{\ms}}{\sqrt{2}}$ are defined by the use of the Eqs.~(\ref{params6}), (\ref{params7}) and (\ref{params8}).~The Eq.~(\ref{omegams1}) is written in terms of the scale independent $\Delta_{x}$ and $\Delta_{y}$ as \\ 
\bqa
\nonumber
&&U(\Delta_{x},\Delta_{y})=\frac{m_{\ms}^2(\Lambda)}{g^2_{\ms}(\Lambda)}(2\Delta_{x}^2+\Delta_{y}^2)-2\frac{h_{x\ms}(\Lambda)}{g_{\ms}(\Lambda)}\Delta_{x}-\sqrt{2}\frac{h_{y\ms}(\Lambda)}{g_{\ms}(\Lambda)}\Delta_{y}-2\frac{c_{\ms}(\Lambda)}{g^3_{\ms}(\Lambda)}\Delta_{x}^2 \ \Delta_{y}+4\frac{\lambda_{1\ms}(\Lambda)}{g^4_{\ms}(\Lambda)}\Delta_{x}^2 \ \Delta_{y}^2 \\ 
&&{\hskip 2 cm}+2\frac{(2 \lambda_{1\ms}+\lambda_{2\ms})}{g^4_{\ms}(\Lambda)} \ \Delta_{x}^{4}+\frac{( \lambda_{1\ms}+\lambda_{2\ms})}{g^4_{\ms}(\Lambda)} \ \Delta_{y}^{4}\;,  \\ \nonumber
&&U(\Delta_{x},\Delta_{y})=\frac{m^2_0}{g^2_0}(2\Delta_{x}^2+\Delta_{y}^2)-2\frac{h_{x0}}{g_0}\Delta_{x}-\sqrt{2}\frac{h_{y0}}{g_0}\Delta_{y}-2\frac{c_{0}}{g^3_{0}}\Delta_{x}^2 \ \Delta_{y}+4\frac{\lambda_{10}}{g^4_{0}}\Delta_{x}^2 \ \Delta_{y}^2 \ \\ 
&&{\hskip 2 cm}+2\frac{(2 \lambda_{10}+\lambda_{20})}{g^4_{0}} \ \Delta_{x}^{4}+\frac{( \lambda_{10}+\lambda_{20})}{g^4_{0}} \ \Delta_{y}^{4}\;, 
\\ \nonumber
&&\Omega_{vac}(\Delta_{x},\Delta_{y})=\frac{m^2_0}{g^2_0}(2\Delta_{x}^2+\Delta_{y}^2)-2\frac{h_{x0}}{g_0}\Delta_{x}-\sqrt{2}\frac{h_{y0}}{g_0}\Delta_{y}-2\frac{c_{0}}{g^3_{0}}\Delta_{x}^2 \ \Delta_{y}+4\frac{\lambda_{10}}{g^4_{0}}\Delta_{x}^2 \ \Delta_{y}^2+2\frac{(2 \lambda_{10}+\lambda_{20})}{g^4_{0}}  \Delta_{x}^{4} \ \\ 
&&{\hskip 2 cm}+\frac{(\lambda_{10}+\lambda_{20})}{g^4_{0}}  \Delta_{y}^{4}+
\frac{2N_c\Delta_{x}^4}{(4\pi)^2}\left[\frac{3}{2}+\ln\left(\frac{\Lambda^2}{\Delta_{x}^2}\right)\right]+\frac{N_c\Delta_{y}^4}{(4\pi)^2}\left[\frac{3}{2}+\ln\left(\frac{\Lambda^2}{\Delta_{y}^2}\right)\right]\;.
\eqa
Expressing the mass parameter and the couplings in terms of the physical masses of mesons,~ the pion decay constant,~the kaon decay constant and Yukawa coupling,~we write
\bqa
\label{rqmeff}
\nonumber
\Omega_{vac}(\Delta_{x},\Delta_{y})&=&\frac{(m^2+                                        
m^2_{\text{\tiny{FIN}}})}{2} \left\lbrace f^2_{\pi}\left(\frac{\Delta_{x}^2}{m^2_u}\right)+\frac{(2f_{K}-f_{\pi})^2}{2}\left(\frac{\Delta_{y}^2}{m^2_s}\right)\right\rbrace-(h_{x}+h_{x\text{\tiny{FIN}}})f_{\pi}\left(\frac{\Delta_{x}}{m_u}\right) 
-(h_{y}+h_{y\text{\tiny{FIN}}})\frac{(2f_{K}-f_{\pi})}{\sqrt{2}}\; \\ \nonumber
&& \left(\frac{\Delta_{y}}{m_s}\right)-\frac{(c+c_{\text{\tiny{FINTOT}}})}{4}f^2_{\pi}(2f_{K}-f_{\pi})\left(\frac{\Delta_{x}^2}{m^2_u}\right)\left(\frac{\Delta_{y}}{m_s}\right)+\frac{(\lambda_{1}+\lambda_{1\text{\tiny{FIN}}})}{4} \
f^2_{\pi} (2f_{K}-f_{\pi})^2\left(\frac{\Delta_{x}^2}{m^2_u}\right) 
\left(\frac{\Delta_{y}^2}{m^2_s}\right) \; \\ \nonumber
&&+\frac{\lbrace 2(\lambda_{1}+\lambda_{1\text{\tiny{FIN}}})+(\lambda_{2}+\lambda_{2\text{\tiny{FIN}}})\rbrace}{8}f^4_\pi\left(\frac{\Delta_{x}^4}{m^4_u}\right)+\frac{\lbrace (\lambda_{1}+\lambda_{1\text{\tiny{FIN}}})+(\lambda_{2}+\lambda_{2\text{\tiny{FIN}}})\rbrace}{16}(2f_{K}-f_{\pi})^4\left(\frac{\Delta_{y}^4}{m^4_s}\right)\\   \nonumber
&&+\frac{2N_c\Delta_{x}^4}{(4\pi)^2}\left[\frac{3}{2}-\ln\left(\frac{\Delta_{x}^2}{m^2_{u}}\right)-\mathcal{C}(m^2_\pi)-m^2_\pi \mathcal{C}^{\prime}(m^2_\pi)\right]+\frac{N_c\Delta_{y}^4}{(4\pi)^2}\left[\frac{3}{2}-\ln\left(\frac{\Delta_{y}^2}{m^2_{u}}\right)-\mathcal{C}(m^2_\pi)-m^2_\pi \mathcal{C}^{\prime}(m^2_\pi)\right]\;. \\
\eqa
\end{widetext}
One notes that due to the dressing of the meson propagator in the on-shell scheme of the RQM model,~the pion decay constant,~the kaon decay constant and Yukawa coupling get renormalized in the vacuum.~But the  Eqs.~(\ref{params6}), (\ref{params7}) and (\ref{params8}) at the scale $\Lambda_0$ give us $g_{\ms}=g_{ren}=g$, ${\x}_{\ms}=f_{\pi,ren}=f_\pi$ and ${\y}_{\ms}=\frac{2f_{K,ren}-f_{\pi,ren}}{\sqrt{2}}=\frac{2f_K-f_\pi}{\sqrt{2}}$.~When the stationarity condition $\frac{\partial \Omega_{vac}(\Delta_{x},\Delta_{y})}{\partial \Delta_{x}}=0$ is applied to the Eq.~(\ref{rqmeff}) in  the non-strange  direction,~one gets  $h_{x0}=m_{\pi,c}^2 \ {\x}_{\ms} =m^2_\pi \left\lbrace 1-\frac{N_cg^2}{(4\pi)^2}m^2_\pi\mathcal{C}^\prime(m^2_\pi)\right\rbrace f_\pi$.~Hence the pion curvature mass $ m_{\pi,c}^2 =m^2_\pi \left\lbrace 1-\frac{N_cg^2}{(4\pi)^2}m^2_\pi\mathcal{C}^\prime(m^2_\pi)\right\rbrace$.~The implementation of the stationarity condition $\frac{\partial \Omega_{vac}(\Delta_{x},\Delta_{y})}{\partial \Delta_{y}}=0$ in the strange direction,~gives $\quad h_{0y}=\left({\frac{{\x}_{\ms}}{\sqrt{2}}+{\y}_{\ms}}\right)m^2_{K,c}-\frac{{\x}_{\ms}}{\sqrt{2}}m^2_{\pi,c}$=$\sqrt{2}f_Km^2_{K,c}-\frac{f_{\pi}}{\sqrt{2}}m^2_{\pi,c}$.~One finds the expression of the kaon curvature mass $m_{K,c}^2$ as given below in the Eq.~(\ref{mkcurve}) by using the expression of 
$h_{y\ms}(\Lambda_0)=h_{y0}$ in the Eq.~(\ref{params5}).~It is worth emphasizing that the pion curvature mass $ m_{\pi,c}$ (as  in Ref.~\cite{fix1})  and the kaon curvature mass are different from their pole masses $m_{\pi} $ and $m_{K}$ due to the consistent on-shell parameter fixing. The minimum of the effective potential remains fixed at ${\x}_{\ms}=f_\pi$ and 
${\y}_{\ms}=\frac{(2f_K-f_\pi)}{\sqrt{2}}$ .
\begin{table*}[!htbp]
    \caption{Parameters of the different model scenarios. The RQM model parameters are obtained by putting the $\Lambda=\Lambda_0$ in the Eqs.~(\ref{params1})--(\ref{params5}).}
    \label{tab:table2}
    \begin{tabular}{p{0.123\textwidth} p{0.123\textwidth}  p{0.123\textwidth} p{0.123\textwidth} p{0.123\textwidth} p{0.123\textwidth} p{0.123\textwidth} p{0.123\textwidth}}
      \toprule 
      Model&$m_{\sigma}(\text{MeV})$&$\lambda_2$&$c(\text{MeV}^2)$ &$\lambda_1$&$m^2(\text{MeV}^2)$& $h_x(\text{MeV}^3)$ & $h_y(\text{MeV}^3)$\\
      \hline 
      \hline
      &$400$&46.43 &4801.82 &-5.89 &$(494.549)^2$ &$(120.73)^3$ &$(336.43)^3$ \\
      QM&$500$&46.43 &4801.82 &-2.69 &$(434.305)^2$ &$(120.73)^3$ &$(336.43)^3$\\
      &$600$&46.43&4801.82 &1.141 &$(342.139)^2 $ &$(120.73)^3$ &$(336.43)^3$\\ \hline
      &$400$&34.88 &7269.20 &1.45 &$(442.447)^2 $&$(119.53)^3$ &$(323.32)^3$ \\
      RQM&$500$&34.88 & 7269.20&3.676 &$(396.075)^2 $&$(119.53)^3$ &$(323.32)^3$\\
      &$600$&34.88 &7269.20 &8.890 & $(256.506)^2$&$(119.53)^3$ &$(323.32)^3$\\
      \hline 
      \hline
    \end{tabular}
\end{table*}
\begin{widetext}
\begin{equation}
\label{mkcurve}
\resizebox{0.95\hsize}{!}{
$m^2_{K,c}=m^2_K\left[1-\frac{N_cg^2}{(4\pi)^2}\left\{\mathcal{C}(m^2_\pi,m_u)+m^2_\pi\mathcal{C}^{\prime}(m^2_\pi,m_u)-\left(1-\frac{(m_s-m_u)^2}{m_K^2}\right)\mathcal{C}(m^2_K,m_u,m_s)+\left(1-\frac{f_\pi}{f_K}\right)\frac{m^2_u-m_s^2+2m_s^2\ln\left(\frac{m_s}{m_u}\right)}{m^2_K}\right\}\right]$}\;.
\end{equation}
Combining the above RQM model vacuum effective potential with the thermal contributions of quarks-antiquarks and the Polyakov loop potential,~we write the grand thermodynamic potential of the RPQM model as the following,
\bqa
\nonumber
\label{rqmomega}
\hspace{-0.8cm}\Omega_{RPQM}(\Delta_{x},\Delta_{y},\Phi,\bar{\Phi},T,\mu)&=&\frac{(m^2+                                        
m^2_{\text{\tiny{FIN}}})}{2} \left\lbrace f^2_{\pi}\left(\frac{\Delta_{x}^2}{m^2_u}\right)+\frac{(2f_{K}-f_{\pi})^2}{2}\left(\frac{\Delta_{y}^2}{m^2_s}\right)\right\rbrace-(h_{x}+h_{x\text{\tiny{FIN}}})f_{\pi}\left(\frac{\Delta_{x}}{m_u}\right)\; \\ \nonumber
&&{\hskip -2 cm}-(h_{y}+h_{y\text{\tiny{FIN}}})\frac{(2f_{K}-f_{\pi})}{\sqrt{2}} \left(\frac{\Delta_{y}}{m_s}\right)-\frac{(c+c_{\text{\tiny{FINTOT}}})}{4}f^2_{\pi}(2f_{K}-f_{\pi})\left(\frac{\Delta_{x}^2}{m^2_u}\right)\left(\frac{\Delta_{y}}{m_s}\right)+\frac{(\lambda_{1}+\lambda_{1\text{\tiny{FIN}}})}{4} \
f^2_{\pi} (2f_{K}-f_{\pi})^2 \; \\ \nonumber
&&{\hskip -2 cm}\left(\frac{\Delta_{x}^2}{m^2_u}\right) 
\left(\frac{\Delta_{y}^2}{m^2_s}\right)+\frac{\lbrace 2(\lambda_{1}+\lambda_{1\text{\tiny{FIN}}})+(\lambda_{2}+\lambda_{2\text{\tiny{FIN}}})\rbrace}{8}f^4_\pi\left(\frac{\Delta_{x}^4}{m^4_u}\right)+\frac{\lbrace (\lambda_{1}+\lambda_{1\text{\tiny{FIN}}})+(\lambda_{2}+\lambda_{2\text{\tiny{FIN}}})\rbrace}{16}(2f_{K}-f_{\pi})^4\left(\frac{\Delta_{y}^4}{m^4_s}\right)\\   \nonumber
&&{\hskip -2 cm}+\frac{2N_c\Delta_{x}^4}{(4\pi)^2}\left[\frac{3}{2}-\ln\left(\frac{\Delta_{x}^2}{m^2_{u}}\right)-\mathcal{C}(m^2_\pi)-m^2_\pi \mathcal{C}^{\prime}(m^2_\pi)\right]+\frac{N_c\Delta_{y}^4}{(4\pi)^2}\left[\frac{3}{2}-\ln\left(\frac{\Delta_{y}^2}{m^2_{u}}\right)-\mathcal{C}(m^2_\pi)-m^2_\pi \mathcal{C}^{\prime}(m^2_\pi)\right]\; \\  
&&{\hskip -2 cm}+\mathcal{U}(T,\Phi,\bar{\Phi})+\Omega_{q\bar{q}} (T,\mu;\Delta_x,\Delta_y,\Phi,\bar{\Phi})\;.
\eqa
\end{widetext}

One gets the nonstrange condensate $\Delta_{x}$,~strange condensate $\Delta_{y}$,~$\Phi$ and $\bar{\Phi}$ in the RPQM model by searching the global minimum of the grand potential in the Eq.~(\ref{rqmomega}) for a given value of 
temperature $T$ and chemical potential $\mu$

\begin{equation}
\frac{\partial \Omega_{RPQM}}{\partial
      \Delta_{x}}=\frac{\partial \Omega_{RPQM}}{\partial\Delta_{y}}=\frac{\partial \Omega_{RPQM}}{\partial\Phi}=\frac{\partial \Omega_{RPQM}}{\partial\bar{\Phi}} =0
\label{EoMMF3}
\end {equation}

In our calculations, we have used the $m_\pi=138.0$ MeV, $m_K=496$ MeV.
Here in the RQM model, fixing the $m_{\eta}^2+m_{\eta^{\prime}}^2=(547.5)^2+(957.78)^2$ and then taking the $\eta$ mass as 527.58 MeV, one gets the ${\eta^{\prime}}$ mass equal to 968.89 MeV. The pole mass  $m_{\eta}=527.58$ MeV and $m_{\eta^{\prime}}=968.89$ MeV have been used for calculating the self energy corrections (for $\eta,\eta^{\prime}$) and fixing of the parameters in the on-shell scheme because it has been checked that when the masses are calculated with the new set of renormalized parameters and respective self energy corrections are added, the same pole masses are reproduced.  

\begin{figure*}[htb]
\subfigure[\ Normalized non-strange chiral condensate.]{
\label{fig5a} 
\begin{minipage}[b]{0.48\textwidth}
\centering \includegraphics[width=\linewidth]{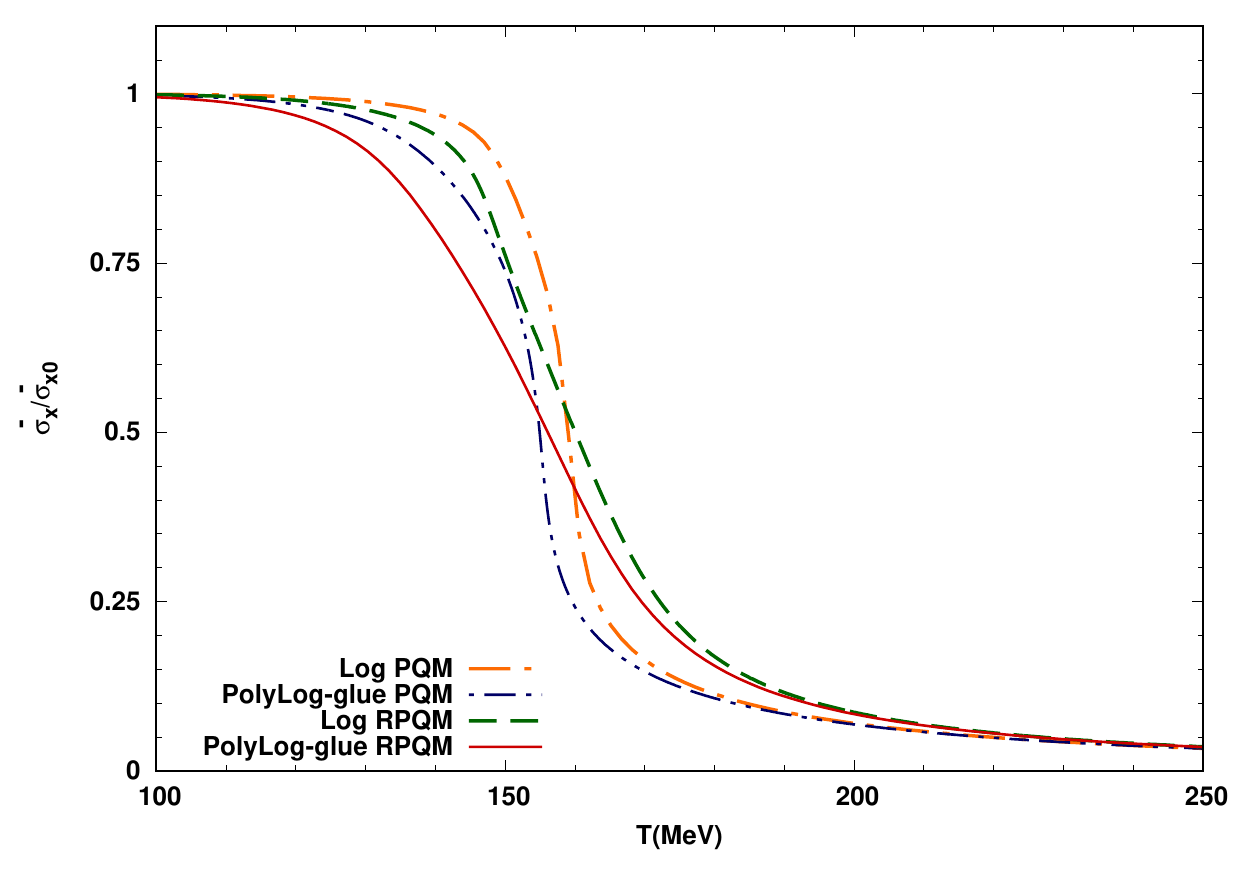}
\end{minipage}}
\hfill
\subfigure[\ Normalized strange chiral condensate.]{
\label{fig5b} 
\begin{minipage}[b]{0.48\textwidth}
\centering \includegraphics[width=\linewidth]{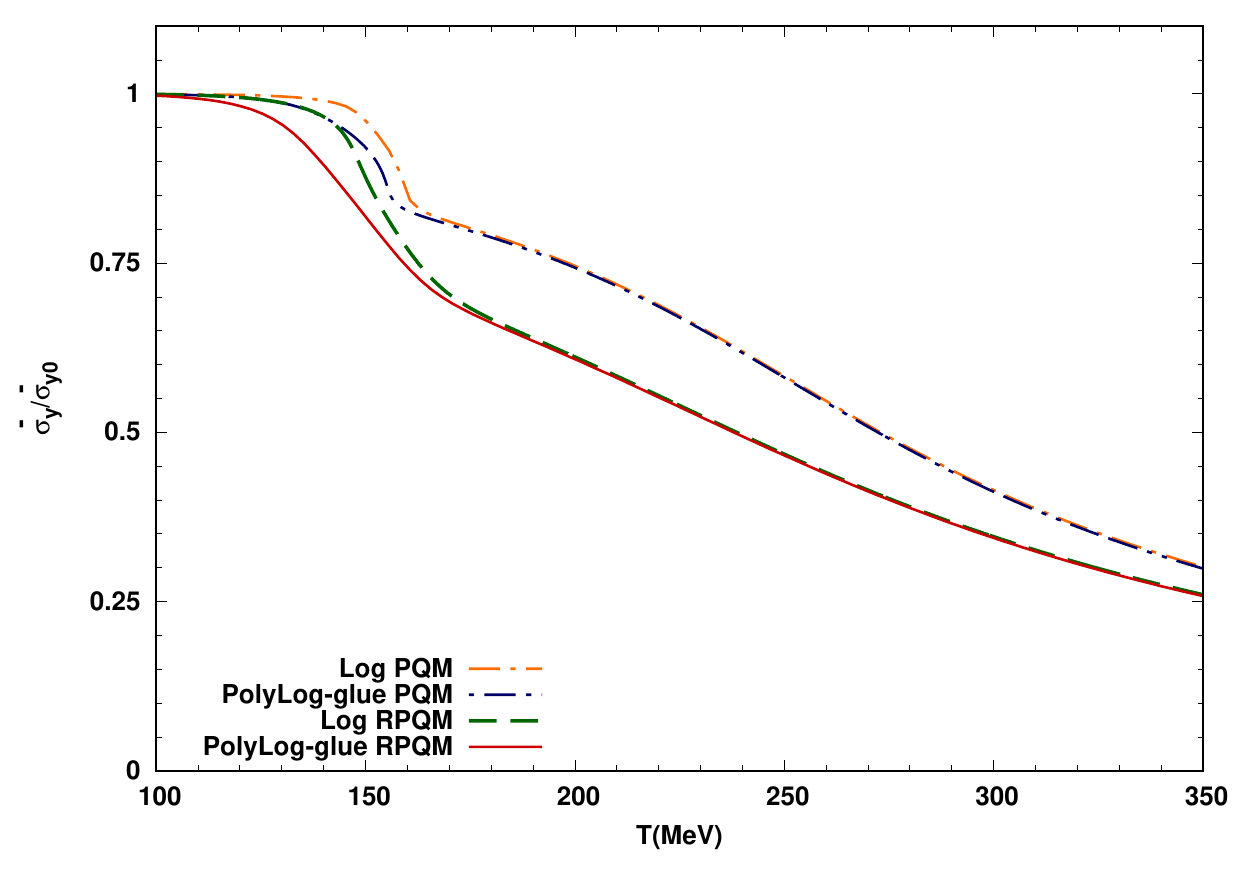}
\end{minipage}}
\caption{Temperature variation of the  $\sigma_x/\sigma_{x0}$  and the $\sigma_y/\sigma_{y0}$ for the $m_\sigma=500$ MeV at $\mu=0$.~The $\sigma_{x0}=f_{\pi}$ and $\sigma_{y0}=\frac{2f_K-f_{\pi}}{\sqrt{2}}$}
\label{fig:mini:fig5} 
\end{figure*}

\section{Results and Discussion}
\label{sec:IV}
The non-strange and strange direction chiral crossover transition and the confinement-deconfinement transition occurring on the temperature axis at $\mu=0$ are being explored here in detail.~The pseudo-critical temperature for the non-strange,~the strange chiral crossover transition, $T^{\chi}_c$,~$T^{s}_c$  and the confinement-deconfinement transition, $T^{\Phi}_c$ are obtained by identifying the peaks in the respective temperature direction variation of the temperature derivative  $\frac{\partial (\x / {\x}_{0})}{\partial T}$,~$\frac{\partial (\y / {\y}_{0})}{\partial T}$ and $\frac{\partial \Phi}{\partial T}$. 

\begin{figure*}[htb]
\subfigure[]{
\label{fig6a} 
\begin{minipage}[b]{0.48\textwidth}
\centering \includegraphics[width=\linewidth]{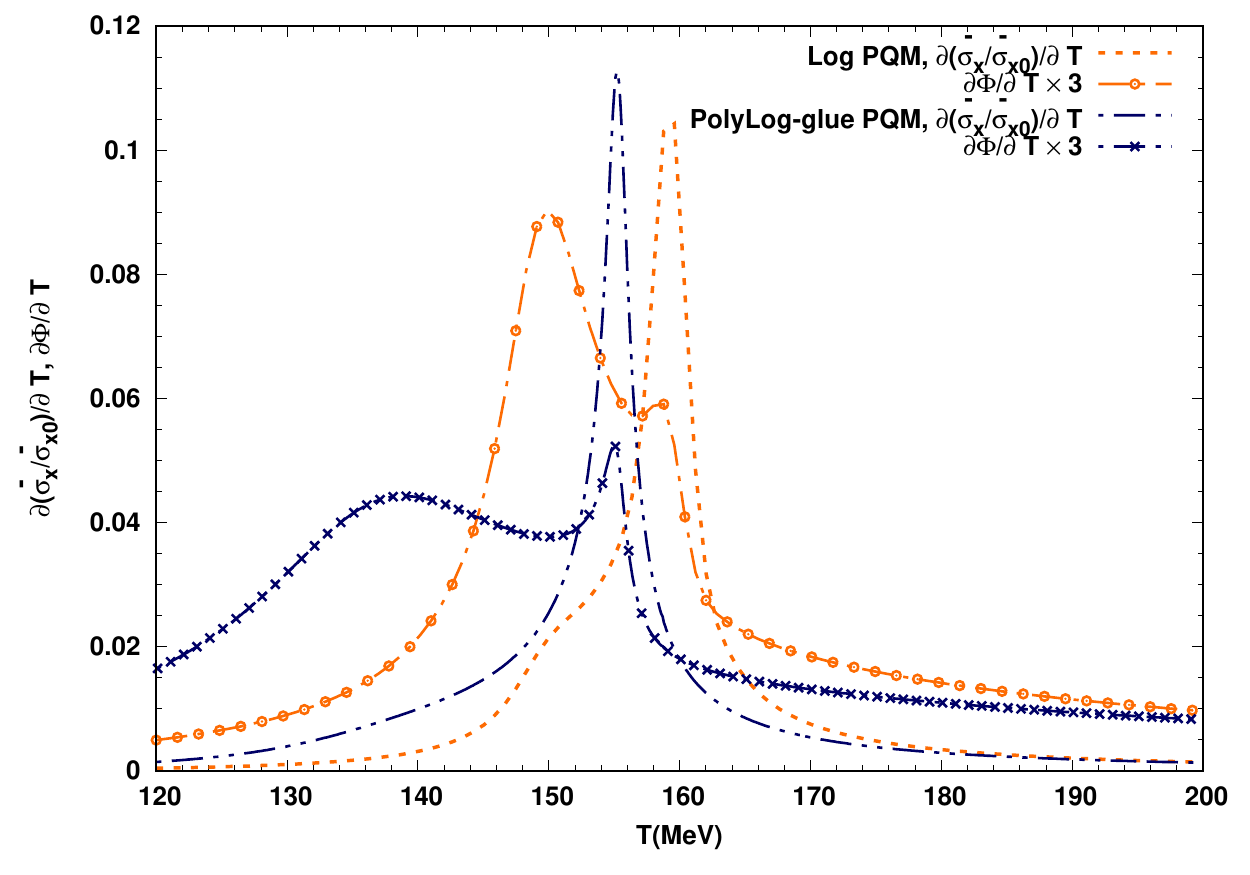}
\end{minipage}}
\hfill
\subfigure[]{
\label{fig6b} 
\begin{minipage}[b]{0.48\textwidth}
\centering \includegraphics[width=\linewidth]{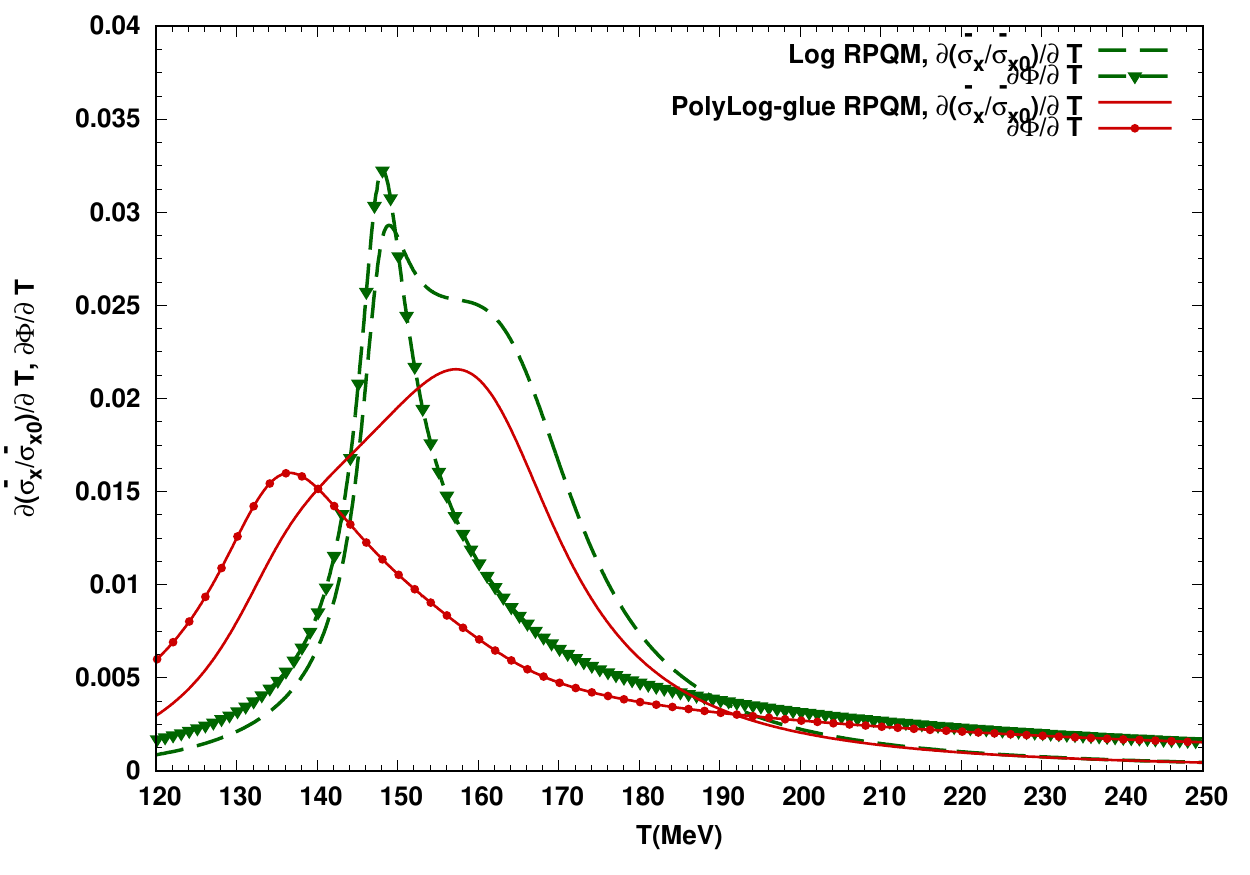}
\end{minipage}}
\caption{Temperature variation of the  $\frac{\partial\x/{\x}_{0}}{\partial T}$ and $\frac{\partial\Phi}{\partial T}$ for the $m_\sigma=500$ MeV at $\mu=0$.}
\label{fig:mini:fig6} 
\end{figure*}

The Fig.\ref{fig5a} and the Fig.\ref{fig5b},~show the respective temperature variations of the normalized non-strange  and strange chiral condensate at $\mu=0$ MeV and $m_{\sigma}=500$ MeV for the PQM and the RPQM model with the Log and the PolyLog-glue form of the Polyakov-loop potential.~The term glue stands for the unquenching of the Polyakov loop potential in the presence of the quark back-reaction at $\mu=0$.~Though,~the temperature variation of the non-strange condensate in the PolyLog-glue PQM model in the Fig.~\ref{fig5a},~shifts early on the temperature scale due to the quark back-reaction,~it is sharper than that of the Log PQM model because,~the early occuring weaker deconfinement transition at the $T_{c}^{\Phi}=138.8$ MeV (from the Table~\ref{tab:table3}) in the PolyLog-glue PQM model,~does not influence its chiral transition (at the $T_{c}^{\chi}=155.2$ MeV) that much while the Log PQM model relatively stronger deconfinement transition which occurs at the $T_{c}^{\Phi}=149.9$ MeV influences the chiral transition at the $T_{c}^{\chi}=159.3$ MeV.~The above behavior can be inferred from the temperature variation of the $\frac{\partial (\x/ {\x}_{0})}{\partial T}$ and the $\frac{\partial \Phi}{\partial T}$ in the Log and the PolyLog-glue PQM model in the  Fig.~\ref{fig6a}.~The temperature variation of the non-strange condensate becomes significantly smoother due to the quark one-loop vacuum correction in the Log RPQM model as shown the Fig~\ref{fig5a} and it also gets influenced by Log form of the Polyakov loop potential which causes a stronger deconfinement transition.~The sharper temperature variation of the $\frac{\partial \Phi}{\partial T}$ and it's peak at the $T_c^{\Phi}=148.1$ MeV has a strong influence on the chiral transition, because the $\frac{\partial (\x/ {\x}_{0})}{\partial T}$ temperature variation peaks at the adjacent $T_c^{\chi}=148.9$ MeV and develops a small very slowly decreasing shoulder like structure thereafter (falling short of developing the second peak).~Therefore the finding of the $T_c^{\chi}$ has an in-built ambiguity of $\sim +9.0$ MeV.~Due to the combined effect of the quark one-loop vacuum correction and the quark back-reaction,~one gets most smooth temperature variation of the non-strange condensate in the Fig.~\ref{fig5a} for the PolyLog-glue RPQM model.~The temperature variation of the $\frac{\partial \Phi}{\partial T}$ is very smooth and well seperated from the quite smooth temperature variation of the $\frac{\partial (\x/ {\x}_{0})}{\partial T}$ and the deconfinement transition occurs early as $\frac{\partial \Phi}{\partial T}$ temperature variation peaks  at $T_c^\Phi=136.6$ MeV while the temperature variation of the $\frac{\partial (\x/ {\x}_{0})}{\partial T}$ peaks later giving $T_c^\chi=157.3$ MeV.  

\begin{table*}[!htbp]
    \caption{Pseudo-critical temperatures  at $\mu=$0.}
    \label{tab:table3}
    \begin{tabular}{p{2cm}| p{2cm} |p{1.33cm} p{1.33cm} p{1.33cm} |p{1.33cm} p{1.33cm} p{1.33cm}|p{1.33cm} p{1.33cm} p{1.33cm} }
      \toprule 
      Polyakov-loop & Models & \multicolumn{3}{|c|}{$m_\sigma=400 \  \text{MeV}$}  & \multicolumn{3}{|c|}{$m_\sigma=500 \  \text{MeV}$} & \multicolumn{3}{|c}{$m_\sigma=600 \  \text{MeV}$} \\
       & & $T^\chi_c(\text{MeV})$ & $T^s_c(\text{MeV})$ & $T^\Phi_c(\text{MeV})$ &  $T^\chi_c(\text{MeV})$ &  $T^s_c(\text{MeV})$& $T^\Phi_c(\text{MeV})$ & $T^\chi_c(\text{MeV})$ &  $T^s_c(\text{MeV})$ & $T^\Phi_c(\text{MeV})$\\
      \hline 
      \hline
      Log & PQM & $149.5$ &$244.3$ &$149.5$& $159.3$ &$252.1$ &$149.9$ & $171.9$ &$261.7$ &$150.5$ \\
             & RPQM & $146.4$ &$222.8$ &$146.0$& $148.9$&$225.6$ & $148.1$ & $181.1$ &$233.8$ &$150.1$\\ \hline
      PolyLog-glue & PQM & $142.5$ & $245.1$ &$142.4$& $155.2$&$252.3$ & $138.8$ & $169.5$ &$261.2$ & $129.1$\\
            & RPQM  & $145.6$ &$225.6$ &$133.6$& $157.3$ &$227.8$ & $136.6$ & $179.6$ &$235.3$ & $138.6$\\
      \hline 
    \end{tabular}
\end{table*}

Strange condensate temperature variations for both the Log and the PolyLog-glue PQM models have a small and sharp kink like structure near the non-strange chiral transition temperature $T_c^{\chi}$.~The above kink like stucture gets smoothed out and hence disappears due to effect of quark one-loop vacuum correction in the RPQM model.~The quark back-reaction present in the PolyLog-glue PQM model causes a noticeable decrease in the strange condensate in the temperature range 120 to 165 MeV.~After $T>165$ MeV,~the strange condensate temperature variation for the PolyLog-glue PQM model becomes degenerate with that of the Log PQM model.~Note that the strange direction explicit symmetry breaking strength $h_y$ (see Table~\ref{tab:table2}) becomes weaker by relatively large amount (in comparison to $h_x$) as the kaon curvature mass $m_{K,c}=467.99$ MeV becomes smaller than it's pole mass $m_K=$ 496 MeV after renormalization of the parameters in the RPQM model.~Hence,~the melting  of the strange condensate gets significantly enhanced due to the effect of quark one-loop vacuum correction in the RPQM model.~In the temperature range 110 to 185 MeV around the non-strange chiral crossover transition temperature ($T^\chi_c$),~the melting of the strange condensate in the Log RPQM model differs significantly from its largest melting noticed in the PolyLog-glue RPQM model.~The strange condensate temperature variation for the Log and the PolyLog-glue RPQM model merge with each other for $T>180$ MeV.~The earliest chiral transition in the strange direction occurs in the Log RPQM model at $T_c^s=225.6$ MeV while the strange direction chiral transition  for the Log PQM model occurs at the $T_c^s$= 252.1 MeV.~The Table~\ref{tab:table3} summarizes the $\mu=0$ transition pseudocritical temperatures $T^\chi_c$, $T^s_c$ and $T_c^{\Phi}$ for $m_{\sigma}=$ 400,~500 and 600 MeV in the PQM and the RPQM model with the Log and the PolyLog-glue form of the Polyakov-loop potential.

\begin{figure*}[!htbp]
\subfigure[\ Subtracted chiral condensate $\Delta_{ls}$.]{
\label{fig7a} 
\begin{minipage}[b]{0.48\textwidth}
\centering \includegraphics[width=\linewidth]{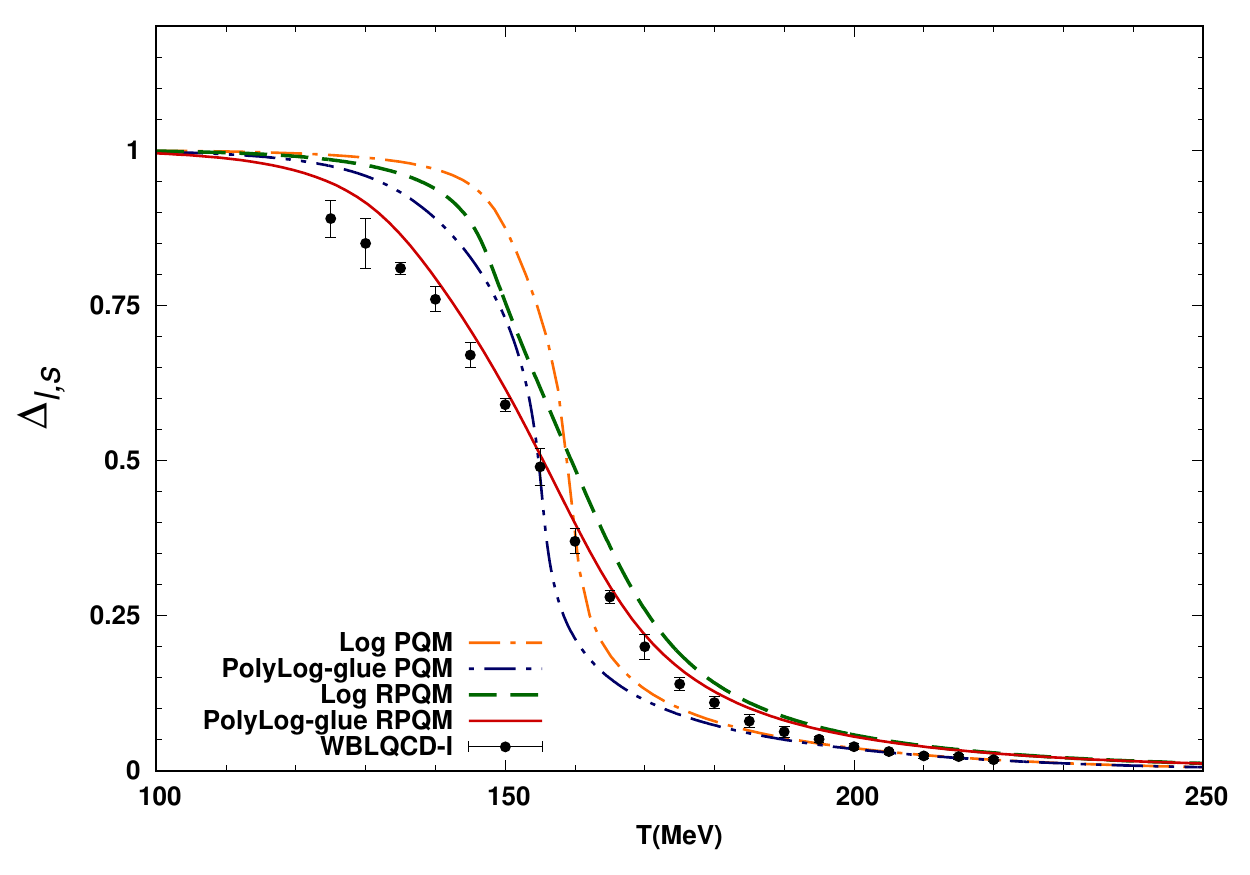}
\end{minipage}}%
\hfill
\subfigure[\ The Polyakov-loop order parameter $\Phi$.]{
\label{fig7b} 
\begin{minipage}[b]{0.48\textwidth}
\centering \includegraphics[width=\linewidth]{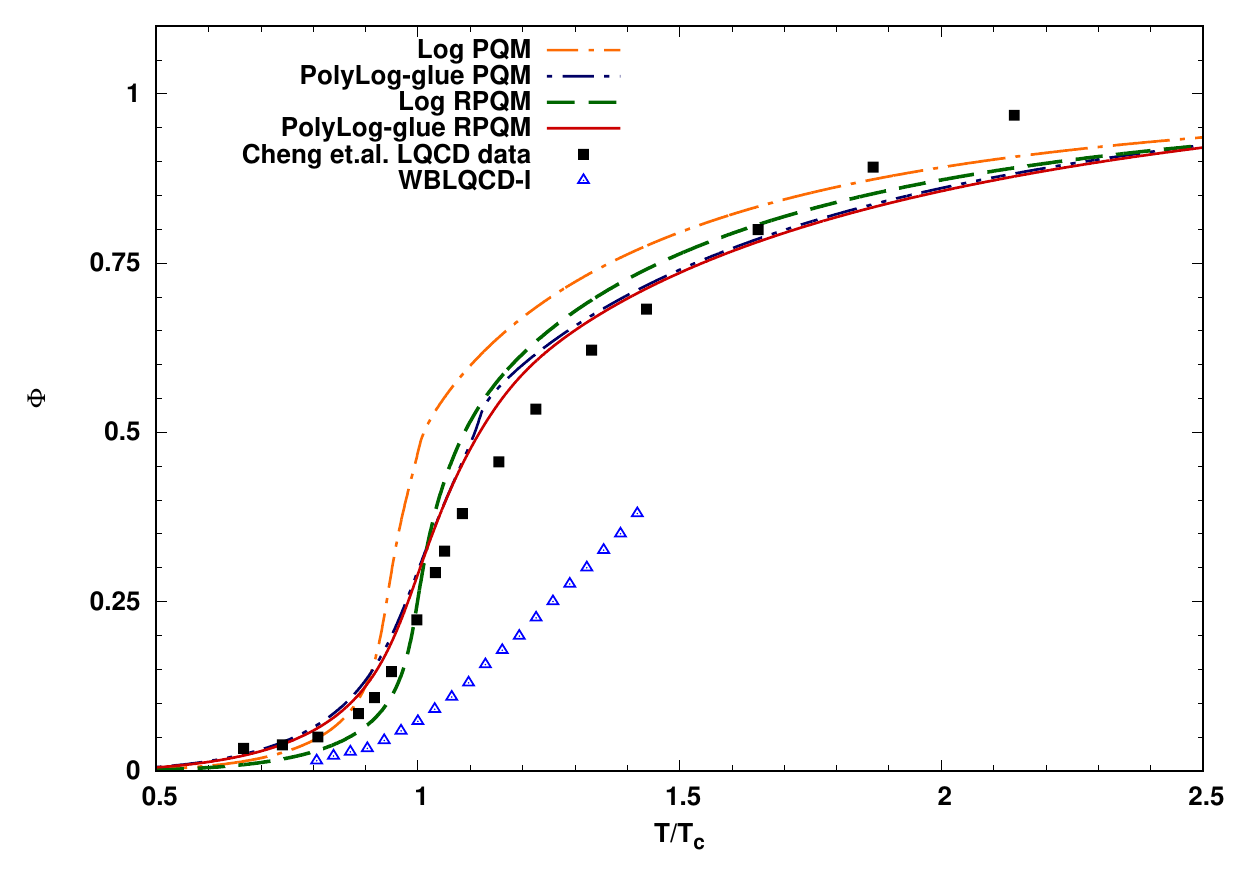}
\end{minipage}}
\caption{Temperature variation of the $\Delta_{ls}$ and the  $\Phi$ for  the $m_\sigma=500$ MeV at $\mu=0$.~The lattice data of the WBLQCD-I for the  $\Delta_{ls}$ and the  $\Phi$,~has been taken from the Ref.~\cite{Wupertal2010} in the continum limit.~The Cheng et.al. LQCD data for the $\Phi$ has been taken from the Ref.~\cite{Cheng:06} for the $N_\tau=6$ in the p4-action.}
\label{fig:mini:fig7} 
\end{figure*}

The lattice data can not be directly compared with the temperature variations of the non-strange and strange condensate,~$<\x>$ and $<\y>$.~A suitable combination of the light and strange quark condensates is calculated to eliminate the quadratic divergences
in the linear quark mass dependent correction to the chiral condensate.~The above quantity is further normalized by the corresponding combination of condensates calculated at $T=0$ MeV \cite{Cheng:08}.~Thus one finds,~the chiral symmetry breaking order parameter called the subtracted chiral condensate $ \Delta_{l,s}(T) $ as given below.  
\begin{equation}
\label{eq:Delta}
\Delta_{l,s}(T)=\frac{<\x>(T)-(\frac{h_{x}}{h_{y}})<\y>(T)}
{<\x>(0)-(\frac{h_{x}}{h_{y}})<\y>(0)}.
\end{equation}
 
The subtracted chiral condensate temperature variations (at $\mu=0$) have been calculated for the PQM and the RPQM model with the Log and the PolyLog-glue form of the Polyakov loop potential,~and compared with the Wuppertal-Budapest collaboration lattice QCD (WBLQCD-I) data of the $\Delta_{ls}$ \cite{Wupertal2010} in the Fig.~\ref{fig7a}.~We find that the calculation of the $\Delta_{ls}$ temperature variation in our PolyLog-glue RPQM model gives the closest fit to the WBLQCD-I data.~The Fig.\ref{fig7b} presents the comparison of the WBLQCD-I \cite{Wupertal2010} and the lattice data in Ref.~\cite{Cheng:08} for the reduced temperature scale variation of the Polyakov loop with the temperature variation of the Polyakov loop condensate obtained in the PQM and RPQM model calculations having the Log and the PolyLog-glue form for the Polyakov loop potential.~In model calculations the reduced temperature $T/T_c$ has been obtained by taking $T_c=T_c^\Phi$.~The lattice data \cite{Cheng:08} for the Polyakov loop in the temperature range 0.7 $T_c$ to 1.7 $T_c$ shows quite a close resemblance with the Polyakov loop temperature variation that we get in our PolyLog-glue RPQM model calculation.~The WBLQCD-I \cite{Wupertal2010} data for the Polyakov loop lie significantly below the temperature variation of the Polyakov loop condensate that one finds in the all model calculations.

\subsection{Comparison of thermodynamic quantities with the lattice QCD data}
\label{sec:IVA}
We will compare the thermodynamic quantities:~pressure,~entropy density,~energy density,~specific heat and speed of sound in the Log PQM model and the RPQM model with the Log and the PolyLog-glue form of the Polyakov-loop potential.~The QCD phase transition has strong influence on these thermodynamic quantities.

The pressure of a QCD system can be defined as negative of the grand potential.
\begin{equation}
 P(T,\mu)=-\Omega(T,\mu)
\end{equation}
the pressure is normalized to zero in the vacuum, i.e. $P(0,0)=0$. 

\begin{figure}[!htbp]
\centering \includegraphics[width=\linewidth]{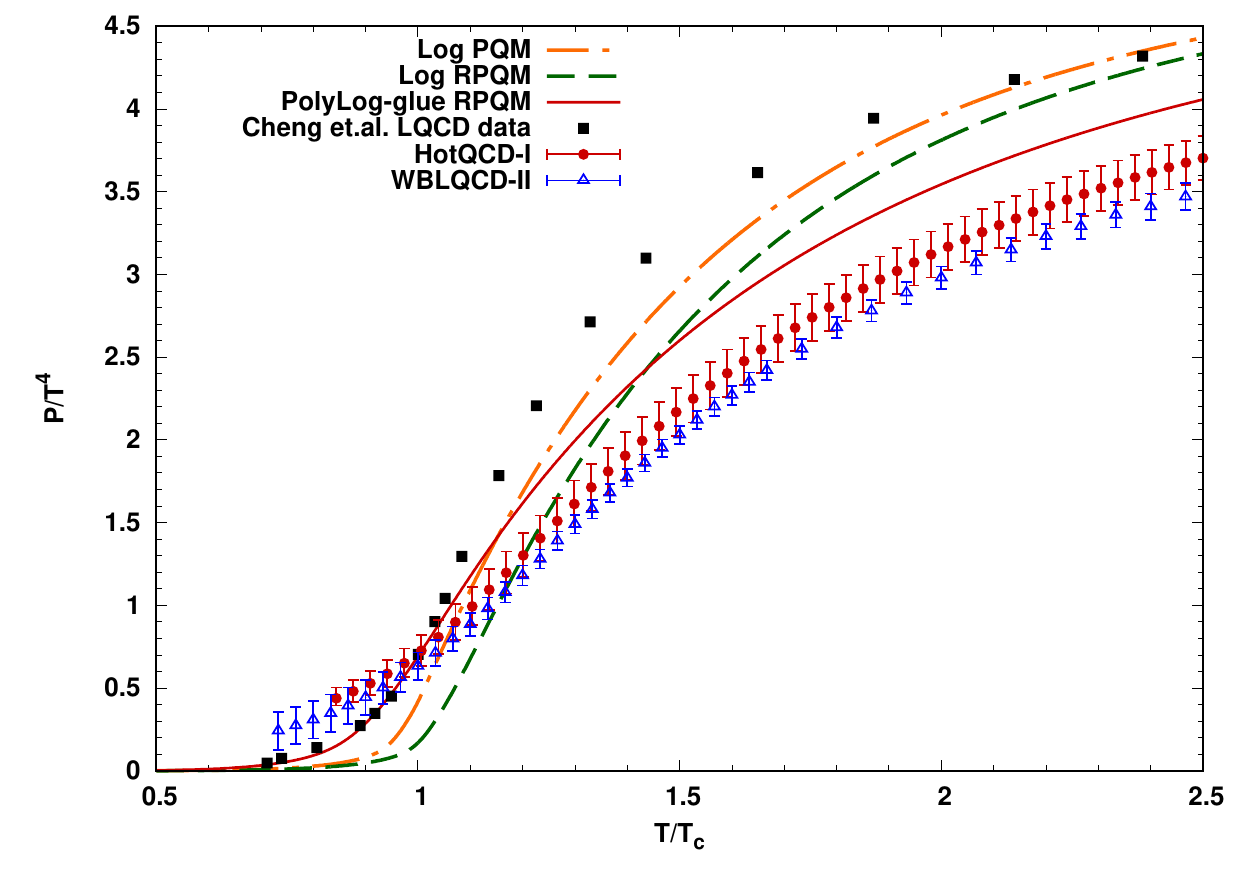}
\caption{Temperature variation of the pressure for the $m_\sigma=500$ MeV at $\mu$=0.~The lattice data of the pressure,~in the continuum limit of the HotQCD-I and WBLQCD-II,~have been taken from the Ref.~\cite{HotQCD2014} and the Ref.~\cite{WB2014} respectively.~The Cheng et.al. LQCD data has been taken from the Ref.~\cite{Cheng:06} for the $N_\tau=6$ in the p4-action.~Here $T_c=T_c^\chi$.}
\label{fig8} 
\end{figure}

The reduced temperature scale variation of the pressure for the PolyLog-glue RPQM model (where the quark back-reaction and the quark one-loop vacuum correction,~both are present) in the Fig.~\ref{fig8},~stands close to the lattice QCD data of the pressure that has been reported by the WBLQCD-II~\cite{WB2014} and the HotQCD-I~\cite{HotQCD2014} collaboration.~The lattice data of Ref.~\cite{Cheng:08} for the pressure, are in exact agreement with the temperature variation of the pressure for the PolyLog-glue RPQM model below $T=1.1 \ T_c$ and above this temperature, the lattice data become closer to the temperature variation of the pressure calculated in the Log PQM model. 

In order to find the thermodynamic quantities other than pressure,~one needs to take temperature derivative of the grand potential.~The entropy density $s$, the energy density $\epsilon$ and the interaction measure $\Delta$ at the zero quark chemical potential, are defined by
\bqa
s&=&-\frac{\partial\Omega}{\partial T}\;,\\
\epsilon &=& -P + Ts\;,\\
\Delta&=&\epsilon-3P\;.
\eqa

\begin{figure*}[!htbp]
\subfigure[\ Energy density]{
\label{fig9a} 
\begin{minipage}[b]{0.48\textwidth}
\centering \includegraphics[width=\linewidth]{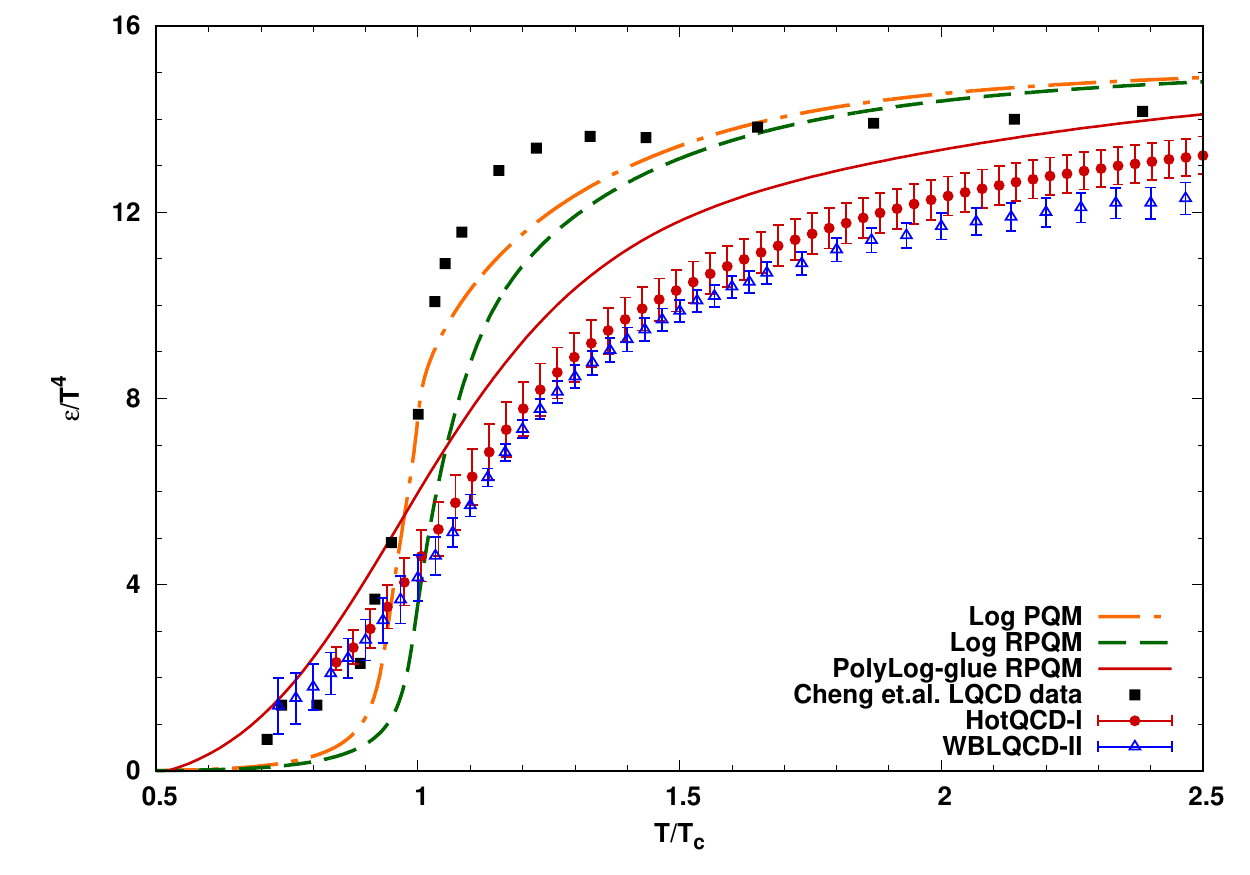}
\end{minipage}}%
\hfill
\subfigure[\ Entropy density]{
\label{fig9b} 
\begin{minipage}[b]{0.48\textwidth}
\centering \includegraphics[width=\linewidth]{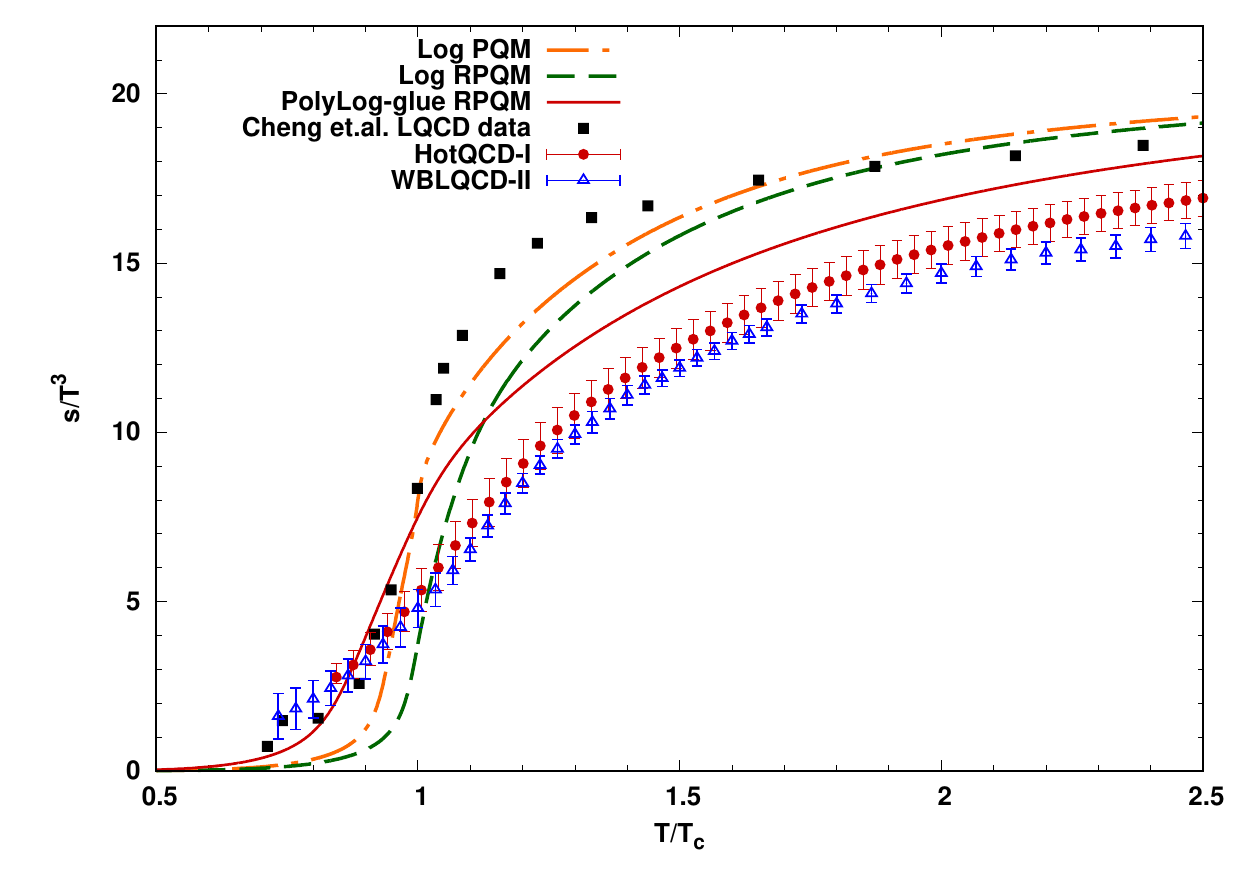}
\end{minipage}}
\caption{Temperature variation of the energy density and entropy density for the $m_\sigma=500$ MeV at $\mu$=0.~The lattice data of the energy density and entropy density,~in the continuum limit of the HotQCD-I and WBLQCD-II,~have been taken from the the Ref.~\cite{HotQCD2014} and the Ref.~\cite{WB2014} respectively.~The Cheng et.al. LQCD data has been taken from the Ref.~\cite{Cheng:06} for the $N_\tau=6$ in the p4-action.~Here $T_c=T_c^\chi$.}
\label{fig:mini:fig9} 
\end{figure*}

The respective temperature variations of the energy density and the entropy density in the Fig.~\ref{fig9a} and the Fig.~\ref{fig9b},~either in the PQM or in the RPQM model with the Log form of the Polyakov-loop potential,~increase very rapidly near the $T_c$ and attain saturation after the temperature 1.5 $T_c$.~The Ref.~\cite{Cheng:08} lattice data of the energy density (entropy density),~are close to the energy density (entropy density) temperature variation of the Log PQM model when the $T<1.6 \ T_c$ and afterwards,~the LQCD data become closer to the energy density (entropy density) obtained in the PolyLog-glue RPQM model calculation.~The lattice QCD data of the WBLQCD-II \cite{WB2014} and the HotQCD-I \cite{HotQCD2014} collaborations for the energy density (entropy density) temperature variations in the Fig.~\ref{fig9a} (Fig.~\ref{fig9b}),~apart from being close to the energy density (entropy density) obtained from the PolyLog-glue RPQM model computation,~show quite a similar rising pattern on the reduced temperature scale. 

\begin{figure*}[!htbp]
\subfigure[\ Interaction measure]{
\label{fig10a} 
\begin{minipage}[b]{0.48\textwidth}
\centering \includegraphics[width=\linewidth]{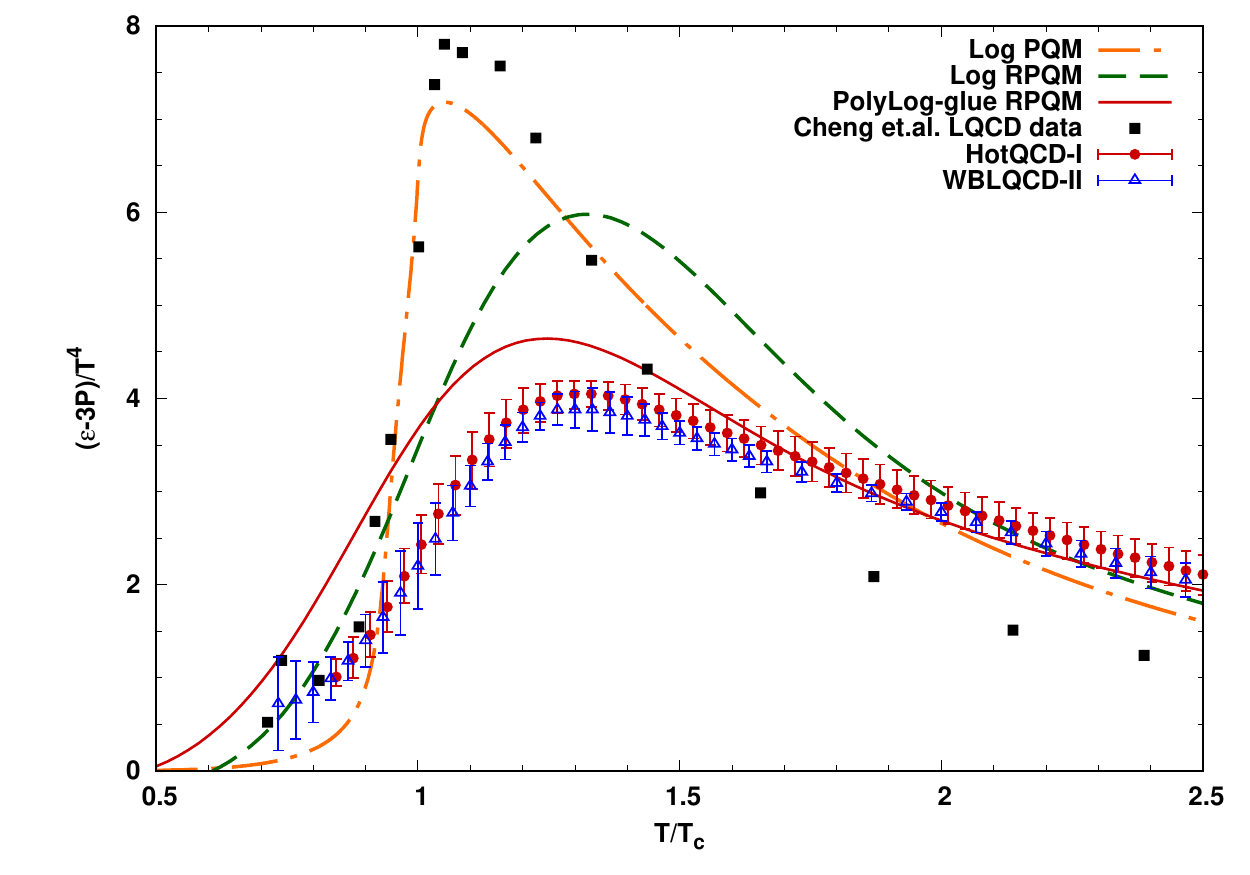}
\end{minipage}}%
\hfill
\subfigure[\ Specific heat]{
\label{fig10b} 
\begin{minipage}[b]{0.48\textwidth}
\centering \includegraphics[width=\linewidth]{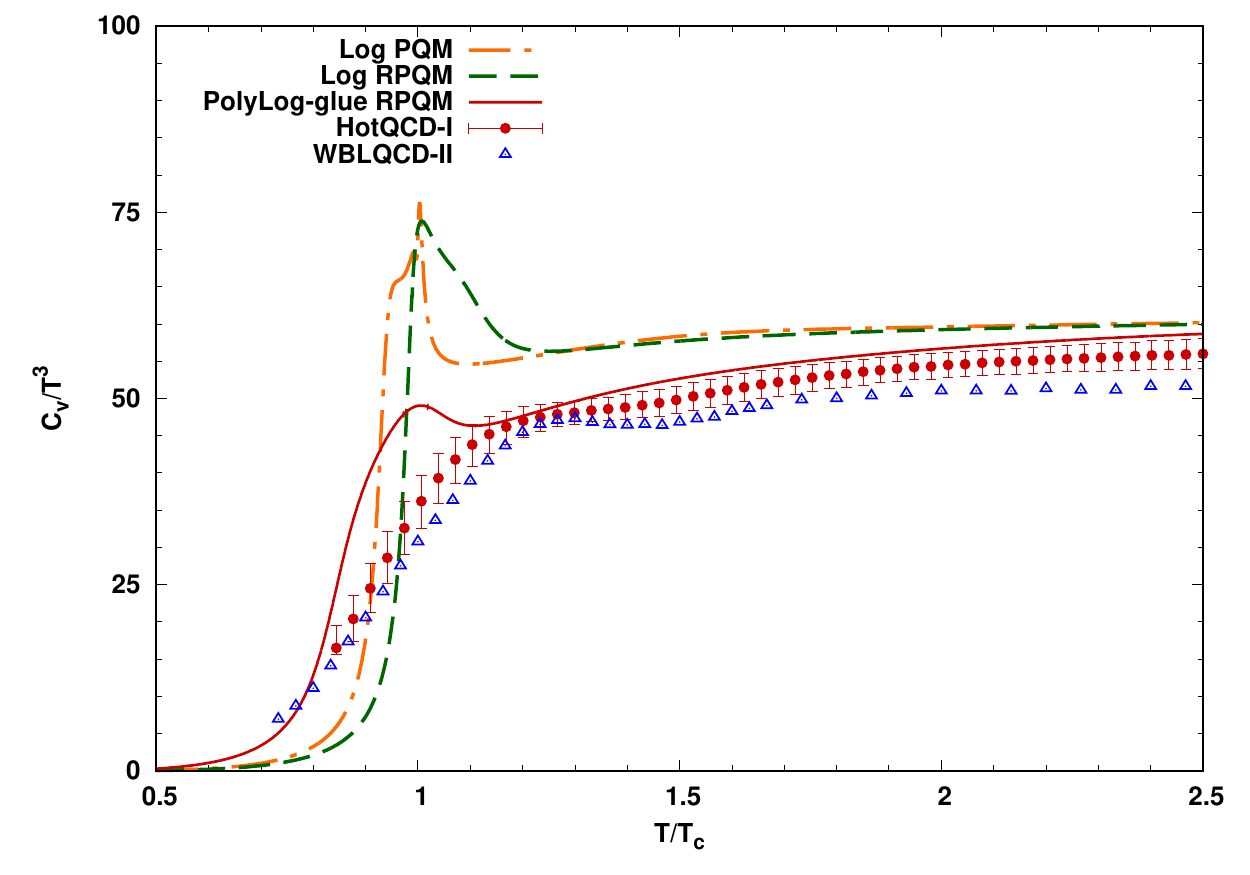}
\end{minipage}}
\caption{Temperature variation of the interaction measure and specific heat for the $m_\sigma=500$ MeV at $\mu$=0.~The lattice data of the interaction measure and specific heat,~in the continuum limit of the HotQCD-I and WBLQCD-II,~have been taken from the the Ref.~\cite{HotQCD2014} and the Ref.~\cite{WB2014} respectively.~The Cheng et.al. LQCD data for interaction measure has been taken from the Ref.~\cite{Cheng:06} for the $N_\tau=6$ in the p4-action.~Here $T_c=T_c^\chi$.}
\label{fig:mini:fig10} 
\end{figure*}

The interaction measure is found to have highest rising temperature variation for the Log PQM model in the Fig.~\ref{fig10a} and it matches well with the several data points of the Ref.~\cite{Cheng:08} lattice data of the interaction measure.~When the quark back-reaction and the quark one-loop vacuum correction,~both are present in the PolyLog-glue RPQM model,~the calculated temperature variation of the interaction measure with a significantly reduced height,~shows quite a close resemblance in its pattern with the latest lattice data for the temperature variation of the interaction measure (also lattice data for the  $T>1.5 \ T_c$ has a good match with the theoretical calculation) reported by the WBLQCD-II \cite{WB2014} and the HotQCD-I \cite{HotQCD2014} collaborations. 

The expression for the specific heat capacity at constant volume is written as 
\bqa
C_{V}=\frac{\partial \epsilon}{\partial T}\bigg{\rvert}_V=-T\frac{\partial^2 \Omega}{\partial T^2}\bigg{\rvert}_V
\eqa

The specific heat (normalized with $T^3$) variation on the reduced temperature scale in the Fig.~\ref{fig10b},~shows the highest and sharpest peak near $T_c$ in the Log PQM model which becomes smoother in the Log RPQM model.~The quark one-loop vacuum correction in the presence of the quark back-reaction in the PolyLog-glue RPQM model,~gives rise to a most smooth temperature variation of the  specific heat that has a well rounded peak structure.~The abovementioned pattern has a close resemblance with the specific heat temperature variation reported in the WBLQCD-II \cite{WB2014} and the HotQCD-I \cite{HotQCD2014} lattice data.~Furthermore,~the PolyLog-glue RPQM model calculation of the specific heat temperature variation for the $T>1.1 \ T_c$ matches very well with the said lattice data for the specific heat.

\begin{figure*}[!htbp]
\subfigure[\ Squared speed of sound ($C_s^2$)]{
\label{fig11a} 
\begin{minipage}[b]{0.48\textwidth}
\centering \includegraphics[width=\linewidth]{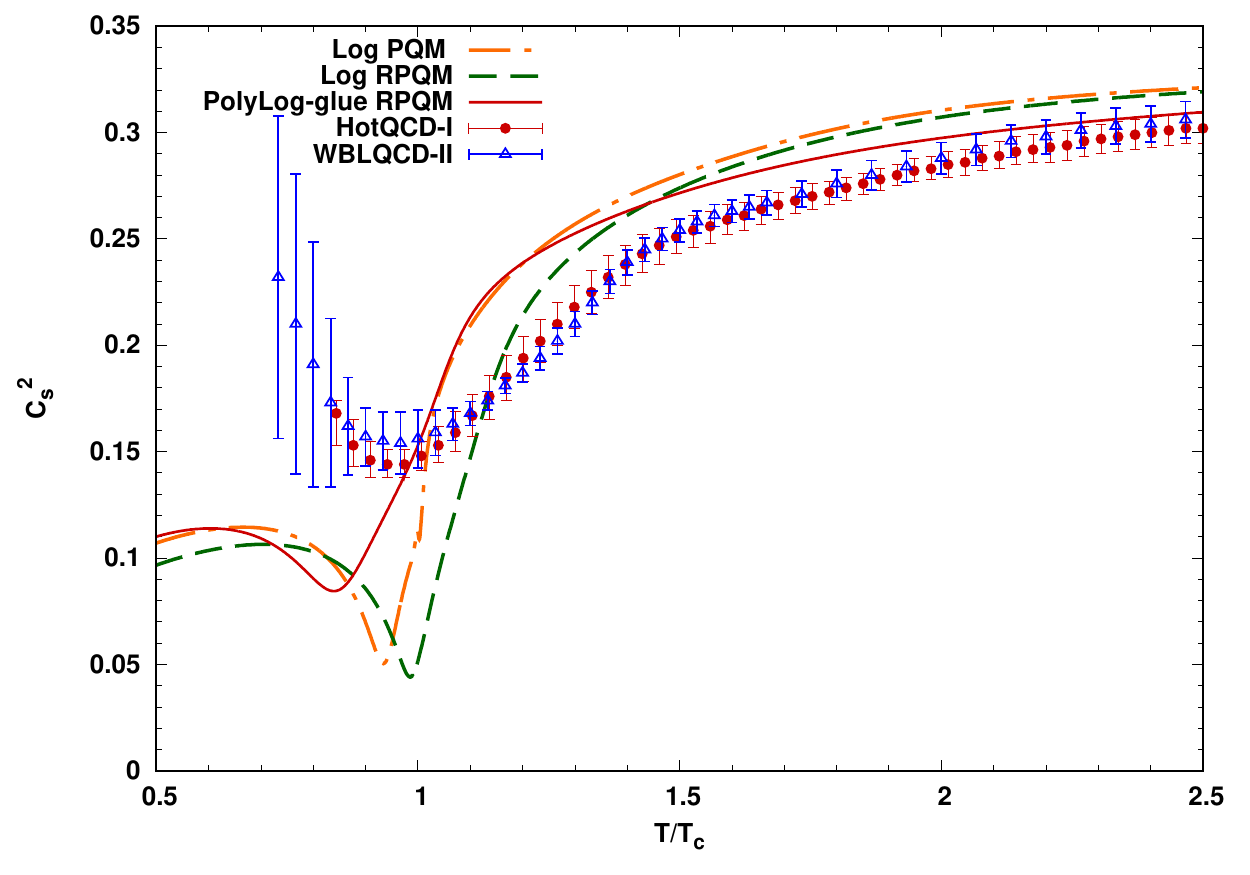}
\end{minipage}}%
\hfill
\subfigure[\ Ratio of pressure with energy ]{
\label{fig11b} 
\begin{minipage}[b]{0.48\textwidth}
\centering \includegraphics[width=\linewidth]{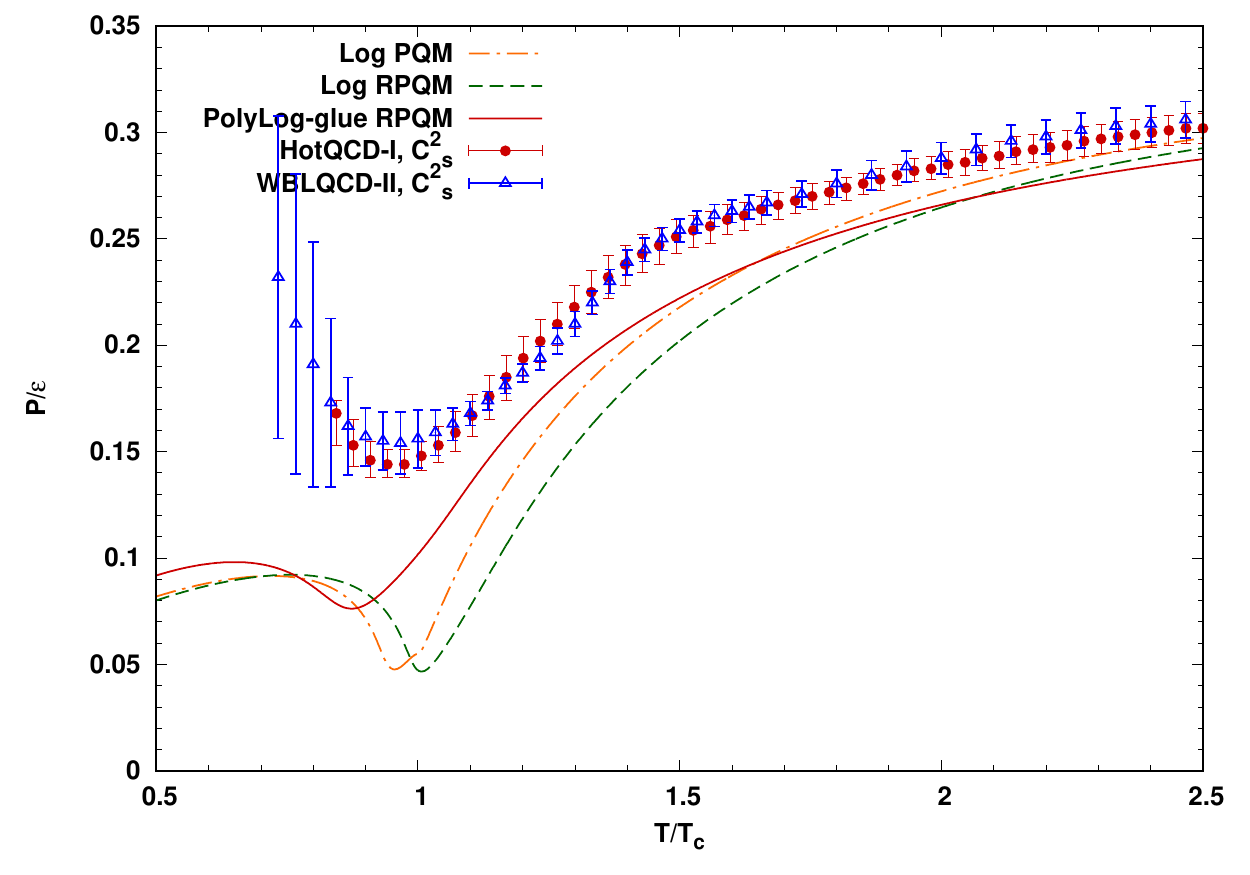}
\end{minipage}}
\caption{Temperature variation of the $C^2_s$ and $P/\epsilon$ for the $m_\sigma=500$ MeV at $\mu$=0.~The lattice data of the $C_s^2$,~in the continuum limit of the HotQCD-I and WBLQCD-II,~have been taken from the the Ref.~\cite{HotQCD2014} and the Ref.~\cite{WB2014} respectively.~In the Fig.(b) model calculations of the $P/\epsilon$ have been compared with the lattice data of $C_s^2$.~Here $T_c=T_c^\chi$.}
\label{fig:mini:fig11} 
\end{figure*}

In the strongly interacting medium,~the speed of sound is a very important quantity and at fix entropy density $s$ the square of the speed of sound is define by
\bqa
C^2_s=\frac{\partial P}{\partial \epsilon}\bigg{\rvert}_s=\frac{\partial P}{\partial T}\bigg{\rvert}_V\bigg{/}\frac{\partial \epsilon}{\partial T}\bigg{\rvert}_V=\frac{s}{C_V}.
\eqa

The reduced temperature scale variations of the $C^2_s$ ($P/\epsilon$) in the Fig.~\ref{fig11a} (Fig.~\ref{fig11b}),~for the Log PQM as well as RPQM and the PolyLog-glue RPQM model,~are compared with the WBLQCD-II \cite{WB2014} and the HotQCD-I \cite{HotQCD2014} lattice data.~The $C^2_s$ temperature variation pattern of the  PolyLog-glue RPQM model in the Fig.~\ref{fig11a},~resembles  closely with the rising trend of the $C^2_s$ that one observes in the lattice QCD data and agreement between the results become quite good after the $T>1.4 \ T_c$ when the lattice data approaches the model result from below.~The lattice data of the $C^2_s$ lies just above the $P/\epsilon$ temperature variations which have been computed in the models.~The $P/\epsilon$ temperature variation pattern in the PolyLog-glue model while closely resembling with the rising pattern of the $C^2_s$ lattice data,~lies just below it upto the temperature 1.7 $T_c$.
 
\begin{table*}[!htbp]
    \caption{Critical end points.}
    \label{tab:table4}
    \begin{tabular}{p{2cm}| p{4cm} |p{3cm} |p{3cm}|p{3cm} }
      \toprule 
      Polyakov-loop & Models & $m_\sigma=400 \  \text{MeV}$ & $m_\sigma=500 \  \text{MeV}$ & $m_\sigma=600 \  \text{MeV}$ \\
       & & $(T_{\cep},\mu_{\cep})$ MeV & $(T_{\cep},\mu_{\cep})$ MeV & $(T_{\cep},\mu_{\cep})$ MeV\\
      \hline 
      \hline
      Log & PQM & $(149.0,26.4)$ & $(146.6,124.7)$& $(141.0,189.8)$\\
             & RPQM & $( 93.8,230.5)$ & $(94.6,252.0)$& $(47.7,316.3)$\\
             & RPQM ($T_0=270$ MeV) & $(104.23,237.8)$ & $-$& $-$\\
  \hline
      PolyLog-glue & PQM & $(138.5,61.1)$ & $(137.7,129.6)$& $(130.4,193.9)$\\
            & RPQM  & $(73.6,227.6)$ & $(70.9,252.7)$& $(25.7,316.1)$ \\
      \hline 
    \end{tabular}
\end{table*}
\subsection{Illustration of the phase diagrams and their CEP positions}
\label{sec:IVB}
\begin{figure}[!htbp]
\centering \includegraphics[width=\linewidth]{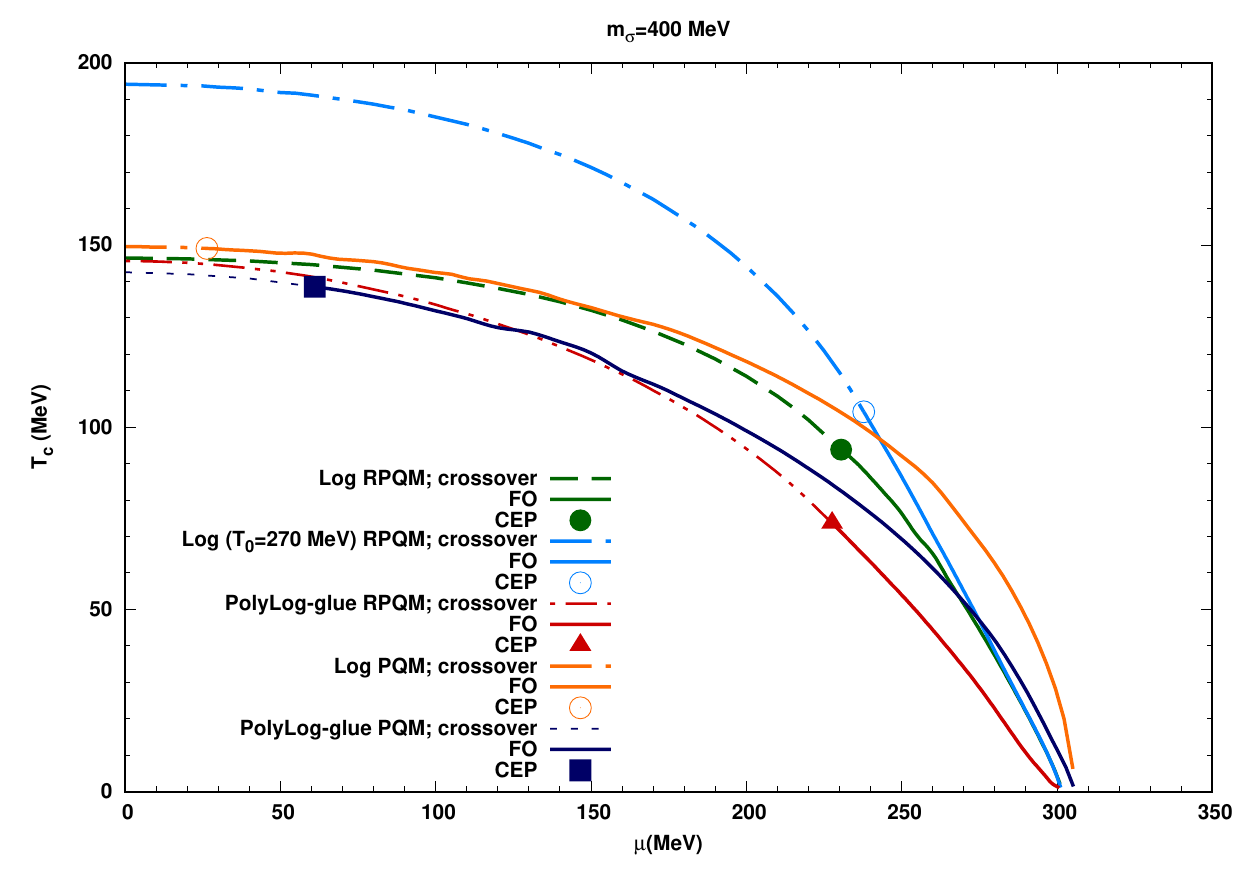}
\caption{Phase diagrams for the $m_\sigma=400$ MeV.}
\label{fig12}
\end{figure}

\begin{figure}
\centering \includegraphics[width=\linewidth]{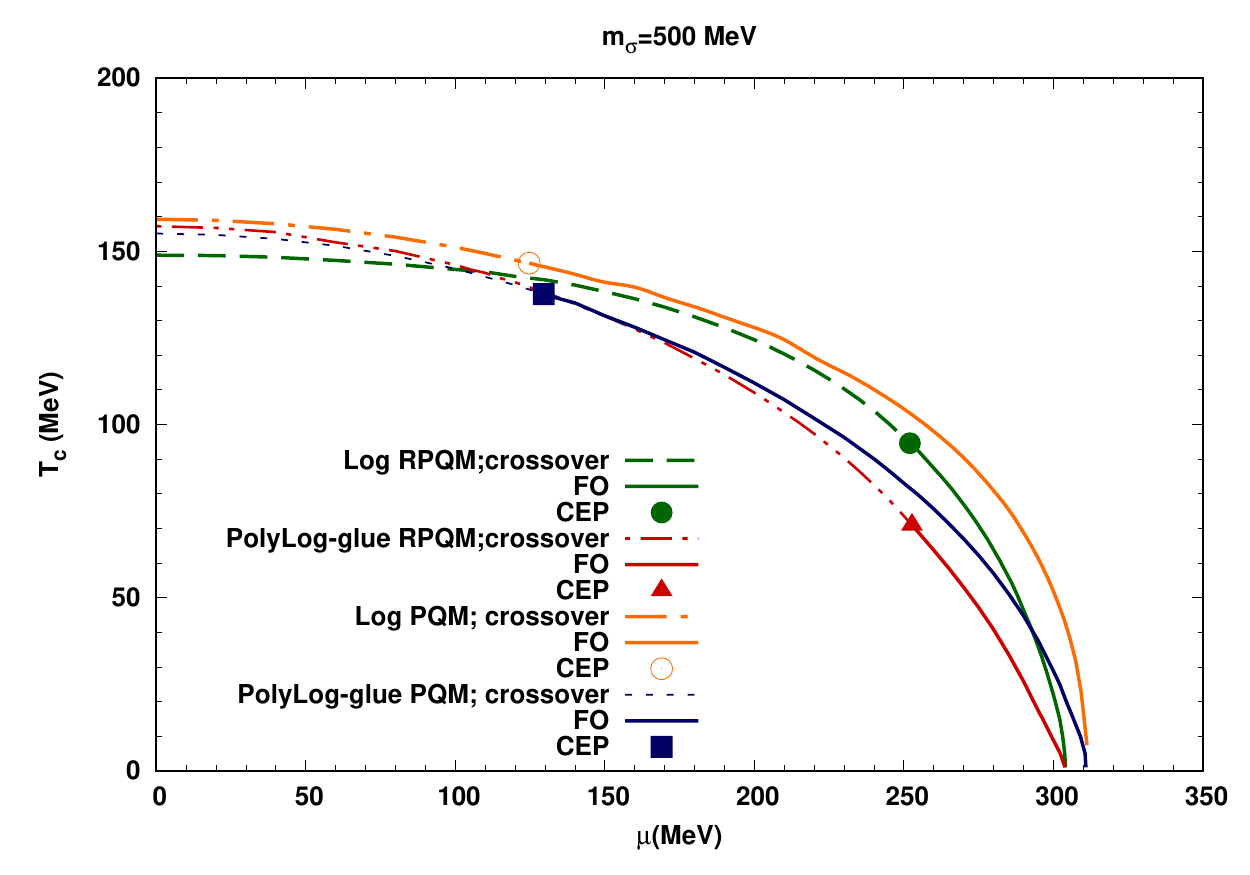}
\caption{Phase diagrams for the $m_{\sigma}=500 $ MeV.} 
\label{fig13} 
\end{figure}

\begin{figure}[!htbp]
\centering \includegraphics[width=\linewidth]{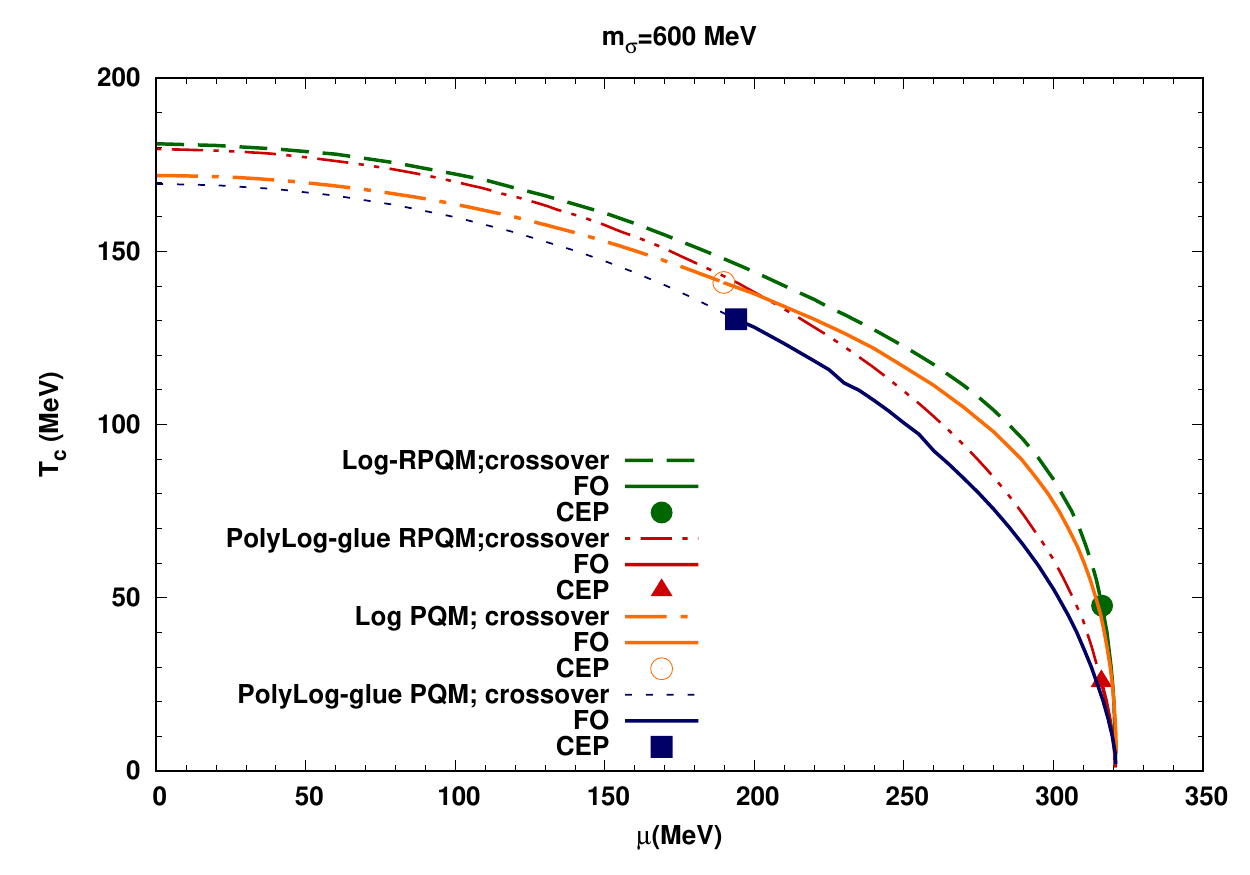}
\caption{Phase diagrams for the $m_\sigma=600$ MeV.}
\label{fig14} 
\end{figure}

The phase diagrams of the PQM and the RPQM model with the Log and the PolyLog-glue form of the Polyakov-loop potential,~have been presented in the Fig.~\ref{fig12},~Fig.~\ref{fig13} and Fig.~\ref{fig14} respectively for the $m_\sigma=$ 400,~500 and 600 MeV.~The table~\ref{tab:table4} presents comparison of the positions of CEP in different model senarios.~In the Fig.~\ref{fig12} for the $m_\sigma=$ 400 MeV and the $T_0=$ 187 MeV,~the PQM model CEP position at the $(T_\cep,\mu_\cep)=(149.0,26.4)$ MeV $\{(T_\cep,\mu_\cep)=(138.5,61.1) \ \mev\}$ for the Log $\{$PolyLog-glue$\}$ Polyakov-loop potential,~shifts to the position $(T_\cep,\mu_\cep)=(93.8,230.5)$ MeV $\{(T_\cep,\mu_\cep)=(73.6,227.6) \ \mev \}$  in the Log $\{$PolyLog-glue$\}$ RPQM model due to the quark one-loop vacuum correction.~Comparing the CEP of the Log RPQM model with that of the  PolyLog-glue RPQM model,~we find that the CEP moves down in the temperature direction by 20.2 MeV due to the effect of the quark back-reaction in the PolyLog-glue RPQM model while it shifts leftward on the chemical potential axis by a small amount (only 2.9 MeV).~In contrast to the above explained behavior,~the temperature down shift in the PolyLog-glue PQM model CEP position is only 10.5 MeV when compared to the CEP of the Log PQM model while its position in the chemical potential direction increases (moves right) by 34.7 MeV.~Our RPQM model result shows the confirmation of the Ref.~\cite{BielichP} observation that the quark back-reaction due to the unquenching of the Polyakov loop potential links the chiral and deconfinement phase transition also at small temperature and large chemical potential.~One also finds that the curvature of the phase transition line increases due to the above effect.

We point out that when the quark one-loop vacuum correction is added to the effective potential of the QM model,~and the curvature masses of the mesons are used to fix the model parameters as in the Refs.~\cite{guptiw,schafwag12,chatmoh1,vkkr12},~the calculated shift in the position of the CEP is quite large and over-estimated as reported in our recent two and 2+1 flavor renormalized quark meson model investigations \cite{RaiTiw,vkkr22,raiti2023} where the consistent and exact parameter fixing have been done using on-shell method.~After including the quark one-loop vacuum correction and fixing the model parameters by the use of curvature masses of the mesons,~Schaefer et.al. \cite{schafwag12},~in their 2+1 flavor PQM model work with the Log Polyakov-loop potential whose parameter $T_0=270$ MeV (for the Pure Yang-Mills $SU_c(3)$ gauge theory),~find the CEP at the $(T_\cep,\mu_\cep)=(90.0,283.0) \  \mev$ when the $m_\sigma=400$ MeV.~It is worth emphasizing that if we take the Log Polyakov-loop potential parameter $T_0=270$ MeV in our on-shell parameterized 2+1 flavor RPQM model,~we find that in our calculation,~the CEP coordinate $(T_\cep,\mu_\cep)=(104.23,237.8) \ \mev$ lies at quite a higher up position in the temperature direction by 14.23 MeV,~having also a significantly robust chemical potential direction shift of  45.2 MeV towards the left in the phase diagram,~when it is compared with the corresponding phase diagram given in the Fig.1(b) of the Ref.~\cite{schafwag12}.

When the $m_\sigma$ increases from 400 to 500 MeV in the Fig.~\ref{fig13},~the CEP moves rightwards on the chemical potential axis.~We find that when the $m_\sigma=$ 500 MeV,~the CEP in the PQM model
at the $(T_\cep,\mu_\cep)=(146.6,124.7) \{(137.7,129.6)\}$ MeV for the Log $\{$PolyLog-glue$\}$ Polyakov-loop potential,~moves to the position $(T_\cep,\mu_\cep)=(94.6,252.0)\{(70.9,252.7) \ \}$ MeV  in the Log $\{$PolyLog-glue$\}$ RPQM model.~Comparing the CEP of the Log RPQM model with that of the  PolyLog-glue RPQM model,~one notes that here  for the $m_\sigma$= 500 MeV case,~the CEP moves down in temperature by a larger amount of 23.7 MeV (compared to the $m_\sigma=$ 400 MeV) due to the effect of the quark back-reaction while its rightward shift in the chemical potential direction is negligibly small (only 0.7 MeV).~In contrast,~the temperature down shift in the PolyLog-glue PQM model CEP position is only 8.9 MeV when compared to the CEP of the Log PQM model while its position in the chemical potential direction shifts rightwards by 4.9 MeV.~Here also we find that the curvature of the phase transition line increases due to the quark back-reaction as in the Ref.~\cite{BielichP}.

The PQM model CEP at the $(T_\cep,\mu_\cep)=(141.0, \ 189.8) \ \{(130.4, \ 193.9)\}$ MeV when the $m_\sigma=600$ MeV in the Fig.~\ref{fig14},~for the Log $\{$PolyLog-glue$\}$ Polyakov-loop potential,~moves to the position $(T_\cep,\mu_\cep)=(47.7, \ 316.3) \ \{(25.7, \ 316.1)\}$ MeV  in the Log $\{$PolyLog-glue$\}$ RPQM model.~Here for the $m_\sigma$= 600 MeV also,~the Log RPQM model CEP shifts,~vertically down on the temperature axis by 22.0 MeV due to the unquenching of the Polyakov-loop potential while the chemical potential is almost the same.~Here it is relevant to emphasize that we are getting a well placed CEP in the phase diagram of the RPQM model even for the case of the $m_\sigma$= 600 MeV while the curvature mass parametrization of the PQM model with the quark one-loop vacuum correction,~generates such an excessively large soothing effect on the chiral phase transition that the CEP altogether disappears from the phase diagram and one gets a complete chiral crossover  transition in the entire $\mu-T$ plane as reported in the Refs.~\cite{guptiw,chatmoh1,schafwag12}.

\section{Summary and Conclusion}
\label{sec:V}
The consistent and improved chiral effective potential of the renormalized 2+1 flavor quark meson model (RQM),~whose parameters are fixed on-shell after the inclusion of quark one-loop vacuum correction,~has been augmented with the different forms of the Polyakov-loop potential with and without quark back-reaction.~The resulting Quantum Chromodynamics (QCD) like framework of the 2+1 flavor renormalized Polykov quark meson (RPQM) model has been used to investigate the interplay of the chiral symmetry restoring and the deconfinement phase transitions.~The results have been compared with the  PQM model where the fermionic vacuum fluctuations are neglected.~The temperature variation of the thermodynamics quantities  ;~pressure,~entropy density,~energy density,~interaction measure,~specific heat,~speed of sound and the ratio $P/\epsilon$ have been computed at $\mu=0$ MeV and compared with the latest 2+1 flavor lattice data  of the Wuppertal-Budapest and HotQCD collaborations \cite{Cheng:08,WB2014,HotQCD2014}.~The PQM and RPQM model phase diagrams for the $m_\sigma=400$, 500, 600 MeV have been computed and compared with/without the quark back-reaction in the Polyakov-loop potential.  

The most smooth temperature variation of the non-strange chiral condensate is observed for the PolyLog-glue RPQM model.~Since the strange direction explicit symmetry breaking strength $h_y$ is reduced by a relatively  large amount after renormalization the melting of the strange condensate is significantly large in the RPQM model.~The temperature variation of the subtracted chiral condensate $\Delta_{ls}$ calculated in the PolyLog-glue RPQM model  at $\mu=$ 0 and the $m_\sigma$= 500 MeV,~shows the best agreement with the WBQCD-I lattice data of the $\Delta_{ls}$.~The lattice data \cite{Cheng:08} for the Polyakov loop condensate in the temperature range 0.7 $T_c$ to 1.7 $T_c$,~shows quite a close resemblance with the Polyakov loop temperature variation that we get in our PolyLog-glue RPQM model calculation.

The  lattice data in Ref.~\cite{Cheng:08},~for the temperature variation of the pressure when the $ T \ < 1.1 \ T_c$ are close to the PolyLog-glue RPQM model result and above this temperature the pressure computed in the Log PQM model agrees well with the data.~The recent  WBLQCD-II \cite{WB2014} and the HotQCD-I \cite{HotQCD2014}    lattice data for the pressure,~entropy density and energy density are quite close to their respective temperature variation that we obtain in  our PolyLog-glue RPQM model calculation.~The lattice data ~\cite{Cheng:08} for the energy density and entropy density,~have closer agreement with their respective temperature variations in the  Log PQM model when the $ T \ < 1.6 \ T_c$ and above this temperature,~their agreement becomes better with the corresponding PolyLog-glue RPQM model results.~The interaction measure peak is smallest in its PolyLog-glue RPQM model temperature variation and agrees well with the $T>1.5\ T_c$.~The very smooth and rounded peak noticed in the specific heat temperature variation of the lattice QCD data,~has quite a close resemblance with our PolyLog-glue RPQM results.~The lattice data of the $C^2_s$ \cite{WB2014,HotQCD2014} temperature variation is close to the PolyLog-glue RPQM model temperature variations for the $C_s^2$ and the ratio $P/\epsilon$ when the $T>1.4 T_c$.

The CEP of the Log RPQM model moves significantly down ($\Delta T_\cep$= 20.2 - 23.7 MeV)   on the temperature axis due to the presence of the quark back-reaction in the PolyLog-glue RPQM model while the shift in the chemical potential direction is negligible.~This finding confirms the observation of the Ref.~\cite{BielichP} that unquenching of the Polyakov-loop potential links the chiral and deconfinement phase transitions at all temperatures and chemical potentials.~We point out that the smoothing effect of the quark one-loop vacuum correction on the chiral transition phase boundary is excessively large and over estimated in the earlier investigations~\cite{chatmoh1,schafwag12} where the curvature masses of the mesons have been used for fixing the model parameters while the $f_\pi$ and $f_K$ are taken as constant.~In these studies, one gets a very large crossover and a smaller first order region.~Therefore,~the CEP gets located in the right-most lower corner of the $\mu-T$ plane at smaller temperatures and very high chemical potentials.~We emphasize that the smoothing effect of the quark one-loop vacuum correction becomes moderate in our consistent on-shell parametrized RPQM model studies because we find comparatively larger first order region in the $\mu-T$ plane of the phase diagram for the $m_\sigma=$ 400 and 500 MeV and therefore the CEP moves higher up in the phase diagram towards the temperature axis.~Note that we are getting a well placed CEP in the phase diagram of the RPQM model even for the case of the $m_\sigma$= 600 MeV while the CEP altogether disappears and one gets a complete chiral crossover  transition in the entire $\mu-T$ plane of the phase diagram reported in the studies~\cite{chatmoh1,schafwag12} where the quark one-loop vacuum correction is included in the PQM model but the parameters are fixed using the curvature masses of the mesons.

\section*{Acknowledgments}

The authors would like to thank people of India.

\appendix
\section{INTEGRALS }
\label{appenA}
The divergent loop integrals are regularized by incorporating dimensional regularization.
\bqa
\int_p=\left(\frac{e^{\gamma_E}\Lambda^2}{4\pi}\right)^\epsilon\int \frac{d^dp}{(2\pi)^d}\;,
\eqa
where $d=4-2\epsilon$ , $\gamma_E$ is the Euler-Mascheroni constant, and $\Lambda$ is renormalization scale associated with the $\overline{\text{MS}}$.

\bqa
\nonumber
\mathcal{A}(m^2_f)&=&\int_p \frac{1}{p^2-m^2_f}=\frac{i m^2_f}{(4\pi)^2}\left[\frac{1}{\epsilon}+1\right. \\
\nonumber
&&\left.+\ln(4\pi e^{-\gamma_E})+\ln\left(\frac{\Lambda^2}{m^2_f}\right)\right]\;,
\eqa

we rewrite this after redefining $\Lambda^2\longrightarrow \Lambda^2\frac{e^{\gamma_E}}{4\pi}$.

\bqa
\label{aint1}
\mathcal{A}(m^2_f)&=&\frac{i m^2_f}{(4\pi)^2}\left[\frac{1}{\epsilon}+1+\ln\left(\frac{\Lambda^2}{m^2_f}\right)\right]\;,
\eqa

\bqa
\label{bint1}
\nonumber
\mathcal{B}(p^2,m_f)&=&\int_k \frac{1}{(k^2-m^2_f)[(k+p)^2-m^2_f)]} \\
&=&\frac{i}{(4\pi)^2}\left[\frac{1}{\epsilon}+\ln\left(\frac{\Lambda^2}{m^2_f}\right)+\mathcal{C}(p^2,m_f)\right]\;,
\eqa

\bqa
\label{bprimeint1}
\mathcal{B}^\prime(p^2,m_f)&=&\frac{i}{(4\pi)^2}\mathcal{C}^\prime(p^2,m_f)\;,
\eqa

\begin{widetext}
\begin{equation}
\label{eq:cp1}
\mathcal{C}(p^2,m_f)=2-2\sqrt{\dfrac{4 m^2_f}{p^2}-1}\arctan\left(\dfrac{1}{\sqrt{\dfrac{4 m^2_f}{p^2}-1}}\right); \quad  \mathcal{C}^{\prime}(p^2,m_f)=\frac{4 m^2_f}{p^4\sqrt{\dfrac{4 m^2_f}{p^2}-1}}\arctan\left(\dfrac{1}{\sqrt{\dfrac{4 m^2_f}{p^2}-1}}\right)-\frac{1}{p^2}\;,
\end{equation}
\begin{equation}
\label{eq:cp2}
\mathcal{C}(p^2,m_f)=2+\sqrt{1-\dfrac{4 m^2_f}{p^2}}\ln\left(\dfrac{1-\sqrt{1-\dfrac{4 m^2_f}{p^2}}}{1+\sqrt{1-\dfrac{4 m^2_f}{p^2}}}\right); \quad \mathcal{C}^{\prime}(p^2,m_f)=\frac{2 m^2_f}{p^4\sqrt{\dfrac{4 m^2_f}{p^2}-1}}\ln\left(\dfrac{1-\sqrt{1-\dfrac{4 m^2_f}{p^2}}}{1+\sqrt{1-\dfrac{4 m^2_f}{p^2}}}\right)-\frac{1}{p^2}\;,
\end{equation}
The Eqs.(\ref{eq:cp1}) and (\ref{eq:cp2}) are valid with the constraints  ($p^2<4m^2_f$) and ($p^2>4m^2_f$) respectively. 
\begin{align}
\label{bint1}
&\mathcal{B}(p^2,m_u,m_s)=\int_k \frac{1}{(k^2-m^2_s)[(k+p)^2-m^2_u)]}=\frac{i}{(4\pi)^2}\left[\frac{1}{\epsilon}+\ln\left(\frac{\Lambda^2}{m^2_u}\right)+\mathcal{C}(p^2,m_u,m_s)\right]\;,\\
&\mathcal{C}(p^2,m_u,m_s)=2-\frac{1}{2}\Biggl[ 1+\frac{m_s^2-m_u^2}{p^2}\Biggr]\ln\left(\frac{m_s^2}{m_u^2}\right)-\frac{\mathcal{G}(p^2)}{p^2}\Biggl[\arctan\left(\frac{p^2-m^2_s+m^2_u}{\mathcal{G}(p^2)}\right)+\arctan\left(\frac{p^2+m^2_s-m^2_u}{\mathcal{G}(p^2)}\right)\Biggr]\;, \quad \quad\\
&\mathcal{G}(p^2)=\sqrt{\{(m_s+m_u)^2-p^2\}\{p^2-(m_s-m_u)^2\}}\;,\\ \nonumber
&\mathcal{C}^\prime(p^2,m_u,m_s)=\frac{m_s^2-m_u^2}{2p^4}\ln\Biggl(\frac{m^2_s}{m^2_u}\Biggr)+\frac{p^2(m_s^2+m_u^2)-(m_s^2-m_u^2)^2}{p^4\mathcal{G}(p^2)}\Biggl[\arctan\Biggl(\frac{(p^2-m^2_s+m^2_u)}{\mathcal{G}(p^2)}\Biggr)\\
&\qquad \qquad \qquad \quad+\arctan\Biggl(\frac{(p^2+m^2_s-m^2_u)}{\mathcal{G}(p^2)}\Biggr)\Biggr]-\frac{1}{p^2}\;.
\end{align}
\end{widetext}



\begin{thebibliography}{99}
\bibitem{Cabibbo75}
N.~Cabibbo and G.~Parisi,
\href{https://doi.org/10.1016/0370-2693(75)90158-6}{Phys.Lett.\textbf{B 59},67-69(1975)}.

\bibitem{SveLer}
L. D. McLerran and B. Svetitsky, \href{https://doi.org/10.1103/PhysRevD.24.450}{Phys. Rev. D 24, 450
(1981)}; B. Svetitsky, \href{https://doi.org/10.1016/0370-1573(86)90014-1}{Phys. Rep. 132, 1 (1986)}.
\bibitem{Mull}
B. Muller, \href{https://doi.org/10.1088/0034-4885/58/6/002}{Rep. Prog. Phys. 58, 611 (1995)}.
\bibitem{Ortms}
H. Meyer-Ortmanns, \href{https://doi.org/10.1103/RevModPhys.68.473}{Rev. Mod. Phys. 68, 473 (1996)}.
\bibitem{Riske}
D. H. Rischke, \href{https://doi.org/10.1016/j.ppnp.2003.09.002}{Prog. Part. Nucl. Phys. 52, 197 (2004)}.
\bibitem{AliKhan:2001ek}
A.Ali~Khan et~al.
\href{https://doi.org/10.1103/PhysRevD.64.074510}{Phys. Rev. {\bf D 64}, 074510 (2001)}.
\bibitem{Digal:01}
S. Digal, E. Laermann and H. Satz,
\href{https://doi.org/10.1007/s100520000538}{Eur. Phys. J. {\bf C 18}, 583 (2001)}.
\bibitem{Karsch:02}
F. Karsch,\href{https://doi.org/10.1007/3-540-45792-5_6}{
Lect. Notes Phys. {\bf 583}, 209 (2002)}.

\bibitem{Fodor:03}
Z. Fodor, S. D. Katz, and K. K. Szabo,
\href{https://doi.org/10.1016/j.physletb.2003.06.011}{Phys. Lett. {\bf B 568}, 73 (2003)}.
\bibitem{Allton:05}
C. R. Allton, M. Doring, S. Ejiri, S. J. Hands, O. Kaczmarek, F. Karsch, 
E Laermann and K. Redlich,
\href{https://doi.org/10.1103/PhysRevD.71.054508}{Phys. Rev. {\bf D 71}, 054508 (2005)}.
\bibitem{Karsch:05}
F. Karsch,
\href{https://doi.org/10.1088/0954-3899/31/6/002}{J. Phys. {\bf G 31}, S633 (2005)}.  
\bibitem{Aoki:06}
Y. Aoki, Z. Fodor, S. D. Katz and K. K. Szabo,
\href{https://doi.org/10.1016/j.physletb.2006.10.021}{Phys. Lett. {\bf B 643}, 46 (2006)}.
\bibitem{Cheng:06}
M. Cheng et al., 
\href{https://doi.org/10.1103/PhysRevD.74.054507}{Phys. Rev. {\bf D 74}, 054507 (2006)}.
\bibitem{Cheng:08}
M. Cheng et al., 
\href{https://doi.org/10.1103/PhysRevD.77.014511}{Phys. Rev. {\bf D 77}, 014511 (2008)}.
\bibitem{JLange}
J. Langelage, S. Lottini, and O. Philipsen, \href{https://doi.org/10.1007/JHEP02(2011)057}{J. High Energy Phys. {\bf 02}
(2011) 057}.
\bibitem{Alf}
M. G. Alford, A. Schmitt and K .Rajagopal, 
\href{https://doi.org/10.1103/RevModPhys.80.1455}{Rev. Mod. Phys. {\bf 80}, 1455 (2008)}. 
\bibitem{Fukhat}
K. Fukushima, and T. Hatsuda,
\href{https://doi.org/10.1088/0034-4885/74/1/014001}{Rep. Prog. Phys. {\bf 74}, 014001 (2011)}.
\bibitem{tHooft:76prl}
G. 't Hooft, 
Phys. Rev. Lett. {\bf 37}, 8 (1976);


\bibitem{Rischke:00}
J. T. Lenaghan, D. H. Rischke and J. Schaffner-Bielich,
\href{https://doi.org/10.1103/PhysRevD.62.085008}{Phys. Rev {\bf D 62}, 085008 (2000)}.
J. T. Lenaghan, D. H. Rischke,
\href{https://doi.org/10.1088/0954-3899/26/4/309}{J. Phys. {\bf G 26}, 431 (2000)}.

\bibitem{Schaefer:09}
B. J. Schaefer and M. Wagner,
\href{https://doi.org/10.1103/PhysRevD.79.014018}{Phys. Rev. {\bf D 79}, 014018 (2009)}.


\bibitem{Roder}
D. Roder,J. Ruppert and D. H. Rischke,
\href{https://doi.org/10.1103/PhysRevD.68.016003}{Phys. Rev. {\bf D 68}, 016003 (2003)}.
\bibitem{fuku11}
K. Fukushima,K. Kamikado and B. Klein,
\href{https://doi.org/10.1103/PhysRevD.83.116005}{Phys. Rev. {\bf D 83}, 116005 (2011)}.
\bibitem{grahl}
M. Grahl  and D. H. Rischke,
\href{https://doi.org/10.1103/PhysRevD.88.056014}{Phys. Rev. {\bf D 88}, 056014 (2013)}.



\bibitem{jakobi}A. Jakovac, A. Patkos, Z. Szep, and P. Szepfalusy, 
\href{https://doi.org/10.1016/j.physletb.2004.01.008}{Phys. Lett. {\bf B 582}, 179 (2004)}.

\bibitem{Herpay:05}
T. Herpay, A. Patk\'{o}s, Zs. Sz\'{e}p and P. Sz\'{e}pfalusy,
\href{https://doi.org/10.1103/PhysRevD.71.125017}{Phys. Rev. {\bf D 71}, 125017 (2005)}.


\bibitem{Herpay:06}
T. Herpay and Zs. Sz\'{e}p, 
\href{https://doi.org/10.1103/PhysRevD.74.025008}{Phys. Rev. {\bf D 74}, 025008 (2006)}.


\bibitem{Herpay:07}
P. Kov\'acs and Zs. Sz\'{e}p,
\href{https://doi.org/10.1103/PhysRevD.75.025015}{Phys. Rev. {\bf D 75}, 025015 (2007)}.


\bibitem{Kovacs:2006ym}
P.~Kovacs and Zs. Szep,
Phys. Rev. {\bf D 75}, 025015 (2007).

\bibitem{kahara}
T. Kahara and K. Tuominen, \href{https://doi.org/10.1103/PhysRevD.78.034015}{Phys. Rev. {\bf D 78}, 034015
(2008)}; \href{https://doi.org/10.1103/PhysRevD.80.114022}{80, 114022 (2009)}; \href{https://doi.org/10.1103/PhysRevD.82.114026}{82, 114026 (2010)}.

\bibitem{Bowman:2008kc}
E.~S. Bowman and J.~I. Kapusta,
\href{https://doi.org/10.1103/PhysRevC.79.015202}{Phys. Rev. {\bf C 79}, 015202 (2009)};
J.~I. Kapusta, and E.~S. Bowman,
\href{https://doi.org/10.1016/j.nuclphysa.2009.10.118}{Nucl.\ Phys.\  {\bf A 830}, 721C (2009)}.

\bibitem{Fejos}
G. Fejos, A. Patkos,
\href{https://doi.org/10.1103/PhysRevD.82.045011}{Phys. Rev. {\bf D 82}, 045011 (2010)}.


\bibitem{Jakovac:2010uy}
A.~Jakovac and Zs. Szep, 
\href{https://doi.org/10.1103/PhysRevD.82.125038}{Phys. Rev. {\bf D 82}, 125038, (2010)}. 

\bibitem{koch}
L. Ferroni, V. Koch, and M. B. Pinto, 
\href{https://doi.org/10.1103/PhysRevC.82.055205}{Phys. Rev. {\bf C 82}, 055205 (2010)}.

\bibitem{marko}
G. Marko and Zs. Szep, \href{https://doi.org/10.1103/PhysRevD.82.065021}{Phys. Rev. {\bf D 82}, 065021 (2010)}.

\bibitem{scav}	
O. Scavenuius, A. Mocsy, I. N. Mishustin, and D. H. Rischke,
\href{https://doi.org/10.1103/PhysRevC.64.045202}{Phys. Rev. C {\bf 64}, 045202 (2001)}.


\bibitem{mocsy}A. Mocsy, I. N. Mishustin, and P. J. Ellis, 
\href{https://doi.org/10.1103/PhysRevC.70.015204}{Phys. Rev. {\bf C 70}, 015204 (2004)}.

\bibitem{bj}
B.-J. Schaefer and J. Wambach, 
\href{https://doi.org/10.1016/j.nuclphysa.2005.04.012}{Nucl. Phys. {\bf A 757}, 479 (2005)}. 

\bibitem{Schaefer:2006ds}
B.-J. Schaefer and J. Wambach,
\href{https://doi.org/10.1103/PhysRevD.75.085015}{Phys. Rev. {\bf D 75}, 085015 (2007)}.



\bibitem{rob} 
R. D. Pisarski and F. Wilczek, \href{https://doi.org/10.1103/PhysRevD.29.338}{Phys. Rev. D {\bf 29}, 338 (1984)}.

\bibitem{hjss}
A. Halasz, A. D. Jackson, R. E. Shrock, M. A. Stephanov,
and J. J. M. Verbaarschot, \href{https://doi.org/10.1103/PhysRevD.58.096007}{Phys. Rev. D {\bf 58}, 096007 (1998)}.

\bibitem{vac} 	
V. Skokov, B. Friman, E. Nakano, K. Redlich, and 
B.-J. Schaefer, 
\href{https://doi.org/10.1103/PhysRevD.82.034029}{Phys. Rev. D {\bf 82}, 034029 (2010)}.

\bibitem{lars}
R. Khan and L. T. Kyllingstad, 
\href{https://doi.org/10.1063/1.3575076}{AIP  Conf. Proc. {\bf 1343}, 504 (2011)}.

\bibitem{guptiw}
U. S. Gupta, V. K. Tiwari,
\href{https://doi.org/10.1103/PhysRevD.85.014010}{Phys. Rev. D {\bf 85},  014010 (2012)}.

\bibitem{schafwag12}
B.-J. Schaefer and M. Wagner, \href{https://doi.org/10.1103/PhysRevD.85.034027}{Phys. Rev. D {\bf 85}, 034027 (2012)}.

\bibitem{chatmoh1}
S. Chatterjee and K. A. Mohan, \href{https://doi.org/10.1103/PhysRevD.85.074018}{Phys. Rev. D {\bf 85}, 074018 (2012)}.


\bibitem{vkkr12}
V. K. Tiwari, \href{https://doi.org/10.1103/PhysRevD.86.094032}{Phys. Rev. D {\bf 86}, 094032 (2012)}.

\bibitem{TranAnd}
J. O. Andersen and A. Tranberg, \href{https://doi.org/10.1007/JHEP08(2016)045}{J. High Energy Phys. {\bf 08}
(2012) 002}.

\bibitem{chatmoh2}
S. Chatterjee and K. A. Mohan, \href{https://doi.org/10.1103/PhysRevD.86.114021}{Phys. Rev. D {\bf 86}, 114021 (2012)}.

\bibitem{vkkt13}
V. K. Tiwari, \href{https://doi.org/10.1103/PhysRevD.88.074017}{Phys. Rev. D {\bf 88}, 074017 (2013)}.
\bibitem{Herbst}
T. K. Herbst, J. M. Pawlowski, and B.-J. Schaefer, \href{https://doi.org/10.1103/PhysRevD.88.014007}{Phys.
Rev. D {\bf 88}, 014007 (2013)}.

\bibitem{Weyrich}
J. Weyrich, N. Strodthoff, and L. von Smekal, \href{https://doi.org/10.1103/PhysRevC.92.015214}{Phys. Rev. C
{\bf 92}, 015214 (2015)}.

\bibitem{kovacs}
P. Kovács, Zs Szép, Gy Wolf,\href{https://doi.org/10.1103/PhysRevD.93.114014}{Phys. Rev. D {\bf 93}, 114014 (2016)}.
\bibitem{zacchi1}
Andreas Zacchi and Jürgen Schaffner-Bielich, \href{https://doi.org/10.1103/PhysRevD.97.074011}{Phys. Rev. D {\bf 97}, 074011 (2018)}. 
\bibitem{zacchi2}
Andreas Zacchi and Jürgen Schaffner-Bielich, \href{https://doi.org/10.1103/PhysRevD.100.123024}{Phys. Rev. D {\bf 100}, 0123024 (2019)}. 
\bibitem{Rai}
S. K. Rai and V. K. Tiwari, \href{https://doi.org/10.1140/epjp/s13360-020-00851-5}{Eur. Phys. J. Plus {\bf 135:844}, (2020)}.
\bibitem{laine}
K. Kajantie, M. Laine, K. Rummukainen, and M. E. Shaposhnikov
\href{https://doi.org/10.1016/0550-3213(95)00549-8}{Nucl. Phys. B {\bf 458}, 90 (1996)}.
\bibitem{Adhiand1}
P. Adhikari, J. O. Andersen and P. Kneschke, \href{https://doi.org/10.1103/PhysRevD.95.036017}{Phys. Rev. D {\bf 95}, 036017 (2017)}.

\bibitem{BubaCar}
S. Carignano, M. Buballa and B-J Schaefer 
\href{https://doi.org/10.1103/PhysRevD.90.014033}{Phys. Rev. D {\bf 90}, 014033 (2014)}.
\bibitem{Naylor}
J. O. Andersen, W. R. Naylor, and A. Tranberg, \href{https://doi.org/10.1103/RevModPhys.88.025001}{Rev. Mod.
Phys. {\bf 88}, 025001 (2016)}.
\bibitem{fix1}
S. Carignano, M. Buballa, and W. Elkamhawy,
\href{https://doi.org/10.1103/PhysRevD.94.034023}{Phys. Rev. D {\bf 94}, 034023 (2016)}.
\bibitem{Adhiand2}
P. Adhikari, J. O. Andersen and P. Kneschke, \href{https://doi.org/10.1103/PhysRevD.96.016013}{Phys.Rev.D {\bf 96}, 016013 (2017)}.
\bibitem{Adhiand3}
P. Adhikari, J. O. Andersen and P. Kneschke, \href{https://doi.org/10.1103/PhysRevD.98.074016}{Phys.Rev.D {\bf 98}, 074016 (2018)}.
\bibitem{asmuAnd}
A.Folkestad, J. O. Andersen, \href{https://doi.org/10.1103/PhysRevD.99.054006}{Phys.Rev.D {\bf 99}, 054006 (2019)}. 
\bibitem{RaiTiw}
S. K. Rai and V. K. Tiwari, \href{https://doi.org/10.1103/PhysRevD.105.094010}{Phys.Rev.D {\bf 105}, 094010 (2022)}.
\bibitem{Polyakov:78plb}
A. M. Polyakov,
\href{https://doi.org/10.1016/0370-2693(78)90737-2}{Phys.\ Lett. {\bf B 72}, 477 (1978)}.

\bibitem{benji}
B. Svetitsky and L. G. Yaffe,
\href{https://doi.org/10.1016/0550-3213(82)90172-9}{Nucl. Phys. B {\bf 210}, 423 (1982)}.

\bibitem{BankUka}
T. Banks and A. Ukawa,  
\href{https://doi.org/10.1016/0550-3213(83)90016-0}{Nucl. Phys. B {\bf 225}, 145 (1983)}.

\bibitem{Pisarski:00prd}
R. D. Pisarski,
\href{https://doi.org/10.1103/PhysRevD.62.111501}{Phys. Rev. {\bf D 62}, 111501(R) (2000)}.

\bibitem{fuku}
K. Fukushima, 
\href{https://doi.org/10.1016/j.physletb.2004.04.027}{Phys. Lett. B {\bf 591}, 277 (2004)}.

\bibitem{Vkt:06}
B. Layek, A. P. Mishra, A. M. Srivastava and V. K. Tiwari, 
\href{https://doi.org/10.1103/PhysRevD.73.103514}{Phys. Rev. {\bf D 73}, 103514 (2006)}.

\bibitem{ratti}
C. Ratti, M. A. Thaler, and W. Weise, 
\href{https://doi.org/10.1103/PhysRevD.73.014019}{Phys. Rev. D {\bf 73},
014019 (2006)}.

\bibitem{Roesnr}
S. Roessner, C. Ratti and W. Weise, 
\href{https://doi.org/10.1103/PhysRevD.75.034007}{Phys. Rev. D {\bf 75},
034007 (2007)}.

\bibitem{fuku2}
K. Fukushima, \href{https://doi.org/10.1103/PhysRevD.78.114019}{Phys. Rev. D {\bf 78}, 114019 (2008)}.

\bibitem{SchaPQM2F}
B. J. Schaefer, J. M. Pawlowski, and J. Wambach, \href{https://doi.org/10.1103/PhysRevD.76.074023}{Phys.
Rev. {\bf D 76}, 074023 (2007)}.
\bibitem{SchaPQM3F}
B. J. Schaefer, M. Wagner, and J. Wambach, \href{https://doi.org/10.1103/PhysRevD.81.074013}{Phys. Rev. {\bf D 81},
074013 (2010)}.
\bibitem{Mao}
H. Mao, J. Jin, and M. Huang, \href{https://doi.org/10.1088/0954-3899/37/3/035001}{J. Phys. {\bf G 37}, 035001}.
\bibitem{TiPQM3F}
U. S. Gupta and V. K. Tiwari, \href{https://doi.org/10.1103/PhysRevD.81.054019}{Phys. Rev. {\bf D 81}, 054019
(2010)}.
\bibitem{Kapusta_Gale}
J. I. Kapusta and C. Gale, Finite Temperature Field Theory
Principles and Applications (Cambridge University Press,
Cambridge, England, 2006).
\bibitem{Schaefer:09wspax}
B. J. Schaefer, M. Wagner and J. Wambach,CPOD(2009)017,arXiv:0909.0289

    
\bibitem{Haas}
L. M. Haas, R. Stiele, J. Braun, J. M. Pawlowski and J. Schaffner-Bielich, \href{https://doi.org/10.1103/PhysRevD.87.076004}{Phys. Rev. {\bf D 87}, 076004 (2013)}.

\bibitem{Redlo}
P. M. Lo, B. Friman, O. Kaczmarek, K. Redlich, and C. Sasaki, \href{https://doi.org/10.1103/PhysRevD.88.074502}{Phys. Rev. {\bf D 88}, 074502 (2013)}.

\bibitem{TkHerbst}
T. K. Herbst, M. Mitter, J. M. Pawlowski, and B.-J. Schaefer and R. Stiele,    \href{https://doi.org/10.1016/j.physletb.2014.02.045}{Phys. Lett. B {\bf 731}, 248 (2014)}.

\bibitem{BielichP}
R. Stiele and J. Schaffner-Bielich, \href{https://doi.org/10.1103/PhysRevD.93.094014}{Phys. Rev. {\bf D 93}, 094014 (2016)}.

\bibitem{THerbst2}
T. K. Herbst, J. M. Pawlowski, and B.-J. Schaefer, \href{https://doi.org/10.1016/j.physletb.2010.12.003}{Phys. Lett. B {\bf 696}, 58 (2011)}.

\bibitem{vkkr22}
V. K. Tiwari, \href{https://doi.org/10.48550/arXiv.2211.11686}{arXiv.2211.11686}.

\bibitem{Zantow}
O.Kaczmarek and F.Zantow, \href{https://doi.org/10.1103/PhysRevD.71.114510}{Phys. Rev. {\bf D 71},114510(2005)}.

\bibitem{Wupertal2010}
S. Borsányi, Z. Fodor, C. Hoelbling, S. D. Katz, S. Krieg, C. Ratti, and K. K. Szabó, \href{https://doi.org/10.1007/JHEP09(2010)073}{J. High Energy Phys. 09 (2010) 73}.

\bibitem{Kobes}
R. Kobes, G. Kunstatter, and A. Rebhan, \href{https://doi.org/10.1103/PhysRevLett.64.2992}{Phys. Rev. Lett.
{\bf 64}}, 2992 (1990); \href{https://doi.org/10.1016/0550-3213(91)90300-M}{Nucl. Phys. B {\bf 355}, 1 (1991)}.
\bibitem{Rebhan}
A. K. Rebhan, \href{https://doi.org/10.1103/PhysRevD.48.R3967}{Phys. Rev.  D {\bf 48}, R3967 (1993)}.
\bibitem{SKGhosh2006}
S. K. Ghosh, T. K. Mukherjee, M. G. Mustafa, and R. Ray, \href{https://doi.org/10.1103/PhysRevD.73.114007}{Phys. Rev. D {\bf 73}, 114007 (2006)}.
\bibitem{bedanga}
B. Mohanty and J.-E. Alam, \href{https://doi.org/10.1103/PhysRevC.68.064903}{Phys. Rev. C {\bf 68}, 064903 (2003)}.
\bibitem{McLerran:2007npa}
L. McLerran and R. D. Pisarski \href{http://dx.doi.org/10.1016/j.nuclphysa.2007.08.013}{Nucl. Phys. A {\bf 796}, 83 (2007)}.
\bibitem{McLerran:2009npa}
L. McLerran, \href{http://dx.doi.org/10.1016/j.nuclphysa.2009.10.063}{Nucl. Phys. A {\bf 830}, 709c (2009)}.
\bibitem{Dutra:2013}
M. Dutra, O. Louren{\c{c}}o, A. Delfino, T. Frederico, and M. Malheiro \href{https://doi.org/10.1103/PhysRevD.88.114013}{Phys. Rev. D {\bf 88}, 114013 (2013)}.
\bibitem{WB2014}
S. Borsanyi, Z. Fodor, C. Hoelbling, S. D. Katz, S. Krieg and K. K. Szabo,\href{https://doi.org/10.1016/j.physletb.2014.01.007}{Phys. Lett. B 730 (2014) 99-104}.
\bibitem{HotQCD2014}
A. Bazavov et al.,
\href{https://doi.org/10.1103/PhysRevD.90.094503}{Phys. Rev. D {\bf 90}, 094503 (2014)}
\bibitem{raiti2023}
S. K. Rai and V. K. Tiwari,
\href{https://doi.org/10.48550/arXiv.2305.16180}{arXiv.2305.16180}.



\end{thebibliography}

\bibliographystyle{apsrmp4-1}

\end{document}